%% file: ms.tex
\documentclass[12pt,preprint]{aastex}

\newcommand{\blantonu}{1.60}
\newcommand{\blantong}{1.25}
\newcommand{\blantonr}{1.29}
\newcommand{\blantoni}{1.48}
\newcommand{\blantonz}{1.89}

\newcommand{\blantonue}{0.32}
\newcommand{\blantonge}{0.05}
\newcommand{\blantonre}{0.04}
\newcommand{\blantonie}{0.05}
\newcommand{\blantonze}{0.05}

\newcommand{\snuu}{0.183}
\newcommand{\snug}{0.235}
\newcommand{\snur}{0.227}
\newcommand{\snui}{0.197}
\newcommand{\snuz}{0.156}

\newcommand{\snuuUP}{+0.06}
\newcommand{\snugUP}{+0.07}
\newcommand{\snurUP}{+0.07}
\newcommand{\snuiUP}{+0.06}
\newcommand{\snuzUP}{+0.05}

\newcommand{\snuuLOW}{-0.05}
\newcommand{\snugLOW}{-0.06}
\newcommand{\snurLOW}{-0.06}
\newcommand{\snuiLOW}{-0.05}
\newcommand{\snuzLOW}{-0.04}

\newcommand{\avgnp}{\langle N(\pars) \rangle}
\newcommand{\pars}{{\bf p}}
\newcommand{\sns}{SDSS-II Supernova Survey}
\newcommand{\psne}{photometric SN candidates}
\newcommand{\psn}{photometric SN candidate}

\newcommand{\numonetwenty}{130}
\newcommand{\numonenineteen}{16}

\newcommand{\paina}{2.68}
\newcommand{\painb}{0.61}
\newcommand{\painc}{4.22}

\newcommand{\myomegamatter}{0.3}

\newcommand{\numsamples}{16}
\newcommand{\numsamplep}{1}
\newcommand{\rateeff}{0.77 \pm 0.01}

\newcommand{\vareff}{\epsilon }

\begin{document}

\title{ A Measurement of the Rate of type Ia Supernovae at Redshift  \\
$z\approx$ 0.1  from the First Season of the \sns
}


\email{bdilday@uchicago.edu}

\author{
Benjamin~Dilday,\altaffilmark{1,2}
Richard~Kessler,\altaffilmark{2,3}
Joshua~A.~Frieman,\altaffilmark{2,4,5}
Jon~Holtzman,\altaffilmark{6}
John~Marriner,\altaffilmark{5}
Gajus~Miknaitis,\altaffilmark{5}
Robert~C.~Nichol,\altaffilmark{7}
Roger~Romani,\altaffilmark{8}
Masao~Sako,\altaffilmark{8,9}
Bruce~Bassett,\altaffilmark{10,11}
Andrew~Becker,\altaffilmark{12}
David~Cinabro,\altaffilmark{13}
Fritz~DeJongh,\altaffilmark{5}
Darren~L.~Depoy,\altaffilmark{14}
Mamoru~Doi,\altaffilmark{15}
Peter~M.~Garnavich,\altaffilmark{16}
Craig~J.~Hogan,\altaffilmark{12}
Saurabh~Jha,\altaffilmark{8,17}
Kohki Konishi,\altaffilmark{18}
Hubert~Lampeitl,\altaffilmark{7,19}
Jennifer~L.~Marshall,\altaffilmark{14}
David~McGinnis,\altaffilmark{5}
Jose~Luis~Prieto,\altaffilmark{14}
Adam~G.~Riess,\altaffilmark{19,20}
Michael~W.~Richmond,\altaffilmark{21}
Donald~P.~Schneider,\altaffilmark{22}
Mathew~Smith,\altaffilmark{7}
Naohiro~Takanashi,\altaffilmark{15}
Kouichi~Tokita,\altaffilmark{15}
Kurt~van~der~Heyden,\altaffilmark{11,23}
Naoki~Yasuda,\altaffilmark{18}
Chen~Zheng,\altaffilmark{8}
John~Barentine,\altaffilmark{24,25}
Howard~Brewington,\altaffilmark{25}
Changsu~Choi,\altaffilmark{26}
Arlin~Crotts,\altaffilmark{27}
Jack~Dembicky,\altaffilmark{25}
Michael~Harvanek,\altaffilmark{25,28}
Myunshin~Im,\altaffilmark{26}
William~Ketzeback,\altaffilmark{25}
Scott~J.~Kleinman,\altaffilmark{25,29}
Jurek~Krzesi\'{n}ski,\altaffilmark{25,30}
Daniel~C.~Long,\altaffilmark{25}
Elena~Malanushenko,\altaffilmark{25}
Viktor~Malanushenko,\altaffilmark{25}
Russet~J.~McMillan,\altaffilmark{25}
Atsuko~Nitta,\altaffilmark{25,31}
Kaike~Pan,\altaffilmark{25}
Gabrelle~Saurage,\altaffilmark{25}
Stephanie~A.~Snedden,\altaffilmark{25}
Shannon~Watters,\altaffilmark{25}
J.~Craig~Wheeler,\altaffilmark{24}
and
Donald~York\altaffilmark{3,4}
}

\altaffiltext{1}{
Department of Physics, 
University of Chicago, Chicago, IL 60637.
}
\altaffiltext{2}{
  Kavli Institute for Cosmological Physics, 
   The University of Chicago, 5640 South Ellis Avenue Chicago, IL 60637.
}
\altaffiltext{3}{
Enrico Fermi Institute,
University of Chicago, 5640 South Ellis Avenue, Chicago, IL 60637.
}
\altaffiltext{4}{
  Department of Astronomy and Astrophysics,
   The University of Chicago, 5640 South Ellis Avenue, Chicago, IL 60637.
}
\altaffiltext{5}{
Center for Particle Astrophysics, 
  Fermi National Accelerator Laboratory, P.O. Box 500, Batavia, IL 60510.
}
\altaffiltext{6}{
  Department of Astronomy,
   MSC 4500,
   New Mexico State University, P.O. Box 30001, Las Cruces, NM 88003.
}

\altaffiltext{7}{
  Institute of Cosmology and Gravitation,
   Mercantile House,
   Hampshire Terrace, University of Portsmouth, Portsmouth PO1 2EG, UK.
}

\altaffiltext{8}{
 Kavli Institute for Particle Astrophysics \& Cosmology, 
  Stanford University, Stanford, CA 94305-4060.
}
\altaffiltext{9}{
Department of Physics and Astronomy,
University of Pennsylvania, 203 South 33rd Street, Philadelphia, PA  19104.
}
\altaffiltext{10}{
Department of Mathematics and Applied Mathematics,
University of Cape Town, Rondebosch 7701, South Africa.
}
\altaffiltext{11}{
  South African Astronomical Observatory,
   P.O. Box 9, Observatory 7935, South Africa.
}
\altaffiltext{12}{
  Department of Astronomy,
   University of Washington, Box 351580, Seattle, WA 98195.
}
\altaffiltext{13}{
Department of Physics, 
Wayne State University, Detroit, MI 48202.
}
\altaffiltext{14}{
  Department of Astronomy,
   Ohio State University, 140 West 18th Avenue, Columbus, OH 43210-1173.
}
\altaffiltext{15}{
  Institute of Astronomy, Graduate School of Science,
   University of Tokyo 2-21-1, Osawa, Mitaka, Tokyo 181-0015, Japan.
}
\altaffiltext{16}{
  University of Notre Dame, 225 Nieuwland Science, Notre Dame, IN 46556-5670.
}

\altaffiltext{17}{
Department of Physics and Astronomy, 
Rutgers University, 136 Frelinghuysen Road, Piscataway, NJ 08854.
}
\altaffiltext{18}{
Institute for Cosmic Ray Research,
University of Tokyo, 5-1-5, Kashiwanoha, Kashiwa, Chiba, 277-8582, Japan.
}
\altaffiltext{19}{
  Space Telescope Science Institute,
   3700 San Martin Drive, Baltimore, MD 21218.
}

\altaffiltext{20}{
Department of Physics and Astronomy,
Johns Hopkins University, 3400 North Charles Street, Baltimore, MD 21218.
}
\altaffiltext{21}{
  Physics Department,
   Rochester Institute of Technology,
   85 Lomb Memorial Drive, Rochester, NY 14623-5603.
}
\altaffiltext{22}{
  Department of Astronomy and Astrophysics,
   The Pennsylvania State University,
   525 Davey Laboratory, University Park, PA 16802.
}
\altaffiltext{22}{
Department of Physics and Astronomy,
University of Pennsylvania, 203 South 33rd Street, Philadelphia, PA  19104.
}
\altaffiltext{23}{
Department of Astronomy,
University of Cape Town, South Africa.
}

\altaffiltext{24}{
  Department of Astronomy,
   McDonald Observatory, University of Texas, Austin, TX 78712
}
\altaffiltext{25}{
  Apache Point Observatory, P.O. Box 59, Sunspot, NM 88349.
}
\altaffiltext{26}{
Department of Astronomy,
Seoul National University, Seoul, South Korea.
}
\altaffiltext{27}{
Department of Astronomy,
Columbia University, New York, NY 10027.
}
\altaffiltext{28}{
Lowell Observatory, 1400 Mars Hill Rd., Flagstaff, AZ 86001
}
\altaffiltext{29}{
Subaru Telescope, 650 N. A'Ohoku Place, Hilo, HI 96720
}
\altaffiltext{30}{
  Obserwatorium Astronomiczne na Suhorze,
   Akademia Pedagogicazna w Krakowie,
   ulica Podchor\c{a}\.{z}ych 2, PL-30-084 Krak\'{o}w, Poland.
}
\altaffiltext{31}{
Gemini Observatory, 670 North A'ohuoku Place, Hilo, HI 96720.
}

\begin{abstract}
We present a measurement of the rate of type Ia supernovae (SNe Ia) 
from the first of three seasons of data
from the \sns. For this measurement, we include 17 SNe Ia at 
redshift $z\le0.12$. Assuming a flat cosmology with
$\Omega_m = 0.3=1-\Omega_\Lambda$, we find a 
volumetric SN Ia rate of 
$[2.93^{+0.17}_{-0.04}({\rm systematic})^{+0.90}_{-0.71}({\rm statistical})]
\times 10^{-5}~{\rm SNe}~{\rm Mpc}^{-3}~h_{70}^3~{\rm year}^{-1}$,
at a volume-weighted mean redshift of 0.09. This result
is consistent with previous measurements of the SN Ia
rate in a similar redshift range.
The systematic errors are well controlled, resulting in the most precise 
measurement of the SN Ia rate in this redshift range. We use a maximum 
likelihood method to fit SN rate models to the \sns~data in 
combination with other rate measurements, thereby constraining 
models for the redshift-evolution of the SN Ia rate. 
Fitting the combined data to a simple 
power-law evolution of the volumetric SN Ia rate,
$r_V \propto (1+z)^{\beta}$, we obtain a value of 
$\beta = 1.5 \pm 0.6$, i.e. the SN Ia rate
is determined to be an increasing function of redshift
at the $\sim 2.5 \sigma$ level. 
Fitting the results to a model in which the volumetric SN rate, 
$r_V=A\rho(t)+B\dot \rho(t)$, where $\rho(t)$ is the
stellar mass density and $\dot \rho(t)$ is the star formation rate, we find
$A = (2.8 \pm 1.2) \times 10^{-14} ~\mathrm{SNe} ~\mathrm{M}_{\sun}^{-1} 
~\mathrm{year}^{-1}$,
$B = (9.3^{+3.4}_{-3.1})\times 10^{-4} ~\mathrm{SNe} ~\mathrm{M}_{\sun}^{-1}$.

\end{abstract}

\setcounter{footnote}{0}
\section{Introduction}
\label{sec:intro}
Type Ia supernovae (SNe Ia) 
have gained increasing attention from astronomers, 
primarily due to their remarkable utility as cosmological distance indicators.
There is now broad consensus that a type Ia supernova is the  
thermonuclear explosion of a Carbon-Oxygen white dwarf star that accretes 
mass from a binary companion until it reaches the 
Chandrasekhar mass limit (e.g.~\citet{Branch_95}). However, 
much remains to be learned
about the physics of SNe Ia, and there is active debate about both the
nature of the progenitor systems and the details of the explosion mechanism. 
For example, the binary companion may be 
a main-sequence star, a giant or sub-giant, or a second white dwarf.
The type of the companion star determines in part 
the predicted time delay between the formation of the
binary system and the SN event~\citep{Greggio_05}. The time delay can be 
constrained observationally 
by comparing the SN Ia rate as a function of redshift 
to the star formation history (SFH)~\citep{Strolger_04,Cappellaro_07}. 

The insight into the nature of the progenitor systems that SN Ia rate 
measurements provide can also potentially strengthen the utility of SNe Ia 
as cosmological distance indicators. Although the strong correlation between 
SN Ia peak luminosity and light curve decline rate was found purely 
empirically \citep{Psk_77,Phillips_93}, the physics underlying this 
relation has been extensively 
studied \citep{Hoeflich_95,Hoeflich_96,Kasan_07}. 
There is hope that improved physical understanding and modeling 
of SN Ia explosions, coupled with larger high-quality
observational data sets, will 
lead to 
improved distance estimates  
from SNe Ia. 
As part of this program, deeper
understanding of the nature of the progenitor systems can help narrow 
the range of initial conditions that need to be explored in carrying 
out the costly simulations of SN Ia explosions that in principle predict 
their photometric and spectroscopic properties.

Measurement of the SN Ia rate may also have a more direct impact on 
the determination of systematic errors in SN Ia distance estimates.
\citet{Mannucci_05,ScanBildsten,Neill_06} and \citet{Sullivan_06}  
have argued that a two-component model of the SN Ia rate, in which 
a prompt SN component follows the star formation rate and a second component
follows the total stellar mass,
is strongly favored over a single SN Ia channel. 
In this picture, since the cosmological star formation rate increases sharply with 
lookback time, the prompt component is expected to dominate the total SN Ia 
rate at high redshift.
\citet{Mannucci_05} and \citet{Howell_07} pointed out that this evolution with redshift 
can be a potential source of systematic error in SN Ia distance estimates, 
if the two populations have different properties.

In order to test such a model for the evolution of the SN Ia 
rate, improved measurements of the rate as a function of 
redshift and of host galaxy properties are needed. 
The Supernova Legacy Survey (SNLS) has recently presented 
the most precise measurement of the SN Ia rate at high redshift ($z \sim 0.5$)
based on 73 SNe~Ia \citep{Neill_06,Sullivan_06}.
At low redshifts ($z\sim 0.1$),
SN Ia rate measurements~\citep{Cappellaro_99,Hardin_00,Madgwick_03,Blanc_04}
have suffered from small sample sizes 
and also from systematic errors associated with 
heterogeneous samples \citep{Cappellaro_99}  
and with selection biases due to the targeting of   
known, relatively luminous galaxies \citep{Hardin_00,Blanc_04}. 
The low-redshift measurement of 
\citet{Madgwick_03}, based on SNe discovered fortuitously in SDSS galaxy 
spectra, is affected by different systematic uncertainties than traditional 
photometric searches, e.g., due to the finite 
aperture of the SDSS spectroscopic 
fibers.

In this paper, we present a new measurement of the 
SN Ia rate at low redshift, based 
upon the first season of data from the \sns. The \sns~\citep{Frieman_08} offers  
several advantages for this measurement. It covers a larger spatial volume than 
previous SN surveys, a result of the combination of intermediate-scale 
(2.5-m) telescope aperture, wide field of view (3 square degrees), 
modest effective 
sidereal exposure time (54 sec), and use of drift-scanning to efficiently 
cover a large sky area ($\sim 300$ square degrees). 
The \sns~is a rolling search, with new 
SNe discovered simultaneously with the follow-up of previously 
discovered SNe. 
Unlike SN
searches that target known galaxies,  
the \sns~is not biased against finding SNe in low-luminosity host galaxies. 
Well-calibrated 
photometry in the SDSS {\it ugriz} passbands \citep{Fukugita_96}, with 
a typical interval between observations of four days, yields well-sampled, 
multi-band light curves that enable photometric typing of SNe with high confidence.
Moreover, rapid on-mountain photometric reduction and image processing coupled with 
an extensive spectroscopic follow-up program enable  
spectroscopic confirmation 
of a very high fraction of the low-redshift SN Ia candidates. 

The \sns~was carried out over three three-month seasons, during 
Sept.-Nov. 2005-7. The results presented here are based on the Fall 2005 
season. The \sns~measures light curves for SNe Ia to redshift 
$z \simeq 0.4$, with a median redshift of 
$\langle z \rangle = 0.22$ 
for spectroscopically confirmed SNe Ia. However, in this first paper we limit the 
analysis to low redshift, $z \leq 0.12$, since our spectroscopic follow-up 
is essentially complete over this redshift range, and the uncertainty due 
to spectroscopically unobserved (and untyped) SNe is therefore negligible.
In presenting a SN Ia rate measurement, one must decide whether to include 
peculiar 
SNe Ia, i.e., events that are photometrically and/or spectroscopically 
unusual, since it is not clear that they are members of the same population 
as the ``normal'' SNe Ia. 
Formerly, the peculiar designation included events such as 1991T and 1991bg,
which are highly overluminous and underluminous events, respectively.
However, since these SNe appear to follow the standard peak-luminosity/decline-rate relation, 
they are now generally considered extreme members of the normal SN Ia population \citep{Nugent_95}. 
Other events, such as 2002ic \citep{Hamuy_03} and 2002cx \citep{Li_03}, exhibit 
more pronounced peculiarities and do not fit the luminosity-decline relation.
The first season  of the \sns~included two such truly peculiar 
events at low redshift, 2005hk~\citep{Phillips_07} and 
2005gj~\citep{Prieto_07, Aldering_06}. 
Although these peculiar events may arise from 
the same evolutionary path as normal SNe Ia, which would argue 
for including them in a SN Ia rate measurement, we have chosen to 
include only SNe with light curves that obey the 
standard brightness-decline relation. More specifically, we include 
in our rate measurement sample only SNe with light curves that 
are well described by the  
MLCS2k2 SN Ia light curve model~\citep{Riess_96,Jha_07}, see 
\S \ref{sec:ssample}. 
Regardless of the physical arguments surrounding peculiar events, we 
exclude them from this analysis 
primarily because we do not yet have a robust determination 
of our efficiency for detecting them. 

The rest of the paper is organized as follows. 
In \S\ref{sec:survey} we provide a brief outline of the survey observing 
strategy and operations as they relate to the rate determination.
In \S\ref{sec:sample} we define selection criteria and 
present the sample of SNe Ia used in this 
measurement, based on spectroscopic and photometric measurements. 
In \S\ref{sec:effs} we present estimates of the detection 
efficiency for low-redshift SNe Ia, based on artificial SNe  
inserted into the survey images
and on Monte Carlo simulations.
We present our measurement of the SN Ia rate 
and discuss the SN Ia rate as a function of host galaxy type
in \S\ref{sec:results}.
In \S\ref{sec:rate_models}
we compare our result to other SN Ia rate measurements
and combine rate measurements to fit semi-empirical models
of rate evolution.

\section{SDSS-II Supernova Survey Overview}
\label{sec:survey}

Here we briefly describe the {\sns}, highlighting the features
that are most relevant to a rate measurement. The survey is described in more
detail in~\citet{Frieman_08} and in \citet{Sako_08}. 
A technical summary of the SDSS is given by~\citet{York_00}, and
further details can be found 
in~\citet{Hogg_01,Ivezic_04,Lupton_99,Smith_02,Tucker_06}.

\subsection{Imaging}

The \sns~is carried out on the 2.5m telescope \citep{SDSS_telescope} at 
Apache Point Observatory (APO), using a wide-field CCD camera \citep{Gunn_98} 
operating in time-delay-and-integrate (TDI, or drift scan) mode. 
Observations are obtained nearly simultaneously in the SDSS 
$ugriz$ filter bands \citep{Fukugita_96}.

\begin{figure} [t] 
\begin{center}
\plotone{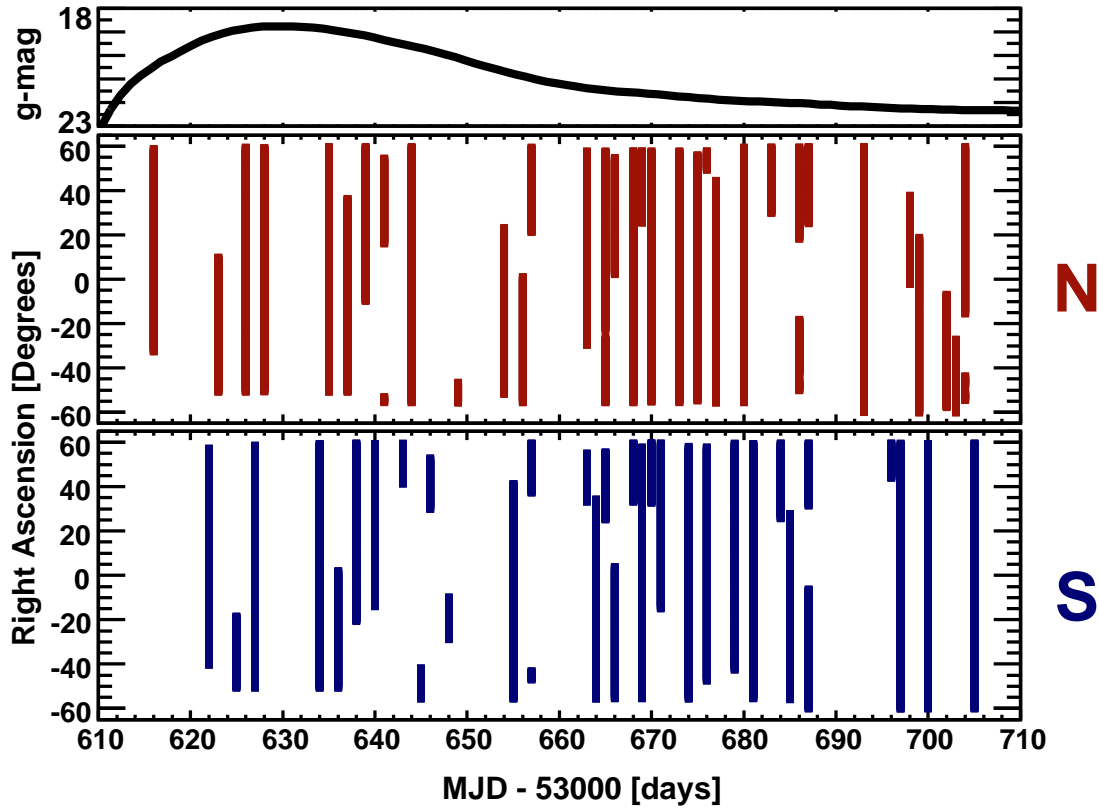}
\end{center}
\caption{
  Right ascension range covered by \sns~imaging runs vs.~epoch.
  The panels labelled N \& S denote the Northern and Southern 
  strips of stripe 82. 
  The regions $\alpha_{J2000} <-51^{\circ}$ and $\alpha_{J2000}>57^{\circ}$
  are not covered early in the season, and these regions are suppressed
  from the rate measurement.
  The top panel shows an example unextincted g-band light curve 
  for a SN Ia at a redshift of 0.12, based on the MLCS2k2 model. 
 }
\label{fig:racov}
\end{figure}

The \sns~covers a region, designated stripe 82, centered on the 
celestial equator in the Southern Galactic hemisphere, bounded by 
$-60^{\circ} < \alpha_{J2000} < 60^{\circ}$, and 
$-1.258^{\circ} < \delta_{J2000} < 1.258^{\circ}$.
Stripe 82 has been imaged multiple times in photometric conditions 
by the SDSS-I survey; co-added images from those runs 
provide deep template images and veto catalogs of variable objects 
for the \sns~transient search.
Due to gaps between the CCD columns on the camera, 
each stripe is divided into northern (N) and southern (S) {\it strips}; 
the \sns~alternates between the N and S strips on subsequent nights. 
Each strip encompasses $\sim 162$ square degrees of sky, 
with a small overlap between them, so that the 
survey covers $\sim 300$ square degrees. 
On average each part of the survey region 
was observed once every four nights during the 2005 season.
Figure~\ref{fig:racov} shows the sky coverage versus survey time,
along with a representative SN~Ia light curve.
\clearpage
\clearpage

\subsection{Supernova Search Pipeline}
\label{sec:search}
There are five main components to the supernova search pipeline: photometric 
reduction, image subtraction, automated object selection, visual inspection, 
and light curve fitting for spectroscopic target selection. 
We describe them briefly in turn.
For a full night of imaging data, the entire pipeline runs in approximately 
20 hours, sufficient for keeping up with the data flow and for rapid 
spectroscopic targeting. 

\roman{enumi}
\roman{enumii}

\renewcommand{\labelenumi}{\roman{enumi}.}

In the first stage of the search pipeline, 
the imaging data is acquired from the camera and processed through the  
the SDSS photometric reduction pipeline, 
known as {\tt PHOTO}~\citep{Lupton_01}.
{\tt PHOTO} produces ``corrected'' images that are
astrometrically calibrated \citep{Pier_03} and provides  
a local estimate of the point spread function (PSF). 
In the second stage, images are processed through the 
difference imaging pipeline.
To run the search pipeline to completion in less than a day
with the available on-mountain computing resources,
only the corrected {\it gri} images are processed beyond the first stage.
The search image is astrometrically and photometrically 
registered to the template image, and the
template image is convolved 
with a kernel chosen to minimize subtraction residuals~\citep{Alard_98}.  
A difference image is then obtained by 
subtracting the convolved template image from the survey image. 
Peaks are detected in the difference image using 
the {\tt DoPHOT} photometry and object detection package~\citep{Schecter_93}.
The signal-to-noise threshold for object 
detection is at $\sim 3.5$, corresponding in typical conditions to
$g \sim 23.2$, $r \sim 22.8$, and $i \sim 22.5$. 
The typical magnitudes at signal-to-noise of 10 for point-like objects  
are $g \sim 21.8$, $r \sim 21.5$, and $i \sim 21.2$.

The third stage of the SN search pipeline comprises a sequence of 
automated filtering operations that select events of potential 
interest from among those detected in the difference images.
We require a close positional match in at least two of the
$gri$ images, which removes cosmic rays, single-band spurious 
noise fluctuations, and a large fraction of asteroids and other rapidly 
moving objects detected by the survey. 
All detections that satisfy these criteria are entered 
into a {\tt MySQL} database 
and are referred to as {\it objects}. 
To reject active galactic nuclei (AGN) and variable stars,
we veto any detection occurring at the position
of a previously cataloged variable, using observations of 
stripe 82 from several previous years. The area corresponding to
previously cataloged variable objects 
represents $\sim 1\%$ of the total survey area.

In addition to SNe, the database of detected objects includes 
a variety of physical and non-physical transients. 
Physical sources include slow-moving asteroids that were
not rejected by the moving object veto, AGN 
and variable stars not 
already cataloged, and high proper-motion stars. Non-physical
sources include improperly masked diffraction spikes from bright stars
and artifacts of imperfect image registration. 
To remove non-physical sources, cut-out images of all objects 
that remain after the automated filtering are visually inspected 
and classified in the fourth stage of the search pipeline. 
Objects visually classified as consistent with a possible SN event 
are flagged for further analysis and are denoted {\it candidates}. Subsequent 
object detections in difference images at the same position are 
automatically associated with the same candidate. 

In the fifth and final stage of processing for the SN search, 
the $gri$ light curve for 
each SN candidate is fit to models of type-Ia, type-Ib/c and type-II SNe.
The non-Ia SN models consist of template light curves constructed
from photometric measurements of individual SNe provided by the 
SUSPECT database\footnote{http://bruford.nhn.ou.edu/\~suspect/index1.html},
coupled with the corresponding SN spectral model provided 
by~\citet{Nugent_02}.
For the SN Ia model, 
a stretch and a wavelength-dependent scale factor is applied to a 
fiducial bolometric light curve 
in a way designed to reproduce the $\Delta m_{15}$ parameterization of 
the peak-luminosity/decline-rate relation~\citep{Hamuy_96a}. 
The time of maximum, $\Delta m_{15}$, redshift, 
and extinction parameter $A_V$ (magnitudes of extinction
in the V-band) are fit parameters that are 
searched on a grid for the set of values that produce the minimum 
value of the $\chi^2$
statistic. 
For some candidates, we additionally carry out difference imaging 
in the $u$ and $z$ passbands in order to better distinguish Type~II 
and Type Ia SNe that tend to have a significantly different 
$u-g$ color at early epochs.
To further constrain the early light curve shape, we carry out 
forced-positional photometry  on difference images 
at the position of the candidate in pre-discovery images.
The relative goodness of fit of candidate $gri$ light curves 
to SNe~Ia and core-collapse SNe models is used as a factor 
in prioritizing spectroscopic follow-up. 
In particular, all SN~Ia candidates 
found before peak and with estimated current $r$-band magnitude $\lesssim 20$ 
are placed on the spectroscopic target list, and our follow-up 
observations are nearly complete out to that magnitude. 
Since the typical peak magnitude for a 
SN~Ia with no extinction at redshift $z=0.1$ is $r \simeq 19.3$, 
we might expect that the spectroscopic SN Ia sample should be 
essentially complete out to roughly this redshift as well; 
we shall see later that this is the case.
This photometric pre-selection of SNe Ia has proven very effective: 
approximately 90\% of the candidates initially targeted as SNe Ia after 
two or more epochs of imaging have resulted in a SN~Ia spectroscopic
confirmation.
The~\sns~photometric classification and spectroscopic target selection
are discussed in full detail in~\citet{Sako_08}.

\subsection{Artificial Supernovae}
\label{sec:fakes}

To measure the SN rate, it is clearly important to understand the
efficiency of the survey for discovering SNe. As part of normal 
survey operations, we insert artificial SNe Ia (hereafter fakes) 
directly into the corrected survey images after the 
photometric reduction ({\tt PHOTO}) but before difference imaging. 
The primary motivation for inserting fakes into the data stream is 
to provide real-time monitoring of the performance of the survey 
software pipeline and of the human scanning of objects.
The fakes provide quantitative information about the efficiency of the 
survey software, human scanning, and the photometric classification of 
SNe Ia that is useful in the rate determination.
Here we describe the basic algorithm for generating fakes and 
inserting them into the data stream; for 
more details, see \citet{Sako_08}.

A fake is a pixel-level simulation of a point source with a light curve 
chosen to closely represent that of a real SN Ia. 
At each epoch for which the fake has a chance of being detected, 
the calculated CCD signal for the fake is directly added to the 
survey image. For the 2005 observing season, we generated a library 
of 874 fake light curves:  each fake light curve is assigned a position, 
redshift, date of peak luminosity (in V-band), 
and an intrinsic luminosity that correlates  with decline rate. 
This resulted in $\sim 7,800$ fake epochs during the season.
The redshift distribution for the fakes was generated by assuming 
that the number of SNe Ia is roughly proportional to the volume element, 
$(dN/dz) \propto z^2$, in the range $0.0 < z < 0.4$.

To model the effect of contamination from host galaxy light on the 
detection efficiency, each fake is placed near a galaxy selected 
from the photometric redshift catalog \citep{Oyaizu_07} 
for SDSS imaging on stripe 82. A host galaxy is drawn at random,
from a distribution proportional to the $r$-band luminosity, 
from galaxies which have a photometric redshift within $\sim 0.01$ 
of the redshift assigned to the fake.  

The SN Ia light curve model used to generate $ugriz$ magnitudes for a fake 
at each epoch is the same model that is used for early light curve fitting 
and photometric typing on the imaging data, but with the light curve 
parameters now chosen from an input probability distribution. 
To generate a point-source image from 
the ideal magnitudes, we use the estimate of the PSF from {\tt PHOTO} at the 
position of the fake at the given epoch. 
We obtain the conversion from magnitudes to 
instrumental units (analog-to-digital units, or ADU)
by running the {\tt DoPHOT} photometry package on a set of 
cataloged stars in the 
survey image for which 
the magnitudes have been previously measured by the SDSS. 
After scaling the PSF model  
to match the computed ADU flux, we add Poisson fluctuations to each pixel.
Finally, the row and column in the field that correspond to the position 
of the fake are taken from the astrometric solution provided by the 
imaging pipeline, and the fake is overlaid on the survey image. 

When a fake is detected in the difference images, its identity 
as a fake is kept hidden while it is scanned by humans.
After scanning, the fakes are revealed so that they are not 
mistakenly targeted for spectroscopic follow-up and so that 
the efficiency of scanners in tagging fakes as SN candidates can 
be monitored.
However, like all candidates, fakes are processed through the 
automated light curve fitter/photometric typing algorithm
so that we can test if they are accurately typed as SNe Ia after a few 
photometric epochs. 
The use of the fakes for measuring 
the survey detection efficiency is discussed in \S~\ref{sec:fakeeffs}.

\subsection{Spectroscopy}
\label{sec:spec}

The classification of SNe is defined by their spectroscopic features. 
In addition, spectroscopy provides a precise redshift determination and, 
in a number of cases, host galaxy spectroscopic-type information.
Spectroscopic follow-up of the \sns~candidates is being 
undertaken by a number of telescopes.
During the 2005 observing season, spectroscopic observations 
were provided by 
the Hobby-Eberly 9.2m at McDonald Observatory, 
the Astrophysical Research Consortium 3.5m at Apache Point Observatory,
the William-Herschel 4.2m, 
the Hiltner 2.4m at the MDM Observatory, 
the Subaru 8.2m and Keck 10m on Mauna Kea, 
and the SALT 11m at the South African Astronomical Observatory.

The classification of SN spectra is performed by comparing the spectral 
data to normal and peculiar supernova spectral templates from the 
work of~\citet{Nugent_02} 
and to a public library of well-measured 
supernova spectra \citep{Matheson_05, Blondin_07}.  
The SN typing in this work is based on visual inspection of the spectra, 
but was guided by applying the cross-correlation 
technique of~\citet{Tonry_79} to the spectrum and the template library.  
The visual inspection relies heavily on the characteristic SN Ia 
features of Si and S absorption, which are usually prominent at 
optical wavelengths for 
this redshift range.

The redshift determination is based on galaxy features when they are 
present; otherwise SN features are used.
In some cases, particularly at low redshift, 
a high-quality spectrum of the SN host galaxy is available from the 
SDSS-I spectroscopic survey. Comparison with those spectra indicate that 
our follow-up spectroscopic redshifts 
are determined to an accuracy of $\sim 0.0005$ when galaxy 
features are used and $\sim 0.005$ when SN features are used.
Further details of the \sns~first-season 
spectroscopic analysis are presented in~\citet{Zheng_08}.

\subsection{Final Photometry}
\label{sec:smp}

To obtain more precise SN photometry than the on-mountain
difference imaging pipeline provides, we re-process the imaging data for all 
spectroscopically confirmed and other interesting SN candidates through 
a final photometry pipeline~\citep{Holtzman_08}. 
In this ``scene-modeling photometry'' (SMP) pipeline, the supernova 
and the host galaxy (the {\it scene}) are modeled respectively 
as a time-varying point-source and a background that is constant in time, 
both convolved with a time-varying PSF. 
This model is constrained by jointly fitting all available images at the 
SN position, including images well before and after the SN explosion.
Since there is no spatial resampling or convolution of the images 
that would correlate neighboring pixels, the error on the flux can be 
robustly determined. The SMP pipeline often provides photometric measurements 
at additional epochs compared to the survey operations pipeline.
The final analysis of SN light curves discussed in this paper is based on SMP; 
in particular, the selection cuts described in \S \ref{sec:sample} are 
made using the SMP pipeline.

\section{Defining the SN Ia Sample for the Rate Measurement}
\label{sec:sample}

The SN~Ia sample for the rate measurement must include
all SNe Ia in the redshift range of interest, not just those for which we have
a confirming spectrum. Although we have 
high efficiency for discovering and 
spectroscopically confirming low-redshift SNe~Ia
(\S \ref{sec:ssample}),
we can take advantage of our rolling search data to carry out an 
extensive post-season hunt for SNe~Ia that may have been missed by the search 
pipeline 
during the survey season (\S \ref{sec:psample}).

\subsection{Spectroscopic SN Sample}
\label{sec:ssample}

In its first season (Fall 2005), the \sns~discovered \numonetwenty\ events
with secure spectroscopic identifications as 
SNe~Ia\footnote{The classification of 2005gj as 
     a SN Ia may be controversial~\citep{Prieto_07}; 
     as noted in \S \ref{sec:intro}, we exclude 
     it from this analysis.} 
and \numonenineteen\ events that are considered probable SNe~Ia 
based on their spectra. 
For SN Ia events satisfying the selection criteria below, 
the spectroscopic follow-up is essentially complete for redshifts 
$z \le 0.12$, so we have chosen to focus on this
redshift range for a first measurement of the SN rate.
For $z \le 0.12$, the sample contains 27 spectroscopically 
confirmed SNe Ia and 2 spectroscopically probable SNe Ia before 
making selection cuts.

For the measurement of the SN Ia rate, 
we impose a number of selection criteria on the SN photometric data, 
with the aim of producing a sample that has a well-characterized 
selection function. 
These criteria are applied to the spectroscopically 
confirmed and probable SNe~Ia with $z\le0.12$. 
For consistency, we will also apply these selection cuts 
to the photometric (i.e., spectroscopically unconfirmed) SN sample 
discussed in \S \ref{sec:psample}. 
The selection cuts for the rate measurement are as follows:

\renewcommand{\labelenumi}{\arabic{enumi}.}
\begin{enumerate}

\item $-51^{\circ} < \alpha_{J2000} < 57^{\circ}$.

Although the \sns~covers the RA range $-60^{\circ}<\alpha_{J2000}<60^{\circ}$, 
early in the Fall 2005 observing season we did not have complete templates
available for the regions $\alpha_{J2000} <-51^{\circ}$ and 
$\alpha_{J2000} > 57^{\circ}$, 
so these RA regions were not initially used for the SN search, 
as shown in Fig. \ref{fig:racov}. 
In principle, we could account for this by modeling the 
time-varying effective search area, but for simplicity we choose 
to excise these RA regions 
from the rate measurement. Furthermore, the calibration
star catalog used by our final photometry pipeline \citep{Ivezic_07}
does not extend below $\alpha_{J2000} \sim -51^{\circ}$, and
we cannot presently simulate light curves for SNe in this region.
This cut removes one confirmed SN~Ia, 2005iu, from
the rate sample.

\item There are photometric observations on at least five separate 
epochs between $-20$ days and $+60$ days 
relative to peak light in the SN rest-frame.

Peak light refers to the date of maximum luminosity in the 
SN rest-frame B-band according to the best-fit MLCS2k2 light curve model. 
This cut requires that the light curve is reasonably well-sampled and 
it is primarily useful for photometrically distinguishing SNe~Ia from other 
SN types with high confidence when there is no SN spectrum available 
(see \S \ref{sec:psample}). Here and below,  
a photometric observation simply means that the survey 
took imaging data at that epoch on that region of sky and 
that SMP reported a SN flux measurement 
(not necessarily significant or even positive) with 
no error flags (see \citet{Holtzman_08}) 
in at least one of the three $gri$ passbands. It 
does {\it not} imply a detection above some signal-to-noise 
threshold. One SN discovered late in the observing season, 
2005lk, fails this cut.

\item At least one epoch with signal-to-noise ratio $> 5$ in each of 
$g$, $r$, and $i$ (not necessarily the same epoch in each passband). 

This cut ensures that there are well-measured points on the light curve,
and is mainly useful for rejecting low signal-to-noise events from 
the photometric sample. 
All spectroscopically confirmed SNe~Ia in the low-redshift 
sample satisfy this cut.

\item At least one photometric observation at least two days 
{\it before} peak light in the SN rest frame.

\item At least one photometric observation at least ten days 
{\it after} peak light in the SN rest frame.

These two cuts require sampling of the light curve 
before and after peak light, ensuring that we have 
a precise determination of the time of peak light. 
These cuts also help remove non-SN Ia contaminants from the 
photometric sample (see \S \ref{sec:psample}). Finally, they  
guarantee that the epoch of peak light occurs during our observing season, 
i.e., between Sept. 1 and Nov. 30, which is one of the criteria 
used in defining the rate measurement in \S \ref{sec:results}. 
Since these cuts are {\it more} restrictive than the requirement 
of peak light during the observing season, they are the main 
contributors to the inefficiency estimated in \S \ref{sec:simeffs}. 
These are the most restrictive 
cuts on the spectroscopic SN~Ia sample, together removing nine events: 
four SNe Ia discovered early in the observing season do not have 
a pre-maximum observation, and five SNe Ia found late in the season 
do not have a photometric observation more than ten days past peak light.

\item MLCS2k2 light curve fit probability $> 0.01$.

The MLCS2k2 light curve fitter \citep{Riess_96,Jha_07} takes 
as input the measured SN magnitudes in each passband at each epoch, 
and the measured SN redshift; it then finds the likelihood as a 
function of the four parameters $\mu$ (the distance modulus), 
$A_V$ (the extinction parameter), 
the time of peak light in rest-frame B-band, 
and the light curve shape/luminosity parameter $\Delta$.  
The MLCS2k2 fit probability is defined by evaluating the usual 
$\chi^2$
statistic for the data and the best fitting 
MLCS model and assuming that this statistic obeys a 
$\chi^2_{n-4}$ probability distribution, 
where $n$ is the number of photometric data points.
The model parameters of the best fitting MLCS2k2 model 
are defined as the mean of the probability distribution
for each corresponding parameter.
This cut on the fit probability provides an automated   
method of removing photometrically peculiar SNe Ia from the sample. 
We find that essentially all of the spectroscopically normal
SNe Ia in our confirmed sample have a fit probability $> 0.1$. However, the
spectroscopically confirmed SN Ia sample is likely to be biased
toward ``high-quality'' light curves, so we place the 
selection cut at a less restrictive value.
Three spectroscopically identified SNe~Ia are rejected by this cut, 
including the peculiar SNe 2005hk ($\chi^2/\mathrm{d.f.}=90/21$) and 
2005gj ($\chi^2/\mathrm{d.f.}=198/45$). The third rejected SN, 
with internal SDSS candidate designation 6968 ($\chi^2/\mathrm{d.f.}=78/27$), 
was classified as a spectroscopically 
probable SN Ia (see \S \ref{sec:spec}) and shows some evidence of 
being spectroscopically similar to 2005hk. 
For the sample of \psne~(\S \ref{sec:psample}), this cut helps 
remove non-SN astrophysical variables, such as AGN and M stars.

\item MLCS2k2 light curve fit parameter $\Delta > -0.4$.

The MLCS parameter $\Delta$ is a measure of the light curve 
shape and intrinsic luminosity. Smaller values of the
$\Delta$ parameter correspond to more slowly-declining,
intrinsically brighter SNe Ia.
This cut requires that $\Delta$ be consistent with
the values observed for 
the low-redshift SNe~Ia that were used to train the 
MLCS2k2 light curve fitter.
For the \psne, this cut helps reject Type~II supernovae, 
which often have a long plateau after the 
epoch of peak luminosity and result in large negative 
fitted values of $\Delta$.

\end{enumerate}

The above selection requirements result in the 
\numsamples~spectroscopically identified SNe~Ia that are
listed in Table~\ref{tab:table1-sn}. 
This sample includes 2005je, which was 
classified as spectroscopically probable.
The spectroscopically confirmed SNe~Ia that are removed from the  
rate-measurement sample 
are listed in Table~\ref{tab:sne_cut}; the last column indicates 
which of the above selection criteria was used to remove each SN.

\clearpage
\input{tab1.tex}
\input{tab2.tex}
\clearpage

\subsection{Photometric SN Sample}
\label{sec:psample}

In addition to the spectroscopically identified SNe Ia 
discussed above, the survey has measured light-curves for  
a few thousand variable objects, 
including possible SNe, for which we did not obtain a classifiable 
spectrum while the source was bright enough 
to identify. We refer
to these spectroscopically unobserved or unclassified 
objects as {\it \psne}. There are a number of reasons for such spectroscopic 
incompleteness, including limited spectroscopic resources, 
targeting errors (e.g., misplacement of a spectroscopic slit), 
poor weather either preventing spectroscopic observations or 
rendering them indeterminate, and possible inefficiencies in the 
spectroscopic target selection algorithm. In order to make a reliable SN Ia 
rate measurement, we must investigate the \psne~to determine the level of 
incompleteness, if any, of the spectroscopic SN~Ia sample. 
This is a challenge, because a sample of purely photometric SN~Ia 
candidates may be heavily contaminated by objects that are not SNe~Ia, 
especially if there are significant numbers of objects with 
multi-band light curves that are not too dissimilar from those of SNe~Ia.  
The combination of selection cuts listed in \S \ref{sec:ssample} 
is designed to meet this challenge, 
by rejecting the majority of non-SN Ia contaminants. 
In addition to the spectroscopically confirmed and spectroscopically
probable SNe Ia discussed in \S 3.1, the \sns~discovered 16 low-redshift
SNe that were spectroscopically confirmed as non-Ia SNe in it's first year.
As a check that our selection cuts are effective at rejecting non-Ia SNe,
we apply the same cuts to this sample of 
16 low-redshift $(z < 0.2)$ spectroscopically confirmed non-Ia SNe.
All but one of these non-Ia SNe are rejected by these selection cuts.
The selection criteria above could be made more restrictive in order to
reduce potential
non-Ia contamination of the photometric SN sample. For example, by requiring
a photometric observation at least 16 (as opposed to 10) days after peak
light
in the SN rest frame, the spectroscopic non-Ia SN above would be eliminated
from the sample. However, we find that such a change would have no impact on
the selection of photometric SN candidates for inclusion in the rate sample.

To determine whether any of the \psne~are genuine SNe~Ia in the 
redshift range of this rate measurement, we must estimate both the SN 
type and redshift for each candidate. 
There are two categories of \psne, 
(a) those for which we have a precise spectroscopic measurement 
of the redshift and 
(b) those for which we do not. 
The redshifts for category (a) candidates come from two sources. 
The first source is the SDSS-I spectroscopic galaxy survey, 
which measured redshifts for $\sim 28,000$ galaxies in our 
survey region at redshifts $z \le 0.12$. 
The second source is from subsequent spectroscopic 
observations of $\sim 80$ host galaxies 
of the highest-quality \psne; these spectra were obtained
in the summer and fall of 2006 and 2007.
Using the sample selection process described below 
in \S \ref{sec:selpsne}, 
we found in our imaging data only one photometric SN~Ia candidate 
that passes the selection criteria in \S \ref{sec:ssample} 
and that has a spectroscopic redshift $z\le0.12$ (category (a)). 
The host galaxy of this SN Ia candidate, which has internal SDSS SN 
designation 9266, has a spectroscopic redshift of $z=0.0361$ 
measured by the SDSS galaxy redshift survey.
This object was not targeted for spectroscopic follow-up during 
the \sns~because it has very high extinction, $A_V \simeq 4$ 
according to the MLCS2k2 fit. This extinction 
value lies outside the range of the $A_V$-search grid for 
the photometric typing algorithm used during the search (\S \ref{sec:search}). 
 
\subsubsection{Redshift estimation for \psne}
\label{sec:photoz}

For each \psn~{\it without} a spectroscopically determined redshift (category 
(b) above), 
we must estimate both the redshift and the SN type from the 
photometric data. We do this using a modification of the standard 
MLCS2k2 light curve fit, in which the redshift 
is included as a parameter in the likelihood function. 
In this instance, the distance modulus $\mu$ is {\it not} treated as a fit 
parameter; instead, we adopt the concordance LCDM cosmology, with 
$\Omega_{m} = 0.3$, $\Omega_{\Lambda} = 0.7$, and dark energy equation 
of state parameter $w=-1$, and fix $\mu(z)$ to its functional form 
for that cosmology. The photometric redshift estimate $z_{phot}$ 
is then obtained by marginalizing over the 
other fit parameters, i.e., the epoch of peak luminosity, 
the extinction $A_V$, and the shape parameter $\Delta$.

Although these SN photometric redshift estimates depend on the 
assumed cosmology, we do not expect them to be extremely 
sensitive to the values of the cosmological parameters, especially 
at the modest redshifts under consideration here.
To test the accuracy of these redshift estimates, we applied the 
MLCS2k2 redshift fit to SNe~Ia light curves that have  
spectroscopically measured redshifts. For this test, we use 
two sets of objects: (i) all spectroscopically confirmed 
SNe Ia with redshift $z \le 0.25$, and (ii) all photometric SN~Ia candidates 
that satisfy the selection criteria in \S \ref{sec:ssample} and that 
have spectroscopic (host galaxy) redshift $z \le 0.25$. 
We include objects in category (ii) because
the spectroscopically confirmed SNe~Ia could represent 
a biased sample of the SN Ia population if there are spectroscopic 
selection biases.
We include objects with redshift $z > 0.12$ to yield a more 
statistically significant test and to check for photometric 
redshift biases that could cause these objects to be erroneously 
included in the $z \leq 0.12$ sample.
There are 61 and 28 events in categories (i) and (ii), respectively.

\clearpage
\begin{figure} [t]
\begin{center}
\plotone{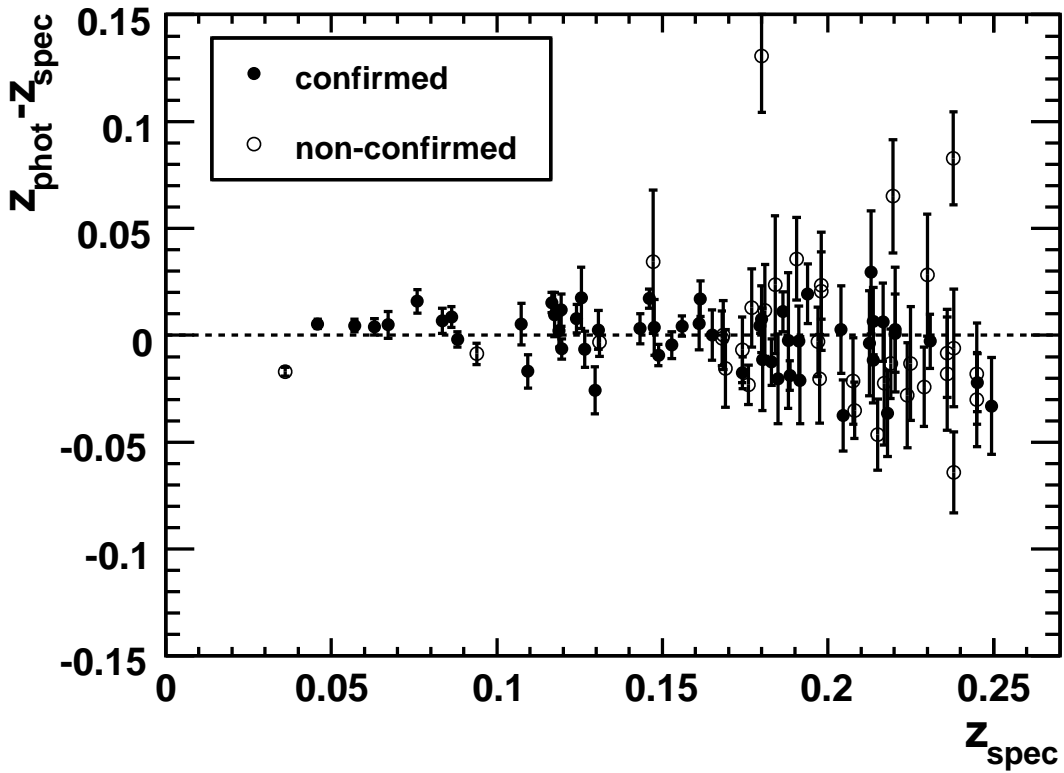}
\end{center}
\caption{
Residuals of the photometric redshift estimates, $z_{phot}-z_{spec}$ 
vs.~$z_{spec}$, for the  
sample of spectroscopically confirmed SNe Ia (black points) 
and for the photometric SN Ia candidates that satisfy the 
rate selection cuts and for which host galaxy redshifts 
are available (open points). 
The marginalized redshift errors reported by the MLCS2k2 
light-curve fits are shown. 
}
\label{fig:photozscatter}
\end{figure}

\begin{figure} [!h]
\begin{center}
\plotone{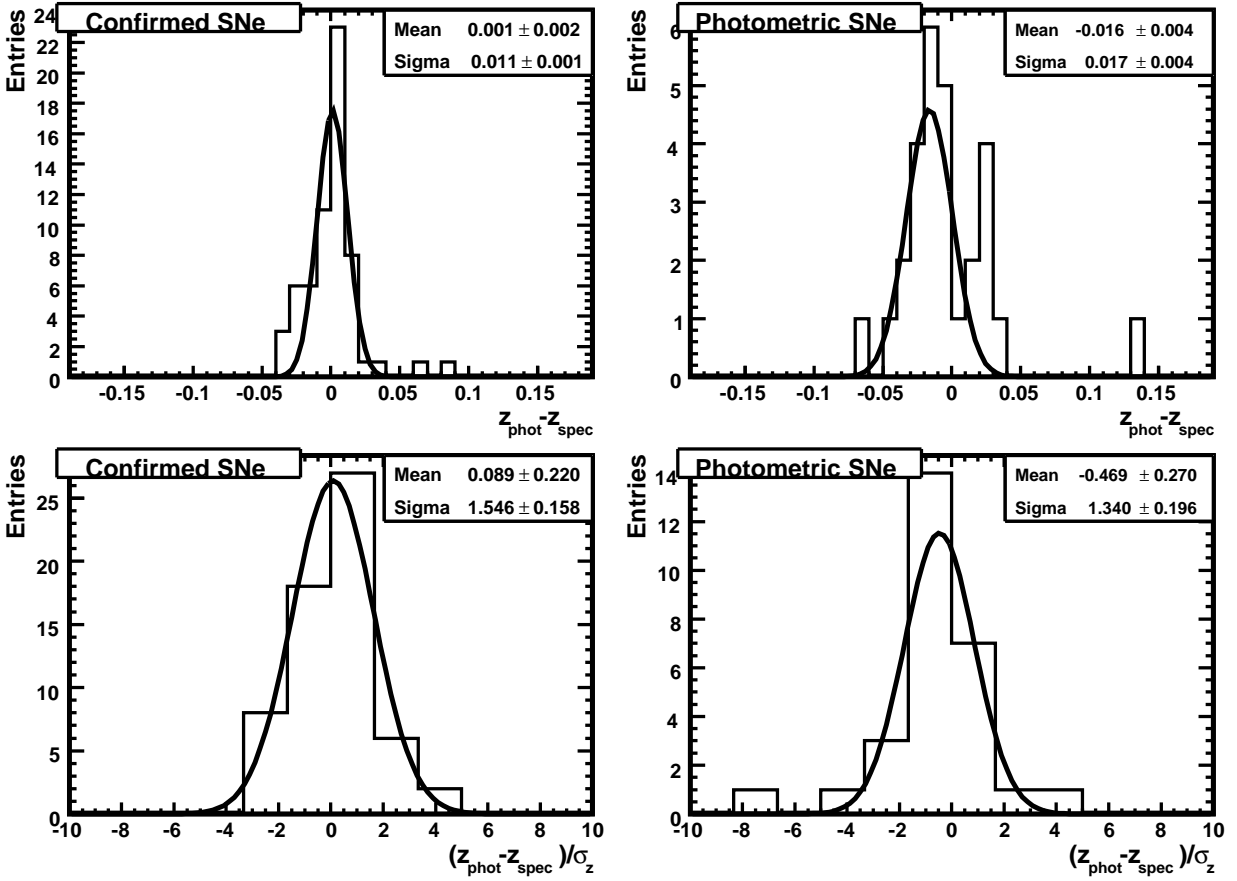}
\end{center}
\caption{
The distribution of photometric redshift residuals
for the spectroscopically confirmed SNe Ia (left panels) 
and the photometric SN Ia candidates (right panels) shown 
in Fig. \ref{fig:photozscatter}. Upper panels show distributions
of the difference between the photometric redshift, $z_{phot}$, 
and the spectroscopic redshift, $z_{spec}$; 
lower panels show distributions of 
$(z_{phot}-z_{spec})/\sigma_z$, where $\sigma_z$ is the 
photometric redshift uncertainty reported by the MLCS2k2 fit.
Inset panels show the inferred mean and dispersion of the Gaussian
fits to each distribution.
}
\label{fig:photozdistsplit}
\end{figure}

\begin{figure} [h]
\begin{center}
\plottwo{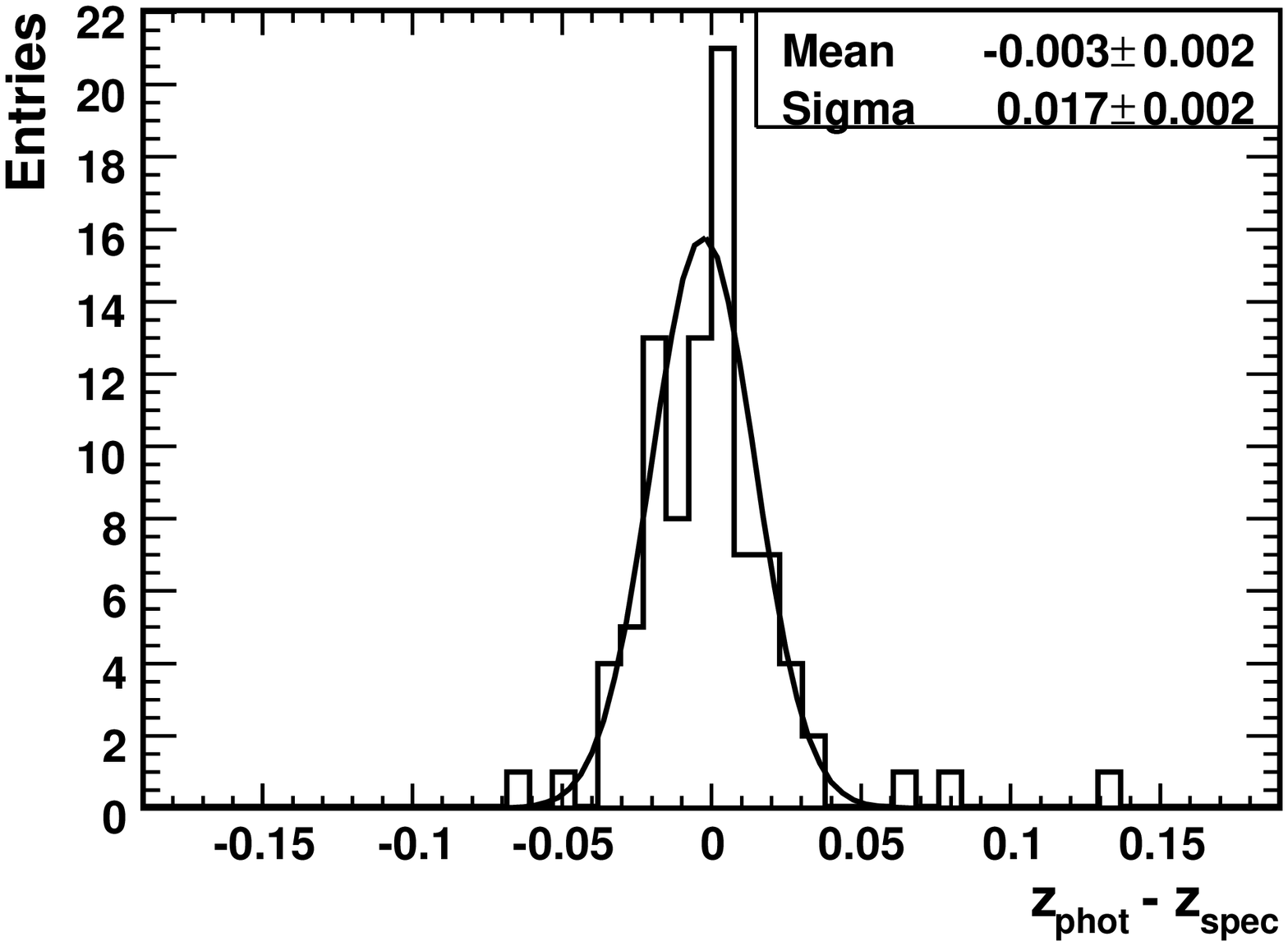}{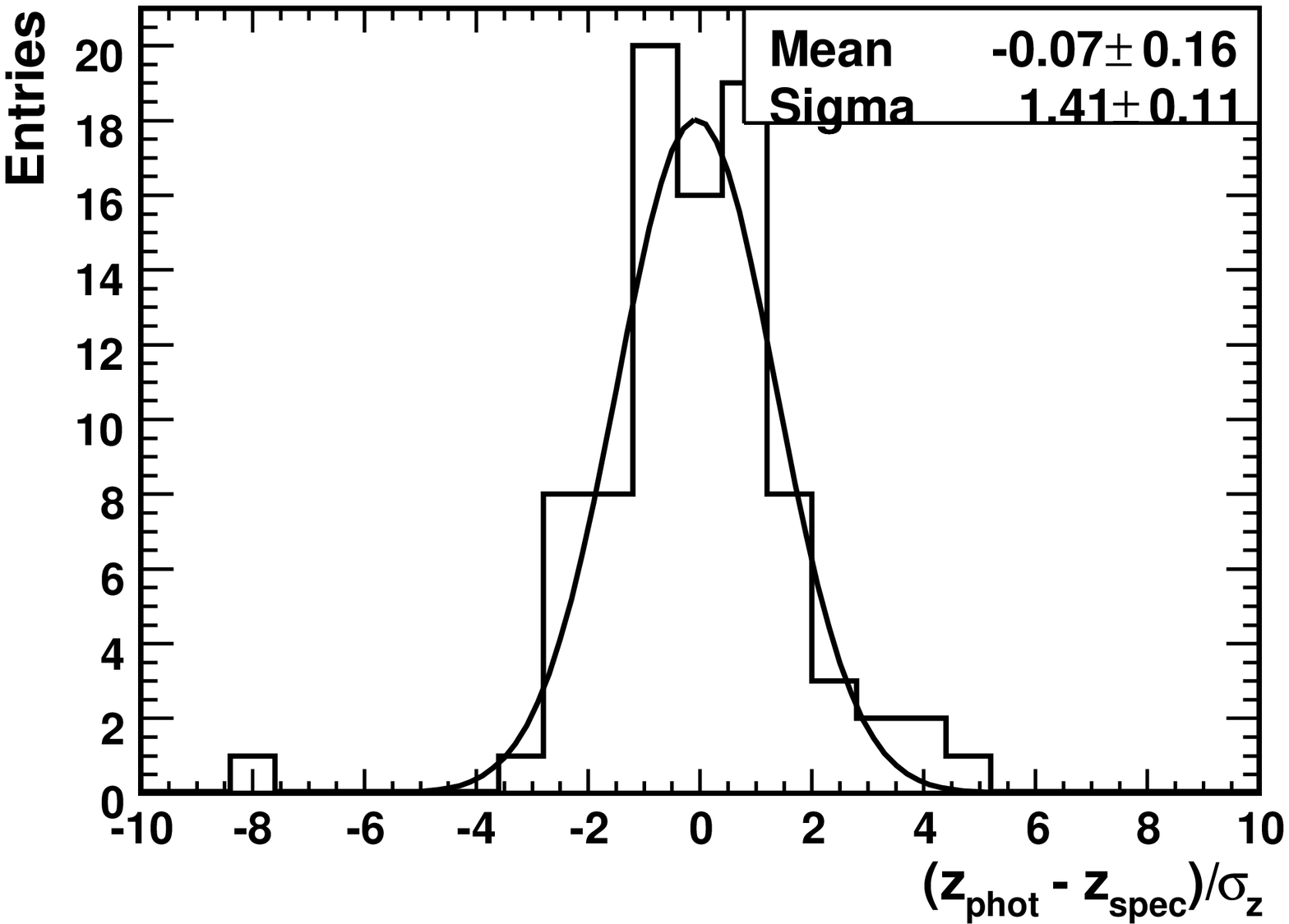}
\end{center}
\caption{
{\it Left panel:} 
The distribution of photometric redshift residuals, $z_{phot} - z_{spec}$,
for the combined spectroscopic+photometric SN Ia samples. 
$z_{phot}$ is the photometric redshift and $z_{spec}$ the spectroscopic
redshift;
{\it Right panel:} distribution of residuals normalized by 
the reported photometric redshift uncertainty, $\sigma_{z}$. 
}
\label{fig:photozdist1}
\end{figure}
\clearpage

The residuals of the SN photometric redshift estimates for this test 
sample of 89 objects are shown in Figure~\ref{fig:photozscatter}. 
The distributions of the residuals are shown 
in Figure~\ref{fig:photozdistsplit} separately for categories (i) and (ii), 
which indicates that 
the distributions for the two samples are consistent.
The fit residuals for the combined sample are shown in 
Figure~\ref{fig:photozdist1}. For the combined sample, 
the mean residual of the photometric redshift estimate is consistent with zero.
The scatter in the photometric redshift estimate for the 
combined test sample is $\sigma_{z}=0.018$, and Fig. \ref{fig:photozscatter}
shows that the scatter increases with redshift. 
The distribution of the residuals normalized by the MLCS-reported 
redshift error is shown in the right panel of Figure~\ref{fig:photozdist1}. 
If the reported redshift errors were accurate and Gaussian, 
this distribution would be a Gaussian with unit variance, $\sigma=1$. 
The distribution appears to be approximately Gaussian, 
but with measured variance $\sigma =1.4$;
we therefore choose to multiply the MLCS-estimated 
photometric redshift error for each candidate by 1.4.

In addition to the SN photometric redshift estimates, we also have  
host galaxy photometric redshift estimates for the majority of 
\psne~\citep{Oyaizu_07}.
Although the galaxy $z_{phot}$ estimates 
have larger scatter than those from the SN light curves,
in principle we could require consistency between these two redshift 
estimates as an additional selection cut on the photometric SN sample. 
Since core-collapse SNe are typically fainter than SNe~Ia, they 
would typically be assigned incorrectly high photometric redshifts 
by the light curve fitter. 
Using the existing selection cuts, however, 
we find no contamination of the rate-measurement sample 
from the \psne~without 
spectroscopic redshifts (see \S \ref{sec:selpsne}).
Therefore a requirement of consistency between 
the supernova- and galaxy-derived photometric redshift estimates 
is not necessary in the present analysis.

\subsubsection{Selection of \psne}
\label{sec:selpsne}

In the Fall 2005 observing season, the software and human data 
processing pipeline described in \S \ref{sec:search} yielded 
11,385 SN candidates, including the 146
spectroscopically confirmed and probable SNe~Ia  
and 20 that were confirmed as other SN types. 
The majority of the remaining candidates ($\sim 60\%$) 
are single epoch events that are most likely to be slow-moving asteroids,
leaving $\sim 4500$ multi-epoch SN candidates. 
To search for photometric 
SNe~Ia among this large set of candidates, we studied two 
subsamples selected according to different criteria. 

The first photometric subsample is designed to exhaust the list 
of candidates that are most likely to be SNe Ia.
This subsample was selected by choosing all SN candidates 
that the survey photometric typing code 
(described in \S \ref{sec:search}) classified as 
SNe~Ia\footnote{More precisely, according to the photometric
typing code, one of the ``type A'' or ``type B'' criteria were
satisfied; see \citet{Sako_08}}
and that were detected in at least 3 epochs 
by the on-mountain software pipeline. The images for the resulting  
subsample of $\sim 420$ candidates were processed through 
the final SMP pipeline, and the resulting light curves were 
fit with the MLCS2k2 program, using the redshift as a fit 
parameter in cases where there was no measured host galaxy 
spectroscopic redshift (the majority of cases). 
One highly extincted SN Ia 
(SDSS-SN 9266, discussed in \S \ref{sec:psample}), with a host galaxy 
redshift measured by the SDSS galaxy survey, was recovered from this
subset of \psne.
No other candidates in this subsample pass the rate selection cuts 
and have a spectroscopic or SN photometric redshift $z\le0.12$.

The second photometric subsample is designed to study the candidates
that are less likely to be SNe Ia.
This subsample was selected by choosing all SN candidates 
with detections at more than two epochs during the search and 
with an estimated time of maximum light, based on the survey 
photometric typing code, in the twenty-day 
interval between modified Julian day (MJD) of 
53660 and 53680 (October 17 and November 6). 
Since the selection criteria for this second subsample 
are looser than for the first (there is no requirement that
a SN Ia light-curve template provides the best-fit), 
the number of candidates it selects would be 
an order of magnitude larger.
Using a restricted time interval provides a manageable number of 
events to study that are 
representative of the population of these lower-quality light curves.
These selection criteria result in 462 candidates, 
which represent $\sim 1/6$ of the multi-epoch SN candidates that have 
not already been included in the samples discussed above.
We find no events in the second subsample that pass the rate measurement 
selection cuts and that have a spectroscopic or SN photometric 
redshift $z\le0.12$. 

Although no other \psne~pass our selection criteria, we must allow 
for uncertainties in the SN photometric redshift estimates from MLCS2k2. 
In the two photometric subsamples, two candidates 
that pass the rate-selection cuts have estimated 
SN photometric redshifts within $\sim 1.5\sigma$ of our cutoff of 
$z=0.12$, using the inflated redshift errors discussed in 
\S \ref{sec:photoz}. 
One of these candidates is from the first photometric subsample 
and has a fitted redshift $0.17 \pm 0.03$ (SDSS-SN internal ID 3077); 
the second is from the second photometric subsample 
and has a fitted redshift of $0.18 \pm 0.04$ (SDSS-SN ID 6861). Efforts are underway 
to obtain spectroscopic redshifts for the host galaxies of these 
events. 
Interpreting the (inflated) photometric redshift errors 
as Gaussian (\S \ref{sec:photoz}), the probability that at least one 
of these two candidates has a redshift $z \leq 0.12$ is significantly 
less than unity. To be conservative, we assign a systematic 
uncertainty of $+1$ SN Ia based on this study.

\subsection{Summary of Rate Sample Selection}

\begin{figure} [h]
\begin{center}
\plotone{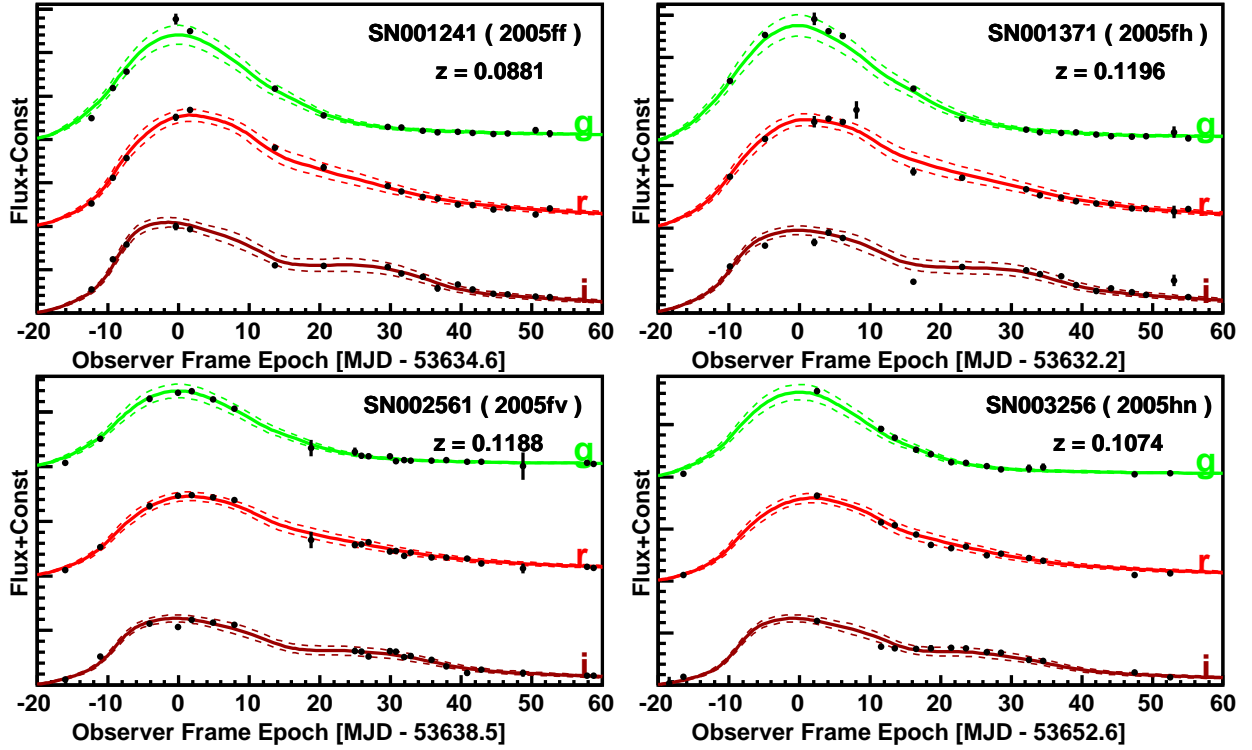}
\end{center}
\caption{$gri$ light curves for SNe Ia used in this rate
measurement. 
Black points show the SDSS SN photometry from SMP. The errors
on the photometry are shown.
Solid curves denote the best-fit SN Ia model
light curves in $g$ (green), $r$ (red), and $i$ (dark red) from
the MLCS2k2 light-curve fitter, and corresponding dotted
curves show the 1-sigma model error range. 
The curves and data for
the different passbands have been vertically offset for clarity. 
The flux offsets are the same for each SN.
}
\label{fig:lcs}
\end{figure}

\begin{figure}[h]
\begin{center}
\plotone{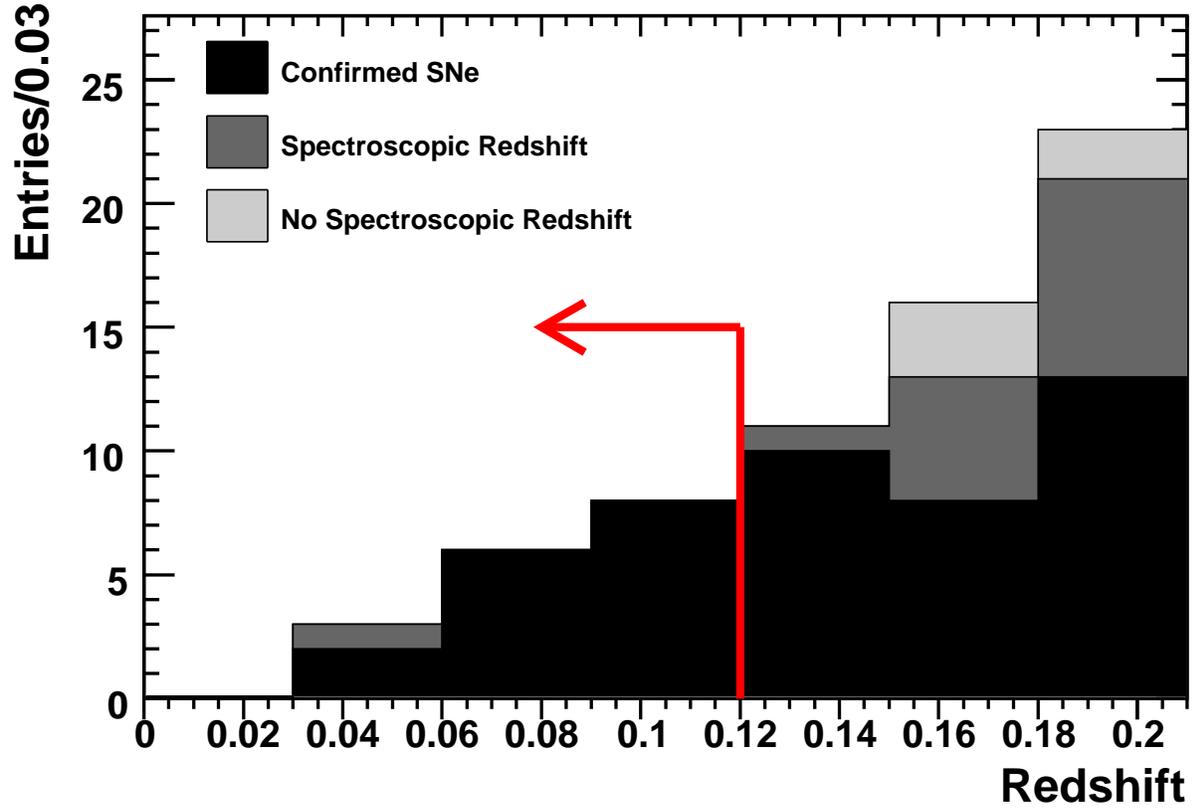}
\end{center}
\caption{
   Redshift distribution for events passing the rate-measurement selection
   requirements in \S \ref{sec:ssample}. Contributions include:
   spectroscopically confirmed SNe~Ia (black), 
   photometric SNe~Ia with host galaxy redshifts (gray),
   and photometric SN~Ia candidates with no spectroscopic 
   redshift (light-gray).
   The arrow shows the redshift cut for this analysis.
}
\label{fig:zsnrate}
\end{figure}

In summary, rate-sample selection requirements have been applied
to SN candidates with $z\le0.12$ from the 2005 observing season.
The resulting sample comprises \numsamples\ spectroscopically identified and 
\numsamplep\ photometrically identified SNe~Ia.
These events are enumerated in Table~\ref{tab:table1-sn}, and their
$gri$ light curves are shown in Figure~\ref{fig:lcs} 
along with the best-fit MLCS2k2 model light-curve.
Figure~\ref{fig:zsnrate} shows the redshift distribution for SNe Ia 
for $z\le0.21$;  the lowest-redshift photometric candidates 
with no spectroscopic 
redshift are in the bin $0.15 < z < 0.18$ and are 
safely above the redshift cut.

\clearpage
\clearpage
{\plotone{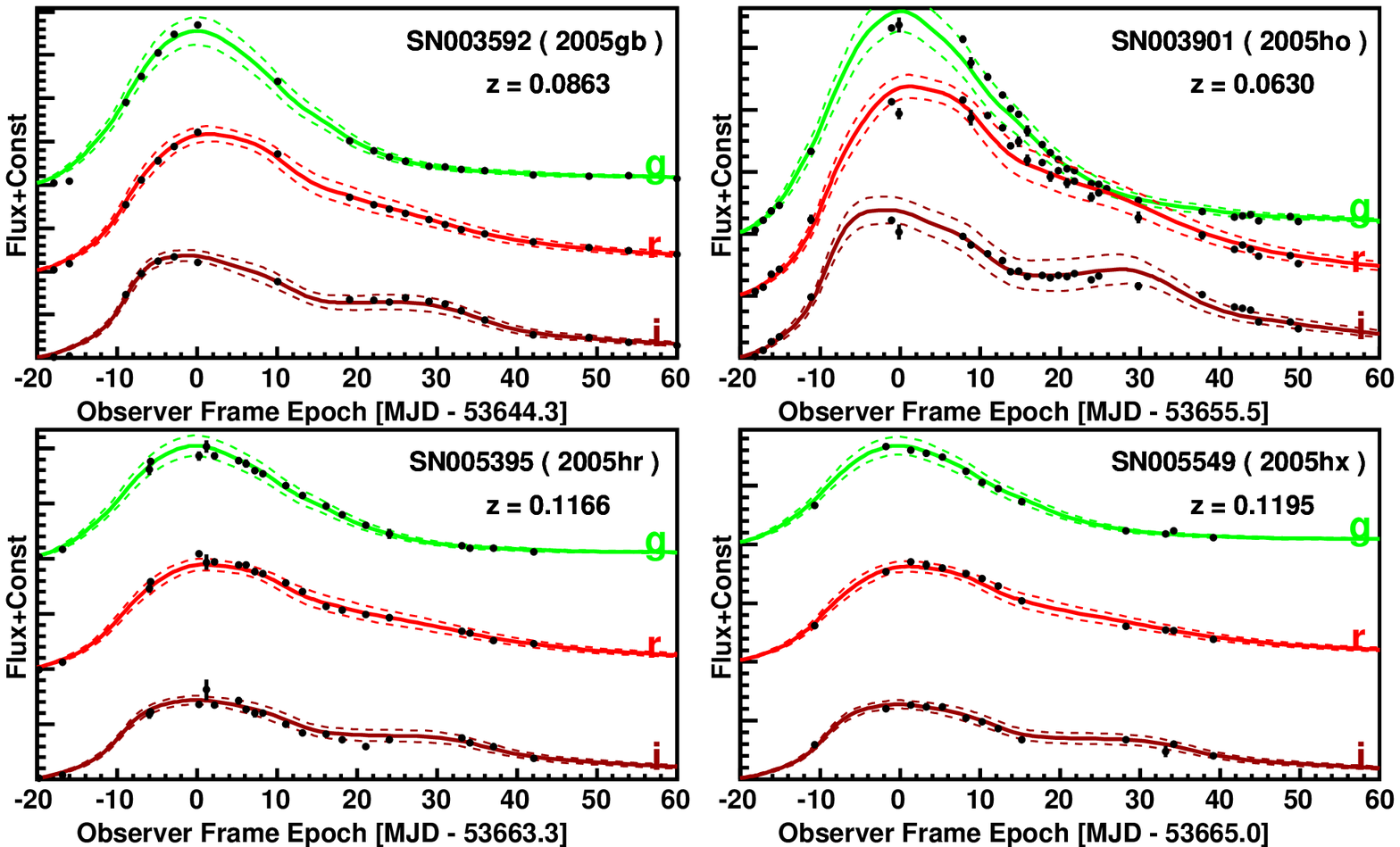}}\\
{\plotone{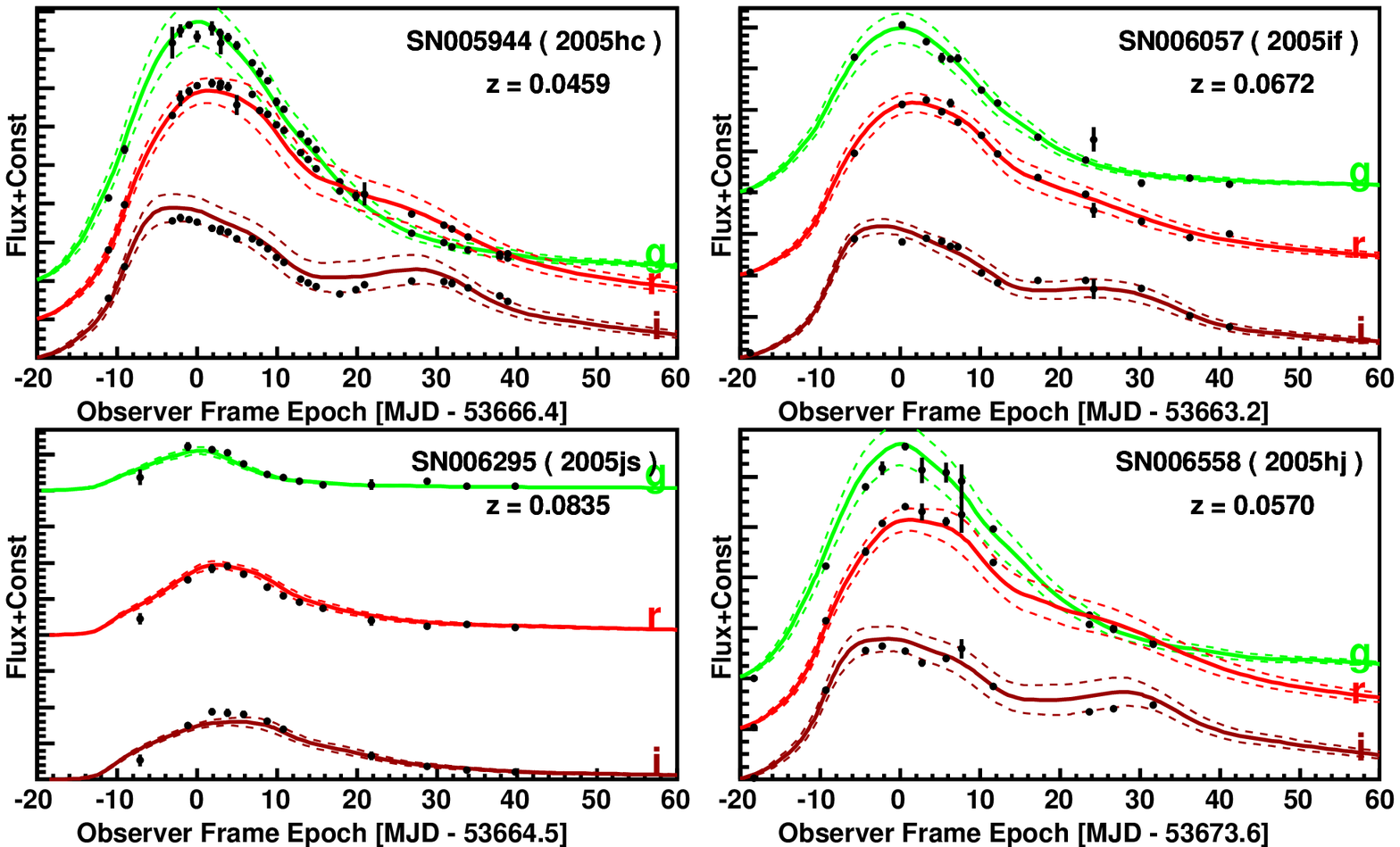}}\\
\centerline{Fig. 5. --- Continued.}
\clearpage
{\plotone{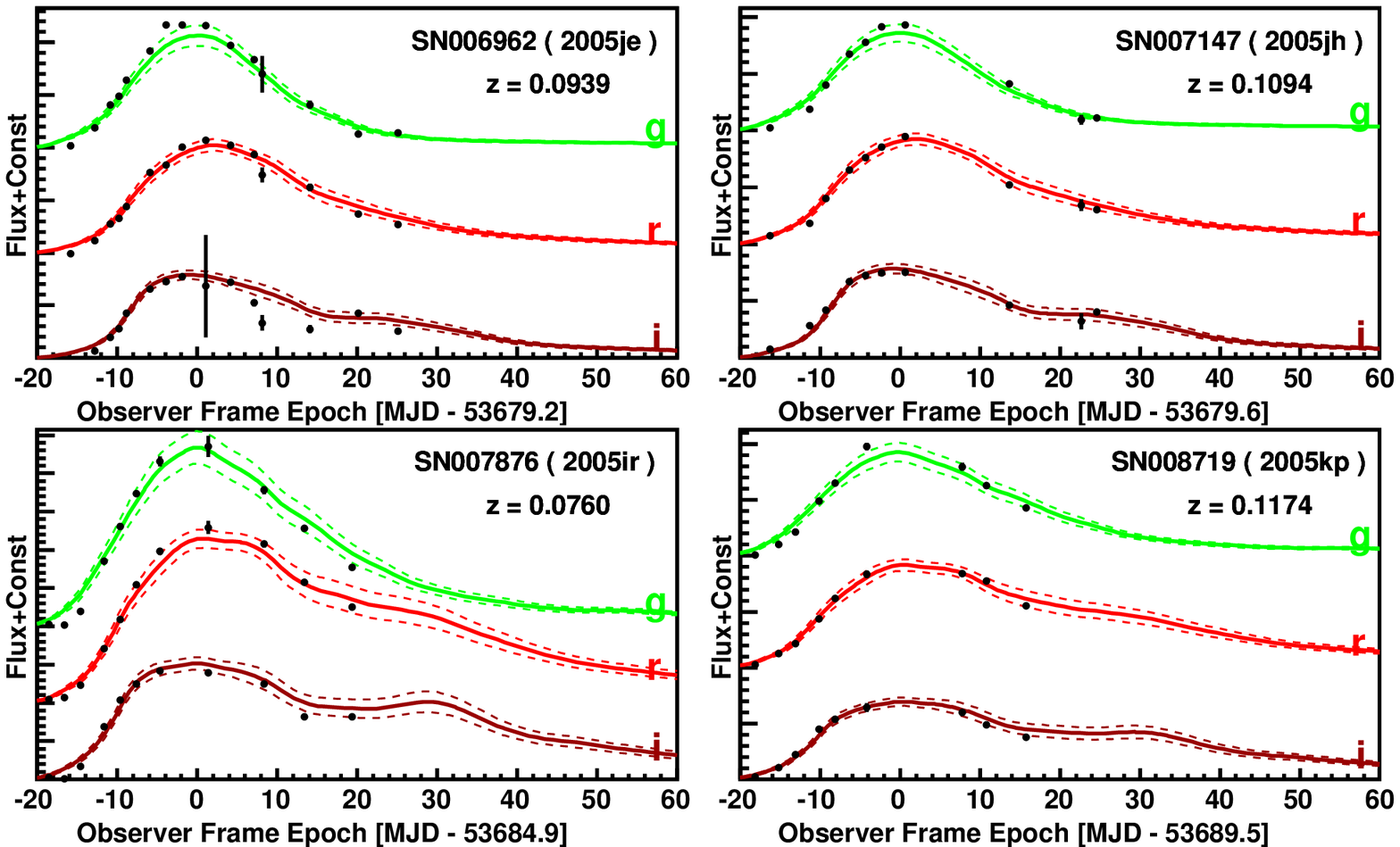}}\\
{\plotone{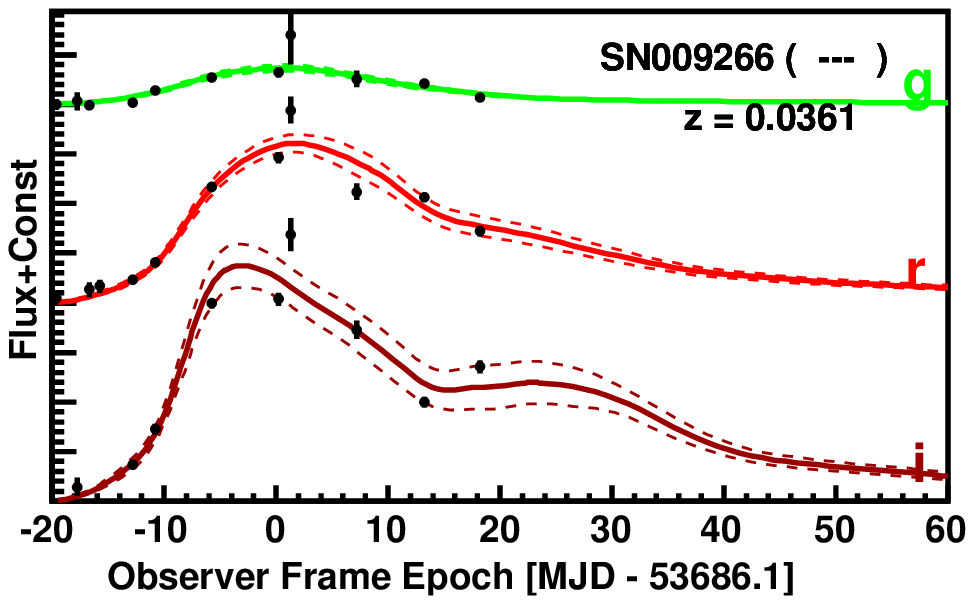}}\\
\centerline{Fig. 5. --- Continued.}

\clearpage
\clearpage

\section{Survey Efficiency}
\label{sec:effs}

To convert the number of discovered SNe Ia into a measurement of the SN rate, 
we must have an estimate of the efficiency for discovering 
SNe Ia at $z \leq 0.12$ that satisfy the sample selection criteria. 
We have two tools at our disposal for this estimate: the 
artificial SN images (fakes) that are inserted into the data stream 
in real time and Monte Carlo simulations of the 2005 observing season.

\subsection{Use of Artificial SN Images}
\label{sec:fakeeffs}

As noted in \S \ref{sec:fakes}, the fake SNe images are used to measure
the efficiency of the on-mountain software pipeline for
point-object detection on a variety of galaxy backgrounds
and observing conditions. The fake SNe are also used 
to measure the efficiency of human scanners for identifying objects as SNe.
While the fakes were designed to model realistic SN~Ia light curves,
the $z^2$ dependence on the redshift distribution
results in only 18 fake light curves with redshift $z < 0.12$.
Although all 18 low-redshift fakes were recovered by the
SN search pipeline, using such a small sample to measure
the pipeline efficiency would result in large statistical and systematic
uncertainties. Furthermore, the fake light curves were
generated with distributions of $A_V$ and $\Delta m_{15}$
that were not realistic, which complicates
the interpretation of discovery efficiency as a function of redshift.
    
To obtain a more reliable determination of the survey efficiency,
we use fake SN~Ia at all redshifts in the following way.
We first use the fakes to measure the object-detection efficiency
as a function of the signal-to-noise ratio (SNR) in the
$g$, $r$, and $i$ passbands. The detection efficiency, 
defined as the ratio of the number of fake epochs 
detected as objects by the on-mountain software pipeline to the number 
of fakes inserted into data images at a given signal-to-noise, 
is shown in Fig. 7. While the
object detection efficiency as a function of magnitude or redshift
is sensitive to observing conditions (seeing, clouds, moon),
the efficiency as a function of SNR is robust against such
variations in conditions. 
As a check that the SNR is an adequate parameterization of the 
point-source detection efficiency, we have split the sample of fakes
into a low-redshift and a high-redshift subsample and determined the
efficiency as a function of SNR for each set independently. We find that the
results are consistent.
With the efficiency as a function of signal-to-noise ratio known,
one can estimate the SN discovery efficiency as a function of redshift
for {\it any} choices of SN Ia light-curve models, observing conditions, and 
population distributions.
These efficiency functions measured with fakes  
are used in the Monte Carlo simulation (\S \ref{sec:simeffs})
to verify that the software pipelines were fully efficient at low redshift. 

\clearpage
\begin{figure}[t]
\begin{center}
\plotone{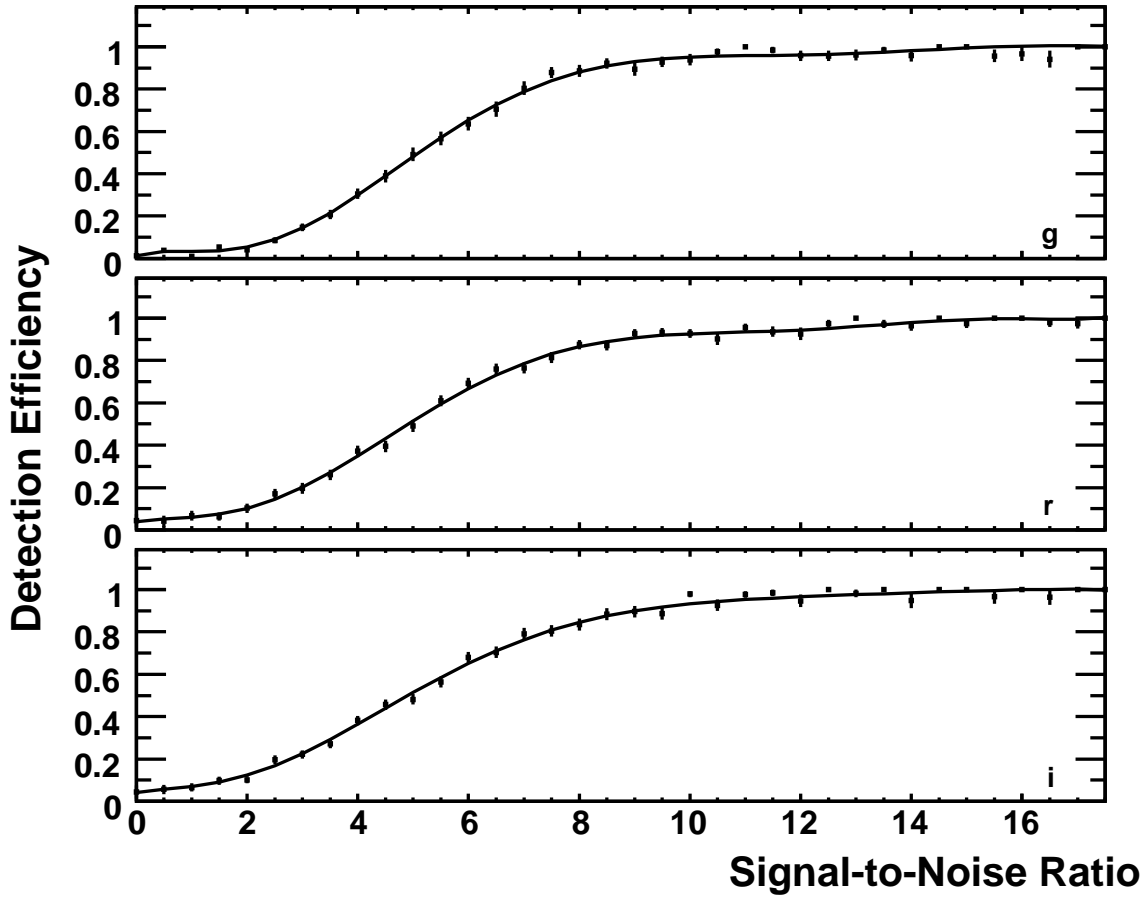}
\end{center}
\caption{
   The mean object detection efficiency as a function of signal-to-noise ratio 
   for SDSS $g$-band (top), $r$-band (middle), and $i$-band (bottom). 
   The efficiency is 
   derived by counting the fraction of fake images detected by the 
   survey pipeline. 
   The binomial errors on the efficiency measurements 
   are shown. 
   The solid lines show the result of a polynomial fit 
   to the efficiency measurements.
   These efficiency functions are used to simulate the  
   difference imaging
   software in the Monte Carlo simulations of the SN light curves.
  }
\label{fig:fakeeffgri}
\end{figure}
\clearpage

The fakes also provide information on the efficiency of the 
human scanners to correctly label as possible SN candidates those fakes 
that were detected as objects by the software pipeline. 
For the 2005 season, $91\%$ of all epochs of fakes visually scanned 
by humans were flagged as SN candidates, 
and $95\%$ of all detected fake SNe were flagged by humans as 
SN candidates at least once.
The 5\% of fakes that were never identified by humans as SN candidates 
were detected on only a single epoch by the software pipeline, 
either because they were at high redshift or because they reached 
peak light well before or well after the observing season. 
Essentially all fakes detected on two or more epochs by the software pipeline  
were flagged by humans as SN candidates at least once. Given the selection 
cuts in \S \ref{sec:ssample}, the human scanning efficiency is 100\% for 
SNe Ia contributing to this low-redshift rate measurement.

\begin{figure} [t]
\begin{center}
\plotone{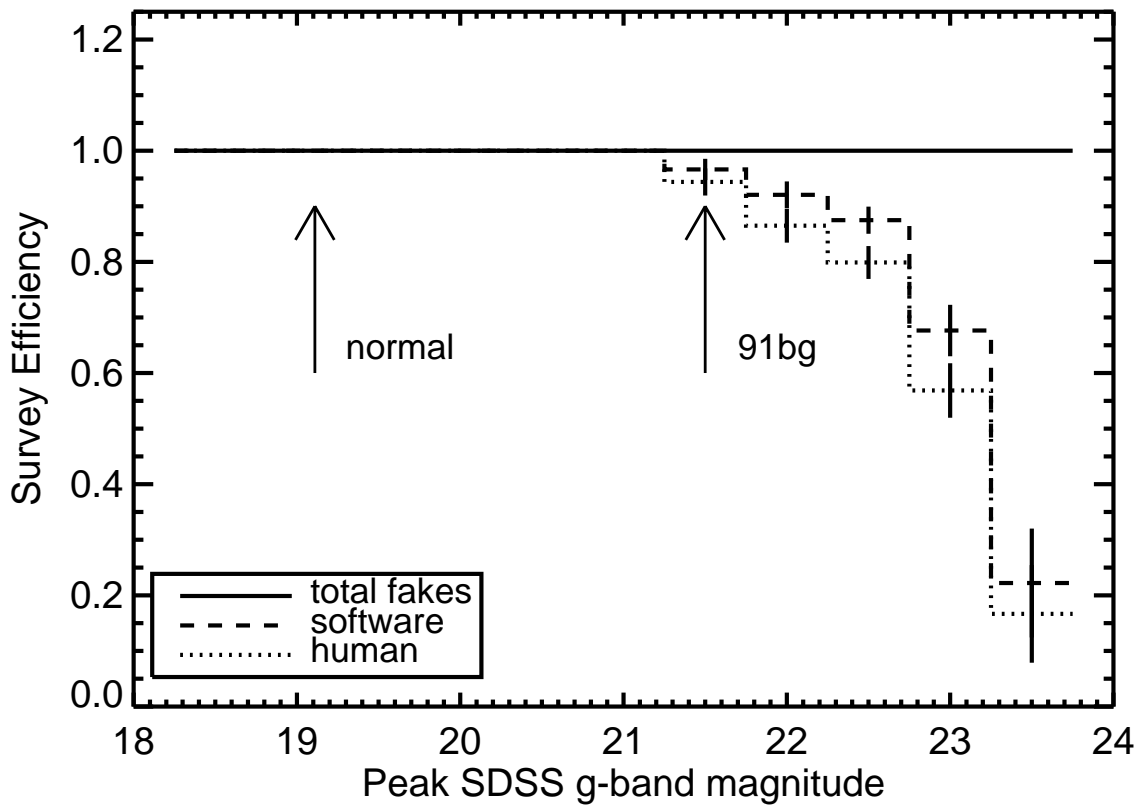}
\end{center}
\caption{
The efficiency for identifying fakes as SN candidates, as a 
function of peak $g$-band magnitude. 
The dashed curve is the efficiency for detection by the software 
pipeline, and
the dotted curve is the efficiency for 
evaluation of the fakes as SN candidates by the human scanners.
The arrows indicate the peak magnitudes for a 
normal and for a 91bg-like SN Ia at a redshift of 0.12 according to  
the MLCS2k2 model.
The binomial errors on the efficiency are shown.}
\label{fig:fakespeakmageff}
\end{figure}

Summary information on the efficiency of the software pipeline 
and the human scanners to detect fakes is presented in 
Figure \ref{fig:fakespeakmageff}, which shows the 
detection efficiencies, i.e., the fraction detected by the 
pipeline and the fraction identified as SN candidates by humans, 
vs.~peak $g$-magnitude. The arrows indicate the peak 
$g$-band magnitudes for an unextincted normal and 
for an unextincted sub-luminous 1991bg-like 
SN Ia at $z=0.12$, according to the MLCS2k2 model. 
This figure indicates that, for the assumed SN~Ia model 
used to generate the distribution of fakes, 
the combined software+human detection efficiency is essentially 
100\% for SNe~Ia in the redshift range $z \leq 0.12$.
\clearpage
\clearpage

\subsection{Monte Carlo Simulations}
\label{sec:simeffs}

To determine the SN~Ia selection efficiency with high
precision and to study 
systematic uncertainties for the rate measurement,
we have developed a detailed Monte Carlo light curve simulator (MC). 
The MC simulates individual light
curve data points based on real observing statistics, but without the
added complexity of adding fake SNe to images.
The MC light curves can be 
generated and analyzed much more rapidly than the fakes, 
so the MC can be used to rapidly 
simulate very large numbers of SN~Ia light curves
to estimate the SN discovery efficiency and the 
uncertainty in the efficiency due to 
assumptions about the SN Ia model distributions. 
The MC code uses the MLCS2k2 model to generate simulated SN~Ia 
light curves instead of the stretch/$\Delta m_{15}$ model 
that was used to generate the fakes.

For each simulated SN~Ia, the following parameters
are randomly drawn from parent distributions: 
\begin{enumerate} 
\item {\bf redshift, $z$:} \\ 
  Drawn from a distribution proportional to the comoving volume element, 
  which assumes a constant SN Ia rate per unit comoving volume.
\item {\bf host galaxy extinction, $A_V$:} \\ 
  Drawn from a distribution
  $P(A_V) \propto e^{-A_V/\tau}$, with $\tau=0.4$. The 
  \citet{ccm89} reddening law, with $R_V=3.1$, 
  is used to extrapolate the extinction
  to other wavelengths. The choice of $\tau = 0.4$ was guided
  by the studies of \citet{Jha_07} and is
  consistent with the inferred extinction distribution
  for spectroscopically confirmed SNe Ia in the SDSS
  SN sample. As we discuss later in this section, the 
  exact choice of $\tau$ makes no practical difference to this
  rate measurement.

   \item {\bf MLCS2k2 light curve shape/luminosity parameter, $\Delta$:} \\
     Drawn from a bimodal Gaussian with a standard deviation 
     of $0.26$ for 
     $\Delta < 0$ and $0.12$ for $\Delta > 0$,
     and truncated to lie within the valid range of the
     MLCS2k2 model, $-0.35 < \Delta < 1.8$.
     The bimodal Gaussian is based on study of the confirmed SNe Ia
     in the \sns~first year data. 

   \item {\bf time of peak light in rest-frame $B$-band:} \\
     Drawn randomly from the interval
     $53616 < \mathrm{MJD} < 53705$ (Sept. 1 - Dec. 1, 2005).
   \item  {\bf sky position:} \\ 
     Drawn randomly from the range 
     of the survey.
   \item {\bf location within host galaxy:} \\
     Drawn from a distribution proportional to the host galaxy
     surface brightness (see below). This 
     variable is used only to determine galaxy background light, not 
     extinction.
\end{enumerate}

\noindent{We note that simulated photometry is generated only at epochs
for which we obtained photometric imaging at the corresponding 
sky location, and therefore the determination of the selection efficiency 
naturally accounts for the temporal inhomogeneity in sky coverage, as can be seen
in Figure~\ref{fig:racov}.}

Using these parameters for each SN, rest-frame 
$UBVRI$ magnitudes are generated from the MLCS2k2 model for 
all dates on which the survey took data at the selected sky position.
These magnitudes are modified according to the 
host galaxy extinction, $K$-corrected to the observed SDSS $gri$ passbands,
and further modified according to the estimated  
Milky Way extinction at that position \citep{Schlegel_98}.
The zero-points from the survey are used to convert the $gri$ 
magnitudes into flux values that would have been measured in ADUs
by the SDSS 2.5~m telescope. 
The CCD gains are then used to determine the number of photo-electrons, 
and hence the signal and noise.
Additional noise is computed for each measurement 
based on the measured observing conditions at each epoch, in each passband,
at the assigned sky location.
Sky noise is simulated by integrating the estimated 
sky noise per pixel over an effective aperture with a size determined 
by the local PSF estimate from {\tt PHOTO}. Noise from the host galaxy 
is simulated by associating 
the SN with a host from the SDSS galaxy 
photometric redshift catalog \citep{Oyaizu_07}  
selected to have a photometric redshift equal to the 
assigned SN redshift. 
In the SDSS DR5~\citep{SDSS_DR5} {\tt photoPrimary} database~\citep{SDSS_EDR}, 
each such galaxy image is fit with both an exponential and a 
de Vaucouleurs surface brightness profile. 
We use the exponential model in the $r$-band 
as a probability distribution from which the SN 
position within the galaxy is drawn. That is, the galaxy noise model 
assumes that the SN Ia rate is roughly proportional to $r$-band 
stellar luminosity. 
The estimated contribution of the galaxy light to the noise in 
each passband is computed by convolving the exponential 
galaxy model with the PSF in the survey image.
In practice, this procedure is computationally expensive, 
so we pre-compute the noise values on a grid of model parameters 
and perform a multi-dimensional linear interpolation
to obtain an estimate of the galaxy noise.

\begin{figure} [!t]
\begin{center}
\plotone{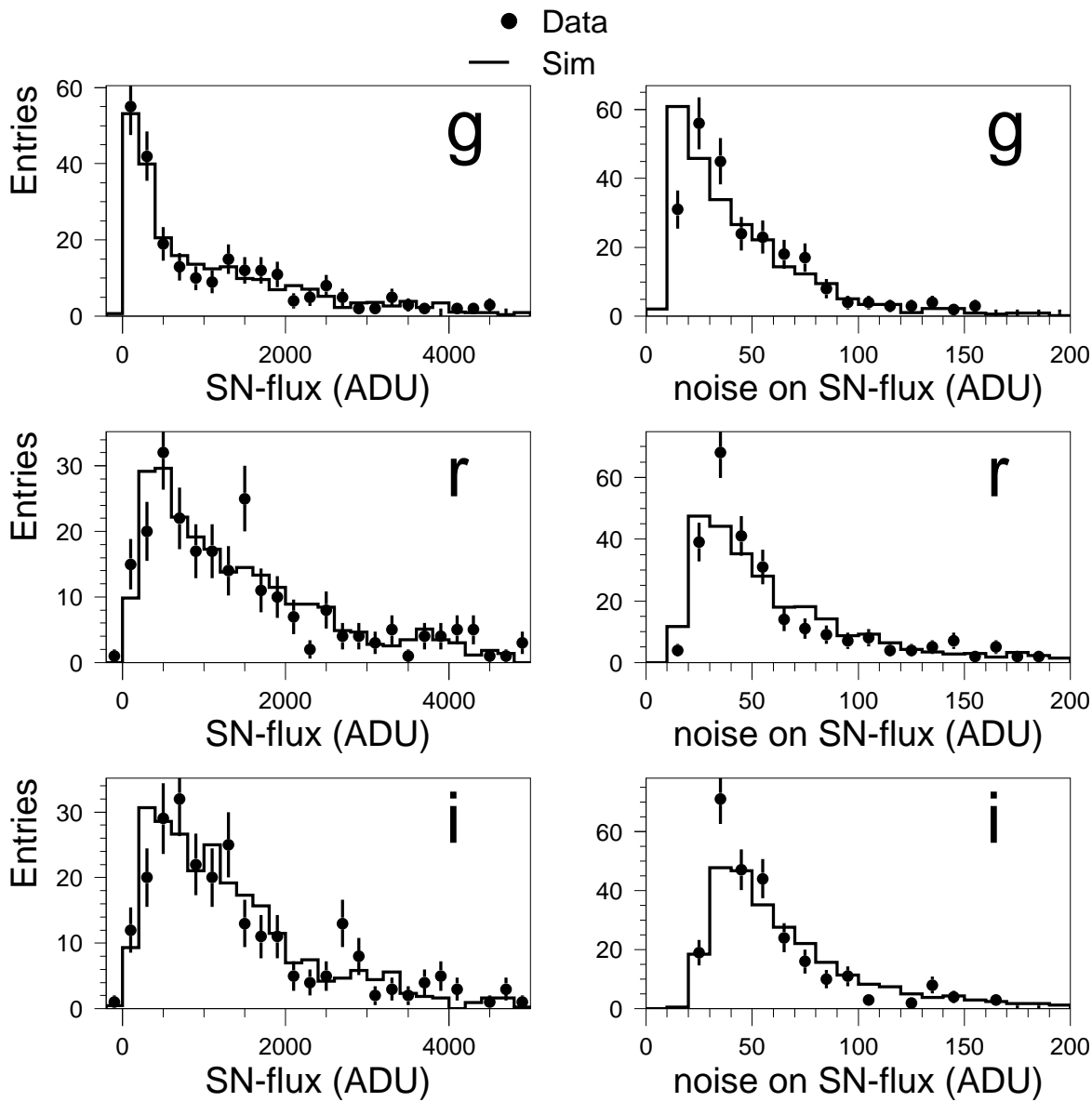}
\end{center}
\caption{
  For SN~Ia that satisfy the selection requirements (\S \ref{sec:sample}),
  the signal-flux and noise distributions are shown for all
  photometric epochs in the $g$,$r$,$i$ filters. Each distribution
  is shown for SDSS data (dots) and for the simulation
  (histogram) that has been scaled to have the same number
  of entries as the data.
}
\label{fig:datasim}
\end{figure}

As a consistency check of the MC as a representation of the SN data, 
we compare
the distributions of signal and noise 
in $gri$ for the MC sample to the signal and noise for
all photometric epochs for 
the low-redshift SNe Ia in the rate-measurement sample. The comparison
of the distributions is shown in 
Fig.~\ref{fig:datasim}.
The distributions of signal and noise are in good agreement, 
indicating that the MC model, and the assumed parameter distributions 
therein, provide a reasonable representation of 
the low-redshift SN~Ia sample.

\clearpage
\clearpage

\begin{figure} [t]
\begin{center}
\plotone{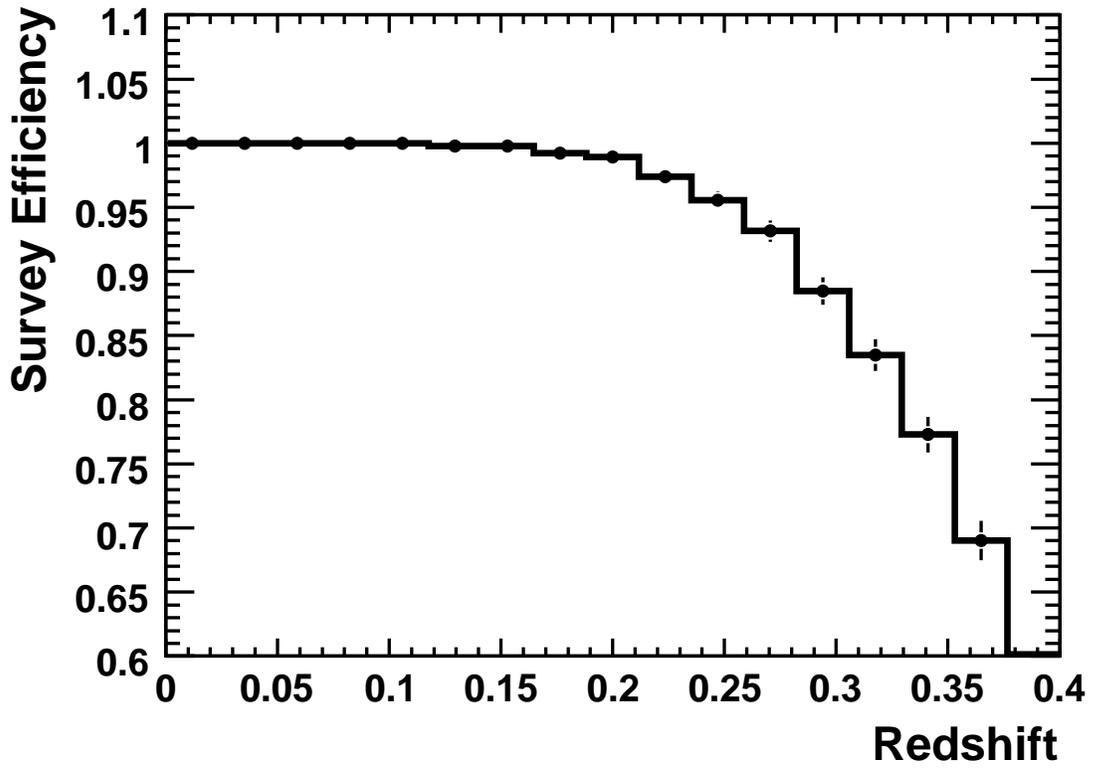}
\end{center}
\caption{
  Search pipeline
  software efficiency as a function of SN Ia redshift, as determined from 
  the Monte Carlo simulation. 
}
\label{fig:snananominal}
\end{figure}

Having shown that the MC generates photometry that is 
consistent with our observed low-redshift SN Ia sample, 
we can use the MC to provide a more reliable determination
of the detection efficiency of the on-mountain search pipeline than 
we could obtain with the relatively small number of low-redshift fakes.
For each epoch generated by the MC, we use 
the efficiency as a function of signal-to-noise ratio
in each passband (from Fig.~\ref{fig:fakeeffgri}) 
to determine the efficiency of the search pipeline
to detect SNe~Ia at all redshifts. 
The resulting 
software pipeline detection efficiency as a function of redshift,
based on a MC study using 15640 total SNe (920 in each of 17
redshift bins) is shown in Figure~\ref{fig:snananominal}; 
the efficiency is 100\% over the redshift range $z\le0.12$. 
Since the fakes tests of \S \ref{sec:fakeeffs} showed that the human 
scanning process causes a negligible loss of efficiency, we conclude 
that the combined efficiency for SN detection by the pipeline and 
identification as a candidate by humans is 100\% over the redshift range 
of interest. This does not guarantee that the efficiency of the 
photometric typing code used for spectroscopic target selection (\S 
\ref{sec:search}) is also 100\%, but the studies of \S \ref{sec:psample} 
indicate that, with the exception of SN 9266, there were no losses 
due to the target selection algorithm.

The final step is to use the MC to compute the survey discovery 
efficiency $\vareff$ for a SN~Ia sample defined by the selection requirements
in \S \ref{sec:ssample}. This efficiency is the ratio of the number of 
SNe Ia that are detected by the pipeline, identified by humans, and that
pass the selection criteria of \S \ref{sec:ssample} to the total 
number of SNe Ia that reach peak light during the survey, i.e., between 
Sept. 1 and Nov. 30. While we have seen that the detection efficiency 
is essentially 100\% out to $z=0.12$, $\vareff$ is less than 100\% 
primarily because the selection requirements on light curve coverage
(cuts 4 and 5 in \S \ref{sec:ssample}) remove some SNe Ia that peaked 
in early September or late November.
Using the MC sample of 15640 light-curves mentioned above and 
fitting a linear function to the resulting selection efficiency 
in the redshift range $0 < z < 0.12$ gives 
\begin{equation}
\label{eqn:lineff}
\vareff(z) = (0.78\pm 0.01) + (-0.13 \pm 0.14)~z
\end{equation}

\noindent That is, the survey efficiency is approximately 
constant at low redshifts, changing by only $\sim 1\%$ 
over the redshift range of the rate measurement. 
The mean SN~Ia discovery efficiency for our rate sample is 
$\langle \vareff \rangle = \rateeff$.

While the data-MC comparison in Fig.~~\ref{fig:datasim} 
indicates that we have made a consistent choice of the parameter 
distributions for the MC model, to estimate the systematic uncertainty
in the discovery efficiency we vary the assumed MC parameter 
distributions and recompute the efficiency. 
We find that varying $\tau$, the parameter controlling the 
extinction distribution, has the largest systematic effect on the 
determination of the discovery efficiency from the MC. 
Varying $\tau$ 
over the range $0.2-0.6$, 
the estimated discovery efficiency 
for the rate-measurement SN~Ia sample changes by less than a percent.

\clearpage
\clearpage

\section{SDSS SN Results}
\label{sec:results}

\subsection{Volumetric SN Ia rate}
\label{sec:volumetric}
For the purpose of interpreting the SN observations 
as a volumetric SN rate, we will assume
a spatially flat cosmology with non-relativistic matter density 
$\Omega_m = \myomegamatter$, dark energy density 
$\Omega_\Lambda=0.7$, and dark energy equation of state 
parameter $w=p/\rho = -1$. For the low-redshift rate 
measurement presented in this section, the dependence 
on cosmological parameters of the survey selection efficiency 
is negligible, and so the uncertainty in the rate due to uncertainty
in cosmology is due entirely to the difference in the volume of the 
survey. A change in $\Omega_{m}$ of 0.02 would lead to 
a $4\%$ change in the rate measurement.

The observed volumetric SN rate, $r_V$,  is defined as

\begin{equation}
\label{eqn:orate}
r_V  = \frac{N}{\widetilde{V T \vareff}} ~,
\end{equation}

\noindent where $N$ is the number of SNe in the sample, 
and $\widetilde{V T \vareff}$ is the effective product of 
the survey volume, $V$, the observer-frame survey duration, $T$,
and the SN discovery efficiency, $\epsilon(z)$, estimated in \S \ref{sec:simeffs},

\begin{equation}
\label{eqn:effthte}
\widetilde{V T \vareff} = (\Theta T ) 
\int_{z_{min}}^{z_{max}}{dz~\vareff(z) u^2(z) \frac{du}{dz} \frac{1}{(1+z)}}~.
\end{equation}

\noindent Here $\Theta$ is the solid angle covered by the survey and 
$u(z)$ is the comoving distance in the Friedmann-Robertson-Walker metric, 

\begin{equation}
u(z) = \int_0^{z}~dz' \frac{c}{H(z')} = \frac{c}{H_0} \int_0^{z}{\frac{dz'}{\sqrt{\Omega_m (1+z')^3 + \Omega_{\Lambda}}}}~.
\end{equation}

\noindent If 
the survey efficiency is independent of redshift,
and if the redshift range covered by the 
SN observations is small, then $\widetilde{VT\vareff}
\sim (V T \vareff)/(1+\langle z \rangle)$, where $V$ is the 
survey volume and 
$\langle z \rangle$ is the volume-weighted mean redshift of the
survey.

For the \sns~we have $N = 17$, $z_{min}=0$, $z_{max}=0.12$, 
$\langle \vareff(z) \rangle = 0.77 \pm 0.01$,
$T = 89$ days $= 0.244 $ years, and
$\Theta = 0.08277*0.98$ steradians.
This value for $\Theta$ is 98\% of the actual sky area covered by the 
survey, 
due to the masking of bright stars and variable sources. Substituting
these values into Eqn. (\ref{eqn:orate}), we find a volumetric
SN Ia rate of

\begin{equation} 
r_V = [2.93^{+0.17}_{-0.04}({\rm systematic})^{+0.90}_{-0.71}({\rm statistical})]
\times 10^{-5} ~{\rm SNe}~{\rm Mpc}^{-3}~h_{70}^{3}~{\rm year}^{-1} ~,
\label{eq:rate1}
\end{equation}

\noindent with $h_{70} \equiv H_{0}~(70~{\rm km}~{\rm s}^{-1} {\rm Mpc}^{-1})^{-1}$
and $H_{0}$ the present value of the Hubble parameter. The statistical 
errors quoted represent the standard frequentist $68.27\%$ 
central confidence interval
on the mean of a Poisson distribution.
The systematic uncertainty represents uncertainty on our determination
of the SN selection efficiency (\S \ref{sec:simeffs}) and on the
number of photometrically identified SNe Ia (\S \ref{sec:psample}).
This measurement 
represents the volume-averaged SN Ia rate at $z\le0.12$. When 
rate measurements are plotted vs.~redshift, it is generally assumed that the 
rate is constant 
over the sampled redshift 
interval. If we assume that the SN Ia rate is constant at $z\le0.12$, 
then Eqn.(\ref{eq:rate1}) can be interpreted as the rate 
at the volume-weighted mean of our redshift range, $\langle z\rangle = 0.09$, 
and we make this assumption when plotting the result.
Our result is shown along with previously reported SN Ia rate 
measurements in Fig.~\ref{fig:ratescomp0}, but we defer discussion 
of comparison and combination with other measurements to \S \ref{sec:evolution}.

\subsection{SN Ia Rate per unit galaxy luminosity}
Early measurements of the SN rate were generally derived from SN
observations that targeted known galaxies; for these surveys, the SN rate 
is most naturally measured as a rate per unit luminosity in some 
passband, traditionally the $B$-band. 
\citet{Blanc_04} have 
converted a number of measurements from the literature 
of the SN Ia rate per unit 
luminosity 
to rates per unit volume,  
and in Table \ref{tab:rcomp} we adopt their values for 
the \citet{Cappellaro_99, Madgwick_03, Hardin_00} rate measurements.

For completeness, we convert our volumetric rate to a rate 
per unit galaxy luminosity in the SDSS passbands.
The galaxy luminosity functions in the SDSS passbands are estimated in 
\citet{Blanton_03}. The corresponding 
luminosity densities in the $ugriz$ passbands, 
at a mean redshift of 
$\langle z \rangle = 0.1$ are 
$\blantonu \pm {\blantonue}$,
$\blantong \pm {\blantonge}$,
$\blantonr \pm {\blantonre}$,
$\blantoni \pm {\blantonie}$,
and 
$\blantonz \pm {\blantonze}$,
in units of $10^{8} L_{\sun} h_{70} \mathrm{Mpc}^{-3}$,
where $L_{\sun}$ is the solar luminosity.
In combination with the volumetric rate measurement of Eqn.(\ref{eq:rate1}), 
this yields the SN Ia rate per unit luminosity in the SDSS passbands, 
$(r_L)_{ugriz}/h_{70}^2 = $
$\snuu^{\snuuUP}_{\snuuLOW}~\mathrm{SNu}_{u}$, 
$\snug^{\snugUP}_{\snugLOW}~\mathrm{SNu}_{g}$, 
$\snur^{\snurUP}_{\snurLOW}~\mathrm{SNu}_{r}$, 
$\snui^{\snuiUP}_{\snuiLOW}~\mathrm{SNu}_{i}$, 
and 
$\snuz^{\snuzUP}_{\snuzLOW}~\mathrm{SNu}_{z}$, 
where 
1 $\mathrm{SNu}_x \equiv 1 ~{\rm SN} ~10^{-10} L^{x}_{\sun} ~(100~{\rm yr})^{-1}$, 
with $L^{x}_{\sun}$ the luminosity in passband $x$, 
in units of solar luminosities.
 
\subsection{SN Ia Rate as a function of host galaxy type}
\label{sec:rate_vs_galtype}
Recent measurements have shown that the specific SN Ia rate  
is higher  
in star-forming galaxies than in passive galaxies. For example,
\citet{Mannucci_05a} found that the SN Ia rate per unit stellar mass is 
$\sim 20-30$ times higher in late-type galaxies than in E/S0 galaxies.
\citet{Sullivan_06} have found a similar trend in the SNLS data.
We will consider the trend of SN rate vs.~star formation activity using the 
\sns~sample in a forthcoming publication.

Here, we consider the low-redshift 
SN Ia rate vs.~host galaxy type. We have 
considered several photometric galaxy-type indicators that are 
accessible through the SDSS DR5 database \citep{SDSS_DR5}, including 
$u-r$ color \citep{Strateva_01}; 
the likelihood of the de Vaucouleurs 
model fit to the galaxy surface brightness profile relative to that of an exponential 
model fit; and the (inverse) 
concentration index \citep{shimasaku_01,yamauchi_05}, 
defined as the ratio of the radii that contain 
50\% and 90\% of the Petrosian flux (see \citet{SDSS_EDR} for definitions of 
these quantities).  
These parameters are listed in   
Table \ref{tab:tab-host} for the host galaxies of the 
SNe included in the rate-measurement sample. Note that the $u-r$ color in 
Table \ref{tab:tab-host} is computed using 
SDSS model magnitudes \citep{SDSS_EDR} 
and is {\it not} $K$-corrected to the galaxy restframe. 
A host galaxy is associated with each SN by determining the nearest object,
based on a measure of the galaxy image size.
Specifically, the SDSS DR5 catalog includes the 
parameters of an iso-photal
ellipse for each galaxy-like object, and for each of the
$ugriz$ filter bands, defined as the ellipse where the
object surface brightness is 25 magnitudes 
arcsec$^{-2}$ \citep{SDSS_EDR}.
We define the distance to a potential host galaxy to be the
semi-major axis of the ellipse that is similar 
(has the same aspect ratio and orientation) to the $r$-band 
iso-photal ellipse and that intersects with the position of the
SN. 
The host galaxy for each SN is defined to be the nearest object
in this measure. For the SNe listed in Table \ref{tab:table1-sn},
the association of SN with host galaxy was confirmed through visual
inspection of the images.

Early-type galaxies generally display red colors (large $u-r$), are reasonably 
well fit by a de Vaucouleurs surface brightness profile (large values of the 
relative de Vaucouleurs likelihood), and show relatively strong central 
light concentration (low values of the inverse concentration index). 
Consequently, these three indicators are strongly correlated and tend to 
give a consistent classification into early and late photometric types, indicated by 
the last column in Table \ref{tab:tab-host}. However,  
the classifications based on the three indicators do not always agree, in which case 
we have made a judgement based on visual inspection of the galaxy image and, 
where available, a high signal-to-noise galaxy spectrum. For these five cases, 
the host type indicated in the last column is marked with an asterisk.
\clearpage
\input{tab3.tex}
\clearpage

Of the three photometric type indicators, $u-r$ correlates most strongly with the 
host type we have assigned to each galaxy in the last column of Table \ref{tab:tab-host}. 
The distribution of SDSS galaxies is approximately bimodal 
in $u-r$ \citep{Strateva_01}, suggesting a natural division between early and late 
types. We therefore use $u-r$ as the galaxy classifier for the purpose of studying 
the SN rate vs.~galaxy type. This is preferable to using the `host type' 
classification in Table \ref{tab:tab-host}, since the subjective human judgement required 
to determine the latter makes it difficult to determine its population properties. 
\citet{Strateva_01} suggest that $u-r = 2.2$ is an optimal separator between 
early and late types. However, our catalog of galaxies with photometric redshifts 
in stripe 82~\citep{Oyaizu_07} appears to be better separated into two subpopulations using $u-r=2.4$. 
Since  
a division at $u-r=2.4$ also provides better agreement with the subjective `host type' 
classification in Table \ref{tab:tab-host}, we use this color cut to separate 
the hosts into early ($u-r>2.4$) and late ($u-r<2.4$) types for the relative 
rate measurement. Using a large sample of galaxies 
from the SDSS DR5 database, we find that the fractional $r$-band
luminosity densities for early and late-type galaxies 
at redshifts $z\le 0.12$ are $54\%$ and $46\%$,
respectively. From Table \ref{tab:tab-host}, we find that the SN Ia rate per unit $r$-band 
luminosity is $\sim 1.68^{+0.52}_{-0.41}$ times 
higher in late-type galaxies 
than in early-type galaxies.
Using the luminosity functions of \citet{Blanton_03}, 
we find that the absolute 
rates per unit luminosity are   
$r_L/h_{70}^2 = 
0.085^{+0.03}_{-0.02}~\mathrm{SNu}_{r}~\mathrm{~(early)~and } 
~0.142^{+0.04}_{-0.03}~\mathrm{SNu}_{r}~\mathrm{~(late)}$. 
The evidence for a larger SN Ia rate in late-type galaxies is statistically marginal with 
the current low-redshift sample. The systematic uncertainty is also significant: if 
we place the host galaxy type cut at $u-r=2.2$, we find no significant difference between 
the rate per unit luminosity in early- and late-type hosts.

\section{Fitting SN Rate evolution models}
\label{sec:rate_models}

As noted in \S \ref{sec:intro}, models for SN Ia progenitors in principle can be 
distinguished by their predictions for the evolution of the SN Ia rate with cosmic 
time. In this section, we present a general maximum likelihood method of fitting 
SN observations to models with a redshift-dependent SN rate. We then apply the method 
to a recently discussed SN Ia rate model, using data from the \sns~and from 
other published rate measurements.

\subsection{Maximum Likelihood method}
\label{sec:maxlike}
In this section we describe a method for fitting SN data to 
models of the SN rate without binning the data. The
method is similar to the
methods described in \citet{Strolger_04}.
and \citet{Strolger_06}. 
A distinguishing feature of our analysis is that 
it allows for combining multiple 
data sets and accounts for systematic errors. 

A general model for the 
volumetric SN rate can be written as $r_V(z;\pars)$, 
where \pars~represents the set of model parameters.  
According to the model, the total number of detected SNe follows a Poisson 
distribution with mean value 

\begin{equation}
\avgnp = \int_{0}^{\infty} dz ~\Theta T \vareff(z)~\frac{r_V(z;\pars)}{(1+z)} 
~u^2(z) ~\frac{du}{dz} ~,
\label{eq:Np}
\end{equation}

\noindent where all symbols were defined in \S \ref{sec:results}. 
The probability 
of detecting a SN at redshift $z$ is given by the integrand of Eqn.(\ref{eq:Np}), 
$P(z_i) \propto d\avgnp/dz$,
giving a likelihood function
for detecting SNe at the $N$ observed redshifts $\{z_i\}$, 

\begin{equation}
L(\{z_i\};\pars) = \frac{e^{-\avgnp} \avgnp^{N}}{N!} 
\prod_{i=1}^{N} \frac{1}{\avgnp} \frac{d \avgnp}{dz}.
\end{equation}

The corresponding log-likelihood function,
suppressing terms
that do not depend on the model parameters, is 

\begin{equation}
\log{L(\{z_i\};\pars)} = -\avgnp 
+  \sum_{i=1}^{N} \log \left(\Theta T \vareff(z_i)~\frac{r_V(z_i;\pars)}{(1+z_i)} ~u^2(z_i) \left[\frac{du}{dz}\right]_{z_i}\right)~.
\end{equation}

\noindent The best-fit model is determined by maximizing the log-likelihood  
with respect to the model parameters, $\pars$.
To incorporate information about the
systematic error in our fits, we weight the contribution 
to the log-likelihood for each data set by 
multiplying each term in the 
log-likelihood function by the factor 
$\sigma^2_{stat}/(\sigma^2_{stat}+\sigma^2_{syst})$, where 
$\sigma_{stat}$ and $\sigma_{syst}$ are the statistical
and systematic errors for the measurement. 
This factor assumes that the systematic errors are approximately
Gaussian and independent of the statistical errors.
We note that auxiliary information about the 
model parameters and uncertainties in 
the survey parameters $\Theta$, $T$, and $\vareff(z)$ could be incorporated 
in a more rigorous way via prior probability distributions. However, this would
require full knowledge of the probability distribution
functions for the efficiency, subject to all 
possible variations of systematic effects, which is in practice unknown.

To combine data from multiple surveys,
the log-likelihood functions for each survey are added together, using 
the appropriate values of
$\vareff(z)$, $\Theta$, and $T$ for each survey. The advantage of this method is 
that it does not involve binning the SN data in redshift; however, it does require 
knowledge of the efficiency function $\vareff(z)$ for each survey.
To evaluate the goodness of fit of a given model, one can use, e.g., 
the Kolmogorov-Smirnov (KS) test
applied to the data and to a large-statistics Monte Carlo 
sample generated from the best-fit model 
parameters.
The code for fitting rate models was tested on large 
MC samples ($\sim 1000$ SNe), and the MC model parameters 
were accurately recovered.

As an illustration of the likelihood method, 
we apply it to the \sns~data, assuming 
a redshift-independent model over the redshift range probed by the data, $r_V(z)=$ constant. 
In this case, the rate that maximizes the likelihood 
can be shown analytically to be given by Eqn. \ref{eqn:orate}.
The probability for this model from the KS test statistic is $p_{KS} =0.42$, 
meaning that if the model is correct, $42\%$ of sample observations drawn 
from the model would have a KS test statistic as large or larger than that
found in comparing this data set to the model. 
In the discussion that follows the
probabilities from the KS test are given as rough estimators of the
goodness of fit only; the distribution of the 
KS test statistic does not in general have an analytic form when 
model parameters are estimated from the data. 

\subsection{SN Rate Models and Star Formation History}
\label{sec:evolution}
As discussed in \S \ref{sec:intro}, measurements of the 
SN Ia rate provide a means to distinguish between 
models of SN Ia progenitor systems.
The connection between the observed SN Ia rate and the 
progenitor systems is made through 
the relation of the SN rate to the cosmic star formation history. 
Sometime after a population of stars form, a fraction of them will end 
up in binary systems that are producing SN Ia explosions. If we denote 
the distribution of delay times between formation of the progenitor 
systems and the SN explosions by $D(t)$, then the volumetric 
SN Ia rate $r_V(t)$ and the cosmic star formation rate $\dot{\rho}(t)$ 
are related by

\begin{eqnarray}
r_V(t) = \int_{0}^{t} dt' \dot{\rho}(t') D(t-t') ~. 
\end{eqnarray}

\noindent We can therefore constrain models for 
the distribution of delay times, $D(t)$, by comparing the SN Ia rate and 
the star formation rate. A discussion of
predicted delay-time distributions for a variety of SN Ia progenitor models is given 
in \citet{Greggio_05}. 
A simple model distribution
that allows for two distinct contributions to the
SN Ia rate is

\begin{equation}
D(t) =  A + B \delta(t) \label{eq:ABmodel}
\end{equation}

\noindent 
where $\delta(t)$ is the Dirac delta function.
This '$A+B$' model was proposed by 
\citet{Mannucci_05} and \citet{ScanBildsten} and it has been used in SN rate
studies by the SNLS \citep{Neill_06} and \citet{Sullivan_06}. The 
SN rate can be written 
$r_V(t) = A\rho(t) + B\dot{\rho}(t)$, where $\rho(t)$ is 
the stellar mass density. 
The $B$ term 
represents an instantaneous or prompt SN Ia 
component and the $A$ term represents an extended 
component in which SNe Ia form
with uniform probability in the time interval following star formation. 
In addition to the '$A+B$' model, we also consider 
a simple model in which $r_V(t)$ evolves as 
a power law in redshift, independent of considerations of 
star formation history.

\subsection{Rate Measurements: Combining Data Sets}
\label{sec:combine}
\begin{figure} [!t]
\begin{center}
\plotone{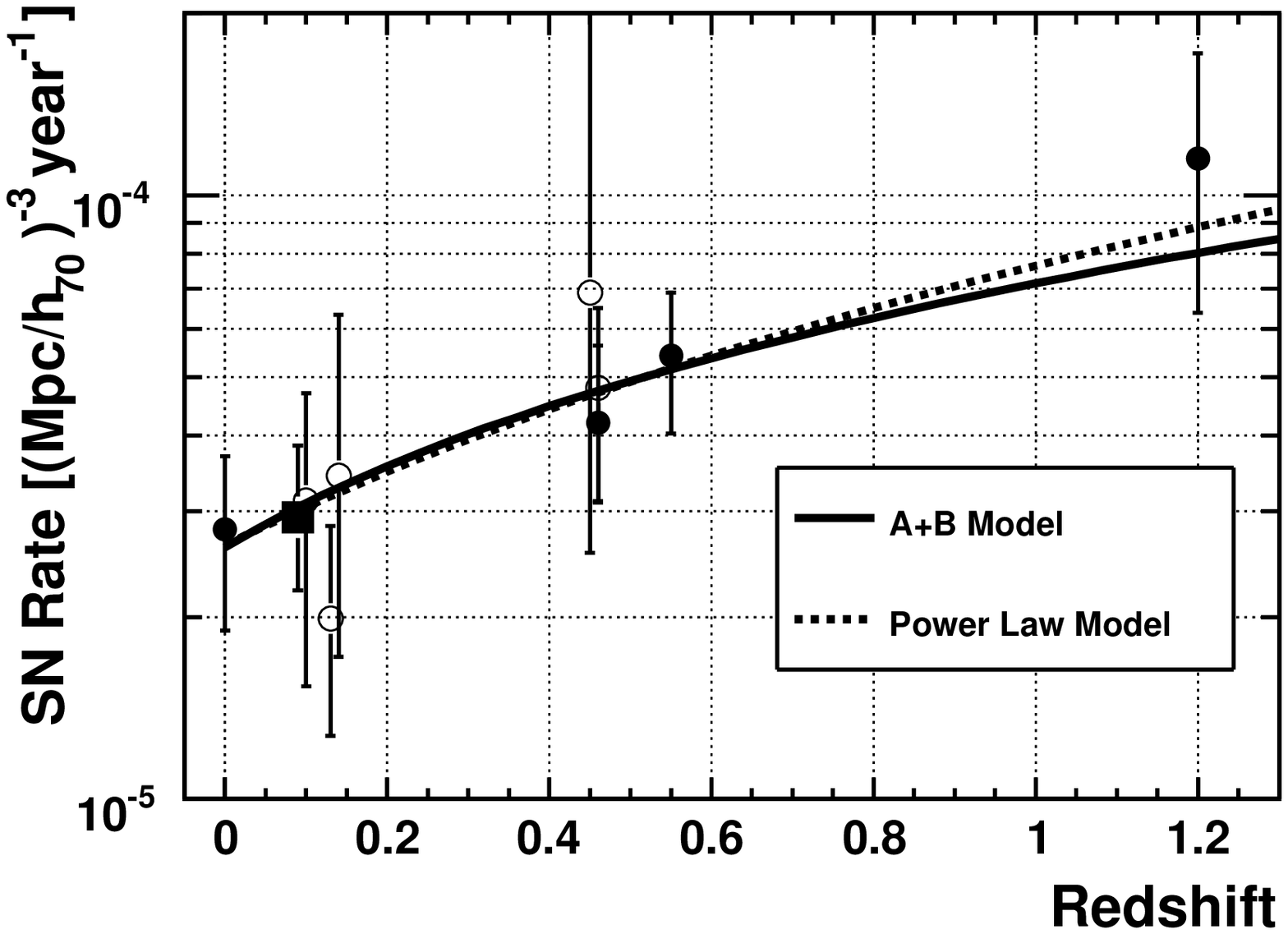}
\end{center}
\caption{Measurements of the SN Ia rate 
discussed in \S \ref{sec:combine}. 
The \sns~measurement in this paper is shown as the solid black square. 
Measurements for which the data is used in the model fits are shown
as solid circles (see Table \ref{tab:rcomp}), 
and measurements not used in the fits as open circles.
To plot each measurement, we have assumed in each case a model in 
which the rate is constant over the redshift range covered by that 
measurement.
The rate as a function of redshift for the best fitting 
'$A+B$' and power law models are overlaid. 
}
\label{fig:ratescomp0}
\end{figure}

The constraints on redshift-dependent models of the SN Ia rate are
improved if one uses SN observations over a wide range of redshifts.
In the following, we combine the low-redshift rate 
measurement from the \sns~with other SN Ia rate measurements in the 
literature. For each data set, 
we require both the SN redshifts and an estimate of the
redshift-dependent selection function $\vareff(z)$, and we therefore
restrict ourselves to using data sets for which it
is straightforward to infer the redshift dependence of the
efficiency. 
We note that several authors, 
including~\citet{Barris_06} 
and~\citet{Poznanski_07} have made SN rate measurements based on
samples of photometrically identified SNe. 
However, in combining data sets for the present 
analysis, we will restrict ourselves to rate measurements
that are based primarily on spectroscopically identified SNe.
Of the nine previously published rate measurements 
that have been based on primarily spectroscopically
identified SNe, shown in Fig. \ref{fig:ratescomp0}, we 
will make use of four, in addition to the one in this work. 
These five rate measurements are shown in bold font in Table \ref{tab:rcomp}.
The weighting factors,
used to account for the systematic uncertainty on each measurement, 
are listed in the last column of Table \ref{tab:rcomp}. 
In cases where
the uncertainty on the measurement is asymmetric, 
we define the 
weighting factor to be the mean of the upper and lower weighting 
factors. Varying the weighting factor between the extremes of using 
the smaller weight, and of using the larger weight, the best-fit 
parameters change by $\sim 5\%$ of the statistical error.
In the subsections below, we briefly describe the data 
from other measurements that we include in the model fits and 
how we describe their efficiency function. 
We also discuss measurements that we exclude from the model fits. 
\clearpage
\input{tab4.tex}
\clearpage

\begin{figure} [!ht]
\begin{center}
\plotone{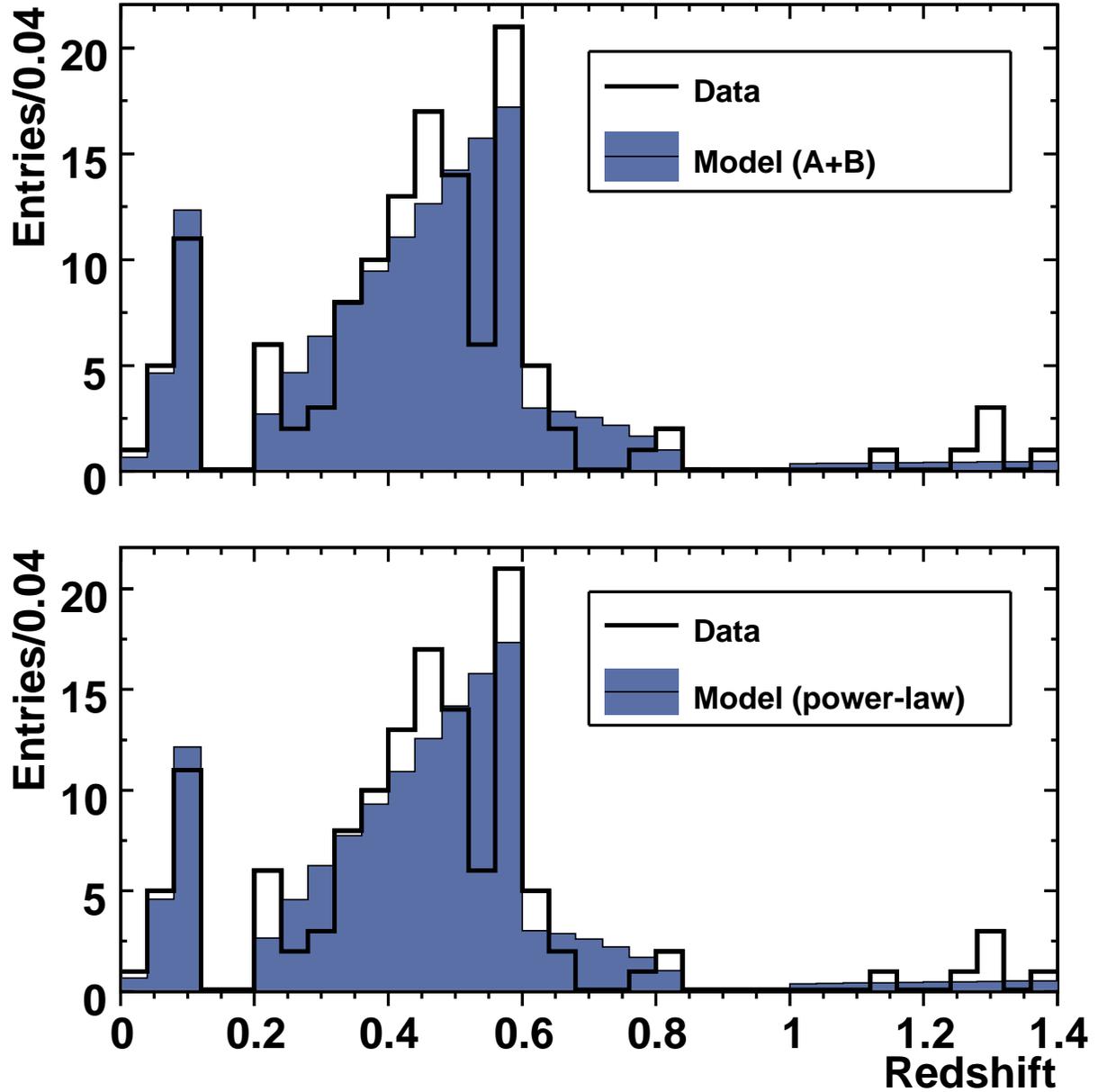}
\end{center}
\caption{Comparison of the observed distribution of SNe
and the predicted distributions for the $A+B$ and power law
rate models. In each panel the shaded region
shows the predicted redshift distribution of the best-fit model. 
The figures include the five highlighted data sets in Table \ref{tab:rcomp}.
}
\label{fig:rateshisto}
\end{figure}
\clearpage

\subsubsection{Other rate measurement data included in the model fits}
\citet{Neill_06} measured the SN Ia rate using 73 SNe Ia from the SNLS.
They state that their sample is spectroscopically complete, i.e., 
that $\vareff(z)$ is constant, to $z=0.6$. 
Including the solid-angle and survey observation time, the factor 
$\Theta T \epsilon(z) = 7.37 \times 10^{-4}$ steradian year.

The measurement of \citet{Pain_02} is based on data from the 
Supernova Cosmology Project, covering about 12 square degrees.
Although \citet{Pain_02} do not give $\vareff(z)$
explicitly, they do provide the redshift distribution
of SNe recovered from their Monte Carlo simulations which 
assumed a constant rate per unit comoving volume. 
With this information, we can compute the relative 
number of MC-generated SNe in each redshift bin and 
thereby the redshift dependence
of their efficiency function. Fitting a quadratic function to this
tabulated efficiency function in the range $0.25 < z < 0.85$ gives 
$\Theta T \epsilon(z) = (\paina +\painb z - \painc z^2)\times 10^{-4}$ 
steradian year.

\citet{Cappellaro_99} measured the SN Ia rate 
for the local Universe by combining data from a number of surveys, 
including visual searches of nearby galaxies. Although they 
do not provide 
an efficiency function or a redshift distribution, 
the redshift range covered by the measurement is so small that
we take the quoted result to be the SN Ia rate at $z=0$. 
We include the \citet{Cappellaro_99} 
rate measurement by adding a standard $\chi^2$ term to the 
log-likelihood function, i.e., a term of 
the form $(r_V(0;\pars)-r_{V,Capp})^2/2\sigma_{Capp}^2$, 
where $r_V(0;\pars)$ is the model prediction at 
redshift zero, $r_{V,Capp}$ is the \citet{Cappellaro_99} measurement 
in Table \ref{tab:rcomp}, and $\sigma_{Capp}$ is the quoted error 
on the measurement.

\citet{Dahlen_04}
measured the SN Ia rate to $z \sim 1.6$ using data from the Great 
Observatories Origins Deep Survey (GOODS) carried out with the 
Advanced Camera for Surveys on the Hubble Space Telescope (HST). 
\footnote{Just before we submitted this paper, \cite{Kuznetsova_07} released
new SN~Ia rate measurements based on analysis of 57 SNe from HST, including the 42 SNe analyzed by
\citet{Dahlen_04}.}
We use their measurement 
in the redshift range $1.0 < z < 1.4$. 
Using their 
scaled efficiency function, as inferred from Figure 14 of \citet{Strolger_04}, 
we fit a function 
$A + B z + C z^{2} + D z^{3}$, valid in the redshift range 
$1.0 < z < 2.0$. The best-fit parameters are
$A = -7.557, B = 55.93, C = -51.07$ and
$D = 12.5$. 
This function is normalized so that the number of expected SNe for 
their survey is equal to six, the number they observed 
in the redshift interval $1.0 < z < 1.4$, giving a value of 
$\Theta T \epsilon(z) = 
[-10.35 + 76.61 z -69.95 z^2 + 17.12 z^3] \times 10^{-4}$
~steradian year.

The redshift dependence of the efficiency function for the present 
data set is discussed in \S \ref{sec:simeffs};
including the solid-angle and survey observation time,
$\Theta T \epsilon(z) = 
[1.54 - 0.025 z] \times 10^{-2}$
~steradian year.

\subsubsection{Rate measurement data not included in the fits}
\label{sec:exclude}
In fitting the models, we choose not to include 
several of the SN Ia rate measurements listed in Table \ref{tab:rcomp}.
We exclude the \citet{Dahlen_04} rate 
measurement in the redshift range $0.2 < z < 0.6$ because 
the efficiency function, given by \citet{Strolger_04}, 
is only plotted for redshifts greater than 1, and because 
the 73 SNe from SNLS \citep{Neill_06} in the same redshift range 
dominate the fit in comparison to the 3 SNe from \citet{Dahlen_04}.
Similar reasoning holds for the measurement 
of \citet{Tonry_03}, which is based on 8 SNe in the redshift range well
covered by the SNLS, and for which the redshifts are not 
explicitly stated. 
Both the \citet{Hardin_00} and the \citet{Blanc_04} 
rate measurements, based on data from the 
EROS microlensing survey, included a requirement that each 
SN be associated with a host galaxy with apparent magnitude 
$R \lesssim 19$, which introduces a bias against faint hosts. 
If SNe Ia occurred at a constant rate per unit $R$-band luminosity
in all galaxies, this would not be an issue. However, as noted above, it has
been shown that the SN Ia rate per unit stellar mass (for which the 
total $R$-band luminosity is a proxy) is a function of SFR  
\citep{Mannucci_05a, Sullivan_06}. 
Finally, the \citet{Madgwick_03}
measurement is based on SNe discovered via principal component analysis 
in spectra obtained by the SDSS galaxy redshift survey.
This measurement
has significant systematic uncertainties that are
different from those in photometric surveys. 
In particular, the SNe discovered by this technique must lie 
within approximately 1.5'' of the cores of their host galaxies, the 
radius of the SDSS spectroscopic fibers. 
To derive a SN rate from these spectroscopic observations, assumptions
are needed about how SNe are distributed within their host galaxies at larger 
galactocentric distances. 

\subsection{Fits to SN Ia Rate Models}
\label{sec:fits}
We now consider fits of the combined SN Ia rate measurements to the rate 
models discussed in \S \ref{sec:evolution}, using the maximum likelihood 
approach of \S \ref{sec:maxlike}. The errors quoted below are 
the values of the fit parameters for which the log-likelihood function
changes by 1/2 compared to its maximum, which assumes that 
the likelihood function is approximately Gaussian.
We use the MINUIT software package \citep{minuit} for the function optimization and
error analysis.

\subsubsection{Power-law redshift evolution of $r_V$}
We first consider a simple two-parameter model that describes power-law 
redshift evolution of the SN rate independent of consideration 
of the star formation history, $r_V(z) = \alpha (1+z)^{\beta}$.
The best-fit power-law model is shown as the 
dashed curve in Fig. \ref{fig:ratescomp0}, and the predicted
redshift distribution is shown in Fig. \ref{fig:rateshisto}.
Fitting this model to the five data sets, we find
\begin{eqnarray*}
\alpha & = & (2.6^{+0.6}_{-0.5}) \times 10^{-5} ~\mathrm{SNe} ~\mathrm{Mpc}^{-3} ~\mathrm{h}_{70}^{3} ~\mathrm{yr}^{-1} \\
\beta & = & (1.5 \pm 0.6) \\
\rho_{\alpha\beta} & = & -0.80
\end{eqnarray*}
\noindent where $\rho_{\alpha\beta}$ is the 
correlation coefficient between the two fitted parameters.
The KS probability for this model is $p_{KS} = 0.63$. 
We emphasize that the fitted value of $\beta$ is greater than 0,
i.e., the rate is determined to be an increasing function of redshift,
at the $\sim 2.5 \sigma$ level.

\subsubsection{The `A+B' model}
We next consider the 
$`A+B'$ model, with $D(t)$ given by Eqn. \ref{eq:ABmodel}.
As discussed by~\citet{Forster_06}, there is still significant
uncertainty on the cosmic star formation rate (SFR), which
is a limitation for placing observational constraints
on SN delay time models. In what follows we choose one 
estimate of the SFR, and do not propagate the systematic 
uncertainties in the SFR.
We follow the approach of 
\citet{Neill_06} and take the star formation rate from 
\citet{Hopkins_06}. The functional form of the star formation rate
is

\begin{equation}
\dot{\rho}(z) = \frac{a + b z}{1 + (z/c)^{d}}
~h_{100} 
~\mathrm{M_{\sun} ~yr^{-1} ~Mpc^{-3}}~,
\label{eq:sfr}
\end{equation}
 
\noindent where $h_{100} =H_0~(100~{\rm km}~{\rm sec}^{-1}~{\rm Mpc}^{-1})^{-1}$, 
$a=0.0118, b=0.08, c=3.3$, and $d=5.2$. For 
the stellar mass density (the $A$ component) we integrate the 
star formation rate over time; as mentioned by \citet{Neill_06}, this can
be expected to overestimate the total stellar mass density 
relative to estimates of the stellar mass density that are based on 
luminosity, as 
it includes a contribution from stars that have burned out.
Performing the fit using the five data sets gives 
\begin{eqnarray*}
A & = & (2.8 \pm 1.2)  \times  10^{-14} ~\mathrm{SNe} ~\mathrm{M}_{\sun}^{-1} 
~\mathrm{yr}^{-1} \\
B & = & (9.3^{+3.4}_{-3.1})  \times  10^{-4} ~\mathrm{SNe} ~\mathrm{M}_{\sun}^{-1} \\
\rho_{AB} & = & -0.78 
\end{eqnarray*}
\noindent where $\rho_{AB}$ is the 
correlation coefficient between the two fitted parameters.
The KS probability for this model is $p_{KS} = 0.71$.
The best-fit '$A+B$' model is shown as the solid 
curve in Fig. \ref{fig:ratescomp0}, and the predicted
redshift distribution is shown in Fig. \ref{fig:rateshisto}.
We note that the uncertainties on the $A$ and $B$ parameters are
$\sim 43\%$ and $\sim 35\%$, respectively. For comparison, if
we perform a fit to the '$A+B$' model, suppressing the SDSS data,
the uncertainties on the fit parameters are 
$\sim 53\%$ and $\sim 38\%$, respectively.
Our analysis here is similar to that presented by 
\citet{Neill_06}, with the primary differences being that
we use a different subset of the available data, and 
we use a maximum likelihood method to fit the data to models of
the SN rate.
For comparison, \citet{Neill_06} found values of 
$A = (1.4 \pm 1.0)\times 10^{-14} ~\mathrm{SNe} ~\mathrm{M}_{\sun}^{-1} 
~\mathrm{yr}^{-1}$ and 
$B = (8.0 \pm 2.6)\times 10^{-4} ~\mathrm{SNe} ~\mathrm{M}_{\sun}^{-1}$.
Both analyses find evidence for two components to the SN rate
with the significance of the '$A$' (extended) component less than 
that of the '$B$' (prompt) component.

We note that 
one cannot  
accurately judge the goodness of fit of
this model using a visual inspection or $\chi^2$ fit to
Fig. \ref{fig:ratescomp0}, 
since the measurements are each plotted {\it assuming} a  
constant-rate model.
A better picture of the goodness of fit 
is given by Fig.  
\ref{fig:rateshisto}, which shows the
observed redshift distribution for the five data sets compared with  
the predicted
redshift distributions for the `$A+B$' and power-law rate models 
convolved with the measured
efficiency functions for the different measurements. The agreement  
between the
predicted distributions for both evolving models 
and that of the data is quite reasonable.

\section{Conclusions}
We have presented a measurement of the SN Ia rate in the redshift range
$0 < z \le 0.12$ from the first season of the \sns. After selection 
cuts, the rate-measurement sample includes a total of 17 SNe Ia, 
of which 16 were spectroscopically 
confirmed. The final SN in the sample is a highly extincted, photometrically identified 
SN Ia with a measured host galaxy redshift. The insertion of artificial 
SNe in the data stream and the use of detailed Monte Carlo simulations of the 
survey efficiency, along 
with the rolling nature of the \sns, have enabled us to obtain a SN Ia 
rate measurement with smaller systematic uncertainties than previous 
measurements in a comparable redshift range. 

We have also applied  
a maximum-likelihood technique, which 
enables us to account for systematic errors and
to fit multiple SN data sets to models of the SN 
rate as a function of redshift.
This maximum likelihood method makes optimal use of the available data,
but requires estimates of the SN detection efficiency,
and its uncertainty, as a function of redshift.
We 
have applied this technique to a combination of recent SN Ia 
data sets, focusing 
on the '$A+B$' model that relates the SN Ia rate  
to the cosmic star formation 
rate. 

Models in which the SN Ia rate evolves with redshift are preferred 
over a model with a constant rate, but the data 
do not distinguish significantly between 
a simple power-law evolution of the SN Ia rate with redshift
and the '$A+B$' model. 
The $A$ and $B$ 
parameter values we obtain are in good
agreement with the results of \citet{Neill_06}. 

In the near future, we will use \sns~data to extend this study in several 
directions, including a higher-statistics measurement of the 
low-redshift rate, measurement of 
the SN Ia rate vs.~host galaxy star-formation rate and other host galaxy 
properties, 
and measurement of the SN Ia rate to $z \sim 0.3$. The Fall 2006 and 
Fall 2007 observing seasons each yielded $\sim 30$ spectroscopically 
confirmed SNe Ia 
at redshift $z\le0.12$, comparable to the size of the 2005 data set 
analyzed here. The final \sns~sample includes 
of order 500 spectroscopically confirmed SNe Ia to $z<0.4$.

\acknowledgements

Funding for the SDSS and SDSS-II has been provided by the Alfred P. Sloan Foundation, the Participating Institutions, the National Science Foundation, the U.S. Department of Energy, the National Aeronautics and Space Administration, the Japanese Monbukagakusho, the Max Planck Society, and the Higher Education Funding Council for England. The SDSS Web Site is http://www.sdss.org/.

The SDSS is managed by the Astrophysical Research Consortium for the Participating Institutions. The Participating Institutions are the American Museum of Natural History, Astrophysical Institute Potsdam, University of Basel, University of Cambridge, Case Western Reserve University, University of Chicago, Drexel University, Fermilab, the Institute for Advanced Study, the Japan Participation Group, Johns Hopkins University, the Joint Institute for Nuclear Astrophysics, the Kavli Institute for Particle Astrophysics and Cosmology, the Korean Scientist Group, the Chinese Academy of Sciences (LAMOST), Los Alamos National Laboratory, the Max-Planck-Institute for Astronomy (MPIA), the Max-Planck-Institute for Astrophysics (MPA), New Mexico State University, Ohio State University, University of Pittsburgh, University of Portsmouth, Princeton University, the United States Naval Observatory, and the University of Washington.

This work is based in part on observations made at the 
following telescopes.
The Hobby-Eberly Telescope (HET) is a joint project of the University of Texas
at Austin,
the Pennsylvania State University,  Stanford University,
Ludwig-Maximillians-Universit\"at M\"unchen, and Georg-August-Universit\"at
G\"ottingen.  The HET is named in honor of its principal benefactors,
William P. Hobby and Robert E. Eberly.  The Marcario Low-Resolution
Spectrograph is named for Mike Marcario of High Lonesome Optics, who
fabricated several optical elements 
for the instrument but died before its completion;
it is a joint project of the Hobby-Eberly Telescope partnership and the
Instituto de Astronom\'{\i}a de la Universidad Nacional Aut\'onoma de M\'exico.
The Apache 
Point Observatory 3.5 m telescope is owned and operated by 
the Astrophysical Research Consortium. We thank the observatory 
director, Suzanne Hawley, and site manager, Bruce Gillespie, for 
their support of this project.
The Subaru Telescope is operated by the National 
Astronomical Observatory of Japan. 
The William Herschel 
Telescope is operated by the 
Isaac Newton Group, 
on the island of La Palma
in the Spanish Observatorio del Roque 
de los Muchachos of the Instituto de Astrofisica de 
Canarias. 
Kitt Peak National Observatory, National Optical 
Astronomy Observatory, is operated by the Association of 
Universities for Research in Astronomy, Inc. (AURA) under 
cooperative agreement with the National Science Foundation. 

This work was supported in part by the Kavli Institute for Cosmological Physics at the University of Chicago through grants NSF PHY-0114422 and NSF PHY-0551142 and an endowment from the Kavli Foundation and its founder Fred Kavli.


\end{document}

%% file: tab1.tex
\begin{deluxetable}{llrrll}
\tablecolumns{6}
\singlespace
\tablewidth{0pc} 
\tablecaption{SNe Ia included in the rate sample. 
\label{tab:table1-sn}
}
\tablehead{
\colhead{SDSS Id} &    
\colhead{IAUC} &    
\colhead{$\alpha$ (J2000.0)} & 
\colhead{$\delta$ (J2000.0)} & 
\colhead{Redshift}  &
\colhead{Redshift}  \\

&
\colhead{Designation} &
&
&
&
\colhead{Source} 
}

\startdata
1241 & 2005ff & 22 30 41.41 & $-$00 46 35.7 & $ 0.088  $ & SN \\
1371 & 2005fh & 23 17 29.71 & $+$00 25 45.8 & $ 0.120  $ & galaxy \\
2561 & 2005fv & 03 05 22.42 & $+$00 51 30.1 & $ 0.119  $ & galaxy \\
3256 & 2005hn & 21 57 04.23 & $-$00 13 24.4 & $ 0.107  $ & galaxy \\
3592 & 2005gb & 01 16 12.58 & $+$00 47 31.0 & $ 0.086  $ & galaxy \\
3901 & 2005ho & 00 59 24.10 & $+$00 00 09.3 & $ 0.063  $ & galaxy \\
5395 & 2005hr & 03 18 33.81 & $+$00 07 24.3 & $ 0.117  $ & SN \\
5549 & 2005hx & 00 13 00.13 & $+$00 14 53.7 & $ 0.120  $ & SN \\
5944 & 2005hc & 01 56 47.94 & $-$00 12 49.1 & $ 0.046  $ & galaxy \\
6057 & 2005if & 03 30 12.87 & $-$00 58 28.5 & $ 0.067  $ & galaxy \\
6295 & 2005js & 01 34 41.51 & $-$00 36 19.4 & $ 0.084  $ & SN \\
6558 & 2005hj\tablenotemark{a} & 01 26 48.40 & $-$01 14 17.3 & $ 0.057  $ & --- \\
6962 & 2005je & 02 35 26.61 & $+$01 04 29.6 & $ 0.094  $ & galaxy \\
7147 & 2005jh & 23 20 04.42 & $-$00 03 19.8 & $ 0.109  $ & galaxy \\
7876 & 2005ir & 01 16 43.80 & $+$00 47 40.7 & $ 0.076  $ & galaxy \\
8719 & 2005kp & 00 30 53.15 & $-$00 43 07.9 & $ 0.117  $ & galaxy \\
9266\tablenotemark{b} & ---    & 03 20 43.16 & $-$01 00 07.2 & $ 0.036 $ & galaxy \\
\enddata
\tablenotetext{a}{SN type confirmed by~\citet{CBET_266}}
\tablenotetext{b}{Photometrically identified SN Ia. 
See \S (\ref{sec:psample}) }
\tablecomments{SDSS Id denotes internal candidate designation.}
\end{deluxetable}

%% file: tab2.tex
\begin{deluxetable}{llrrrr}
\tablecolumns{5}
\singlespace
\tablewidth{0pc} 
\tablecaption{Spectroscopically confirmed SNe Ia with $z \le 0.12$ cut  
from the rate sample.
\label{tab:sne_cut}
}
\tablehead{
\colhead{SDSS Id} &    
\colhead{IAUC} &    
\colhead{$\alpha$ (J2000.0)} & 
\colhead{$\delta$ (J2000.0)} & 
\colhead{Redshift}  &
\colhead{Cut Index}  \\

&
\colhead{Designation} &
}

\startdata
722   & 2005ed & 00 02 49.37 & $+$00 45 04.6 & 0.086 & 4 \\
739   & 2005ef & 00 58 22.87 & $+$00 40 44.6 & 0.107 & 4 \\
774   & 2005ex & 01 41 51.24 & $-$00 52 35.0 & 0.093 & 4 \\
2102  & 2005fn & 20 48 53.04 & $+$00 11 28.1 & 0.095 & 4 \\
4524  & 2005gj & 03 01 11.95 & $-$00 33 13.9 & 0.062 & 6 \\
6773  & 2005iu & 20 20 15.61 & $+$00 13 02.5 & 0.090 & 1 \\
6968  & ---    & 01 18 13.37 & $-$00 54 23.6 & 0.098 & 6 \\
8151  & 2005hk & 00 27 50.88 & $-$01 11 53.3 & 0.013 & 6 \\
10028 & 2005kt & 01 10 58.04 & $+$00 16 34.1 & 0.066 & 5 \\
10096 & 2005lj & 01 57 43.03 & $-$00 10 46.0 & 0.078 & 5 \\
10434 & 2005lk & 21 59 49.43 & $-$01 11 37.3 & 0.103 & 5,2 \\
10805 & 2005ku & 22 59 42.61 & $-$00 00 49.3 & 0.045 & 5 \\
11067 & 2005ml & 02 14 04.42 & $-$00 14 21.1 & 0.119 & 5 \\
\enddata

\tablecomments{SDSS Id denotes internal candidate designation.
See section \ref{sec:ssample} for explanation of cut index.
}

\end{deluxetable}

%% file: tab3.tex
\begin{deluxetable}{llrrccc}
\tablecolumns{7}
\singlespace
\tablewidth{0pc} 
\tablecaption{Host Galaxies for SNe Ia in the rate sample.
\label{tab:tab-host}
}
\tablehead{
\colhead{SDSS Id} &    
\colhead{$\alpha$ (J2000.0)} & 
\colhead{$\delta$ (J2000.0)} & 
\colhead{$u-r$}  &
\colhead{de Vaucouleurs}  &
\colhead{concentration}  &
\colhead{Host Type} \\
& & & &
\colhead{likelihood} &
\colhead{index} &
}

\startdata
1241 &  22 30 41.15 & $-$00 46 34.5 & 2.82 & 0.91 & 0.382 & Early \\
1371 &  23 17 29.70 & $+$00 25 46.8 & 2.97 & 0.00 & 0.377 & Early\tablenotemark{*} \\
2561 &  03 05 22.64 & $+$00 51 35.0 & 2.59 & 0.00 & 0.410 & Late\tablenotemark{*} \\
3256 &  21 57 04.19 & $-$00 13 24.5 & 1.99 & 0.00 & 0.442 & Late \\
3592 &  01 16 12.71 & $+$00 47 26.0 & 2.21 & 0.00 & 0.427 & Late \\
3901 &  00 59 24.11 & $+$00 00 09.5 & 1.40 & 1.00 & 0.418 & Late\tablenotemark{*} \\
5395 &  03 18 33.80 & $+$00 07 24.0 & 1.29 & 0.49 & 0.360 & Late\tablenotemark{*} \\
5549 &  00 12 59.97 & $+$00 14 54.9 & 1.01 & 0.46 & 0.461 & Late \\
5944 &  01 56 48.50 & $-$00 12 45.3 & 2.57 & 0.00 & 0.469 & Early\tablenotemark{*} \\
6057 &  03 30 12.89 & $-$00 58 28.1 & 1.79 & 0.00 & 0.485 & Late \\
6295 &  01 34 41.84 & $-$00 36 15.2 & 2.97 & 1.00 & 0.312 & Early \\
6558 &  01 26 48.46 & $-$01 14 17.3 & 2.23 & 0.00 & 0.427 & Late \\
6962 &  02 35 26.58 & $+$01 04 28.3 & 2.71 & 1.00 & 0.379 & Early \\
7147 &  23 20 04.44 & $-$00 03 20.2 & 3.22 & 1.00 & 0.350 & Early \\
7876 &  01 16 43.87 & $+$00 47 36.9 & 1.71 & 0.00 & 0.532 & Late \\
8719 &  00 30 53.23 & $-$00 43 07.3 & 1.12 & 0.01 & 0.412 & Late \\
9266 &  03 20 43.19 & $-$01 00 08.2 & 2.25 & 0.00 & 0.398 & Late \\
\enddata
\tablenotetext{*}{At least one of the 
three photometric type indicators indicates a different type from that listed.}

\tablecomments{SDSS Id denotes internal candidate designation. $\alpha$ and $\delta$ are the coordinates of the host galaxy of the SN.
The photometric morphology indicators, $u-r$, de Vaucouleurs likelihood, and concentration index  
are described in (\S \ref{sec:rate_vs_galtype}).}
\end{deluxetable}

%% file: tab4.tex
\begin{deluxetable}{rllccc}
\tablecolumns{9}
\tabletypesize{\small}
\singlespace
\tablewidth{0pc} 
\tablecaption{SN Ia Rate Measurements.
\label{tab:rcomp}
}
\tablehead{
\colhead{Reference} & 
\colhead{Redshift} &    
\colhead{Mean} &    
\colhead{N$_{\mathrm{SNe}}$} & 
\colhead{Rate} & 
\colhead{$\sigma^2_{stat}/\sigma^2_{tot}$}  \\

& 
\colhead{Range} &
\colhead{Redshift} & 
&
\colhead{[$10^{-5}$ SNe $h^{3}_{70}$ Mpc$^{-3}$ yr$^{-1}$]} &

}
\renewcommand{\arraystretch}{1.5}
\startdata
{\bf \citet{Cappellaro_99}\tablenotemark{*}} & $\sim 0$            & $\sim 0$ & $70$ &  $ 2.8 \pm 0.9$         &  N/A \\
{\bf This work\tablenotemark{*}}             & $0 - 0.12$          & $0.09$   & $17$ &  $ 2.9^{+0.9}_{-0.7}$   &  0.988 \\
\citet{Madgwick_03}         & $0 - 0.19$          & $0.10$   & $19$ &  $ 3.1 \pm 1.6$         &  N/A \\
\citet{Blanc_04}            & $0 - 0.3 $          & $0.13$   & $14$ &  $ 2.0^{+0.84}_{-0.72}$ &  N/A \\
\citet{Hardin_00}           & $\sim 0.02 - 0.2 $  & $0.14$   & $4$  &  $ 3.4^{+2.9}_{-1.7}$   &  N/A \\
\citet{Dahlen_04}           & $0.2 - 0.6 $        & $0.45$   & $3$  &  $ 6.9^{+15.8}_{-3.7}$  &  N/A \\
{\bf \citet{Neill_06}\tablenotemark{*}}      & $0.2 - 0.6 $        & $0.45$   & $73$ &  $ 4.2^{+1.4}_{-1.1}$   &  0.492 \\
\citet{Tonry_03}            & $\sim 0.25 - 0.6 $  & $0.46$   & $8$  &  $ 4.8 \pm 1.7$         &  N/A \\
{\bf \citet{Pain_02}\tablenotemark{*a}}       & $0.25 - 0.85 $      & $0.55$   & $37$ &  $ 5.4^{+1.5}_{-1.4}$   &  0.643 \\
{\bf \citet{Dahlen_04}\tablenotemark{*}}     & $1.0 - 1.4 $        & $1.2$    & $6$  &  $11.5^{+4.7}_{-5.1}$   &  0.686 \\
\enddata

\tablenotetext{a}{The value of the rate has been corrected to our assumed cosmology, according to 
equation 3 of~\citet{Pain_02}.
}

\tablecomments{Measurements included 
in the model fits are shown in bold face, and are marked with an asterisk.
Mean redshift refers to the mean of the expected SN 
redshift distribution, under the assumption of a constant SN rate.
For \citet{Madgwick_03}, this is estimated as the mean of the
observed SN redshift distribution.
Systematic and statistical errors, 
when reported separately, have been combined in quadrature. 
Rate measurements reported here assume constant volumetric 
rate over the range of each survey.
$\sigma_{stat}$ is the reported statistical error on the measurement. 
$\sigma_{tot}$ is the sum in quadrature of the reported statistical and systematic errors.}

\end{deluxetable}

%% file: ms.bbl
\begin{thebibliography}{70}
\expandafter\ifx\csname natexlab\endcsname\relax\def\natexlab#1{#1}\fi

\bibitem[{{Adelman-McCarthy} {et~al.}(2007){Adelman-McCarthy}, {Ag{\"u}eros},
  {Allam}, {Anderson}, {Anderson}, {Annis}, {Bahcall}, {Bailer-Jones},
  {Baldry}, {Barentine}, {Beers}, {Belokurov}, {Berlind}, {Bernardi},
  {Blanton}, {Bochanski}, {Boroski}, {Bramich}, {Brewington}, {Brinchmann},
  {Brinkmann}, {Brunner}, {Budav{\'a}ri}, {Carey}, {Carliles}, {Carr},
  {Castander}, {Connolly}, {Cool}, {Cunha}, {Csabai}, {Dalcanton}, {Doi},
  {Eisenstein}, {Evans}, {Evans}, {Fan}, {Finkbeiner}, {Friedman}, {Frieman},
  {Fukugita}, {Gillespie}, {Gilmore}, {Glazebrook}, {Gray}, {Grebel}, {Gunn},
  {de Haas}, {Hall}, {Harvanek}, {Hawley}, {Hayes}, {Heckman}, {Hendry},
  {Hennessy}, {Hindsley}, {Hirata}, {Hogan}, {Hogg}, {Holtzman}, {Ichikawa},
  {Ichikawa}, {Ivezi{\'c}}, {Jester}, {Johnston}, {Jorgensen}, {Juri{\'c}},
  {Kauffmann}, {Kent}, {Kleinman}, {Knapp}, {Kniazev}, {Kron}, {Krzesinski},
  {Kuropatkin}, {Lamb}, {Lampeitl}, {Lee}, {Leger}, {Lima}, {Lin}, {Long},
  {Loveday}, {Lupton}, {Mandelbaum}, {Margon}, {Mart{\'{\i}}nez-Delgado},
  {Matsubara}, {McGehee}, {McKay}, {Meiksin}, {Munn}, {Nakajima}, {Nash},
  {Neilsen}, {Newberg}, {Nichol}, {Nieto-Santisteban}, {Nitta}, {Oyaizu},
  {Okamura}, {Ostriker}, {Padmanabhan}, {Park}, {Peoples}, {Pier}, {Pope},
  {Pourbaix}, {Quinn}, {Raddick}, {Re Fiorentin}, {Richards}, {Richmond},
  {Rix}, {Rockosi}, {Schlegel}, {Schneider}, {Scranton}, {Seljak}, {Sheldon},
  {Shimasaku}, {Silvestri}, {Smith}, {Smol{\v c}i{\'c}}, {Snedden}, {Stebbins},
  {Stoughton}, {Strauss}, {SubbaRao}, {Suto}, {Szalay}, {Szapudi}, {Szkody},
  {Tegmark}, {Thakar}, {Tremonti}, {Tucker}, {Uomoto}, {Vanden Berk},
  {Vandenberg}, {Vidrih}, {Vogeley}, {Voges}, {Vogt}, {Weinberg}, {West},
  {White}, {Wilhite}, {Yanny}, {Yocum}, {York}, {Zehavi}, {Zibetti}, \&
  {Zucker}}]{SDSS_DR5}
{Adelman-McCarthy}, J.~K., {Ag{\"u}eros}, M.~A., {Allam}, S.~S., {Anderson},
  K.~S.~J., {Anderson}, S.~F., {Annis}, J., {Bahcall}, N.~A., {Bailer-Jones},
  C.~A.~L., {Baldry}, I.~K., {Barentine}, J.~C., {Beers}, T.~C., {Belokurov},
  V., {Berlind}, A., {Bernardi}, M., {Blanton}, M.~R., {Bochanski}, J.~J.,
  {Boroski}, W.~N., {Bramich}, D.~M., {Brewington}, H.~J., {Brinchmann}, J.,
  {Brinkmann}, J., {Brunner}, R.~J., {Budav{\'a}ri}, T., {Carey}, L.~N.,
  {Carliles}, S., {Carr}, M.~A., {Castander}, F.~J., {Connolly}, A.~J., {Cool},
  R.~J., {Cunha}, C.~E., {Csabai}, I., {Dalcanton}, J.~J., {Doi}, M.,
  {Eisenstein}, D.~J., {Evans}, M.~L., {Evans}, N.~W., {Fan}, X., {Finkbeiner},
  D.~P., {Friedman}, S.~D., {Frieman}, J.~A., {Fukugita}, M., {Gillespie}, B.,
  {Gilmore}, G., {Glazebrook}, K., {Gray}, J., {Grebel}, E.~K., {Gunn}, J.~E.,
  {de Haas}, E., {Hall}, P.~B., {Harvanek}, M., {Hawley}, S.~L., {Hayes}, J.,
  {Heckman}, T.~M., {Hendry}, J.~S., {Hennessy}, G.~S., {Hindsley}, R.~B.,
  {Hirata}, C.~M., {Hogan}, C.~J., {Hogg}, D.~W., {Holtzman}, J.~A.,
  {Ichikawa}, S.-i., {Ichikawa}, T., {Ivezi{\'c}}, {\v Z}., {Jester}, S.,
  {Johnston}, D.~E., {Jorgensen}, A.~M., {Juri{\'c}}, M., {Kauffmann}, G.,
  {Kent}, S.~M., {Kleinman}, S.~J., {Knapp}, G.~R., {Kniazev}, A.~Y., {Kron},
  R.~G., {Krzesinski}, J., {Kuropatkin}, N., {Lamb}, D.~Q., {Lampeitl}, H.,
  {Lee}, B.~C., {Leger}, R.~F., {Lima}, M., {Lin}, H., {Long}, D.~C.,
  {Loveday}, J., {Lupton}, R.~H., {Mandelbaum}, R., {Margon}, B.,
  {Mart{\'{\i}}nez-Delgado}, D., {Matsubara}, T., {McGehee}, P.~M., {McKay},
  T.~A., {Meiksin}, A., {Munn}, J.~A., {Nakajima}, R., {Nash}, T., {Neilsen},
  Jr., E.~H., {Newberg}, H.~J., {Nichol}, R.~C., {Nieto-Santisteban}, M.,
  {Nitta}, A., {Oyaizu}, H., {Okamura}, S., {Ostriker}, J.~P., {Padmanabhan},
  N., {Park}, C., {Peoples}, J.~J., {Pier}, J.~R., {Pope}, A.~C., {Pourbaix},
  D., {Quinn}, T.~R., {Raddick}, M.~J., {Re Fiorentin}, P., {Richards}, G.~T.,
  {Richmond}, M.~W., {Rix}, H.-W., {Rockosi}, C.~M., {Schlegel}, D.~J.,
  {Schneider}, D.~P., {Scranton}, R., {Seljak}, U., {Sheldon}, E., {Shimasaku},
  K., {Silvestri}, N.~M., {Smith}, J.~A., {Smol{\v c}i{\'c}}, V., {Snedden},
  S.~A., {Stebbins}, A., {Stoughton}, C., {Strauss}, M.~A., {SubbaRao}, M.,
  {Suto}, Y., {Szalay}, A.~S., {Szapudi}, I., {Szkody}, P., {Tegmark}, M.,
  {Thakar}, A.~R., {Tremonti}, C.~A., {Tucker}, D.~L., {Uomoto}, A., {Vanden
  Berk}, D.~E., {Vandenberg}, J., {Vidrih}, S., {Vogeley}, M.~S., {Voges}, W.,
  {Vogt}, N.~P., {Weinberg}, D.~H., {West}, A.~A., {White}, S.~D.~M.,
  {Wilhite}, B., {Yanny}, B., {Yocum}, D.~R., {York}, D.~G., {Zehavi}, I.,
  {Zibetti}, S., \& {Zucker}, D.~B. 2007, \apjs, 172, 634

\bibitem[{{Alard} \& {Lupton}(1998)}]{Alard_98}
{Alard}, C. \& {Lupton}, R.~H. 1998, \apj, 503, 325

\bibitem[{{Aldering} {et~al.}(2006){Aldering}, {Antilogus}, {Bailey}, {Baltay},
  {Bauer}, {Blanc}, {Bongard}, {Copin}, {Gangler}, {Gilles}, {Kessler},
  {Kocevski}, {Lee}, {Loken}, {Nugent}, {Pain}, {P{\'e}contal}, {Pereira},
  {Perlmutter}, {Rabinowitz}, {Rigaudier}, {Scalzo}, {Smadja}, {Thomas},
  {Wang}, \& {Weaver}}]{Aldering_06}
{Aldering}, G., {Antilogus}, P., {Bailey}, S., {Baltay}, C., {Bauer}, A.,
  {Blanc}, N., {Bongard}, S., {Copin}, Y., {Gangler}, E., {Gilles}, S.,
  {Kessler}, R., {Kocevski}, D., {Lee}, B.~C., {Loken}, S., {Nugent}, P.,
  {Pain}, R., {P{\'e}contal}, E., {Pereira}, R., {Perlmutter}, S.,
  {Rabinowitz}, D., {Rigaudier}, G., {Scalzo}, R., {Smadja}, G., {Thomas},
  R.~C., {Wang}, L., \& {Weaver}, B.~A. 2006, \apj, 650, 510

\bibitem[{{Barris} \& {Tonry}(2006)}]{Barris_06}
{Barris}, B.~J. \& {Tonry}, J.~L. 2006, \apj, 637, 427

\bibitem[{{Blanc} {et~al.}(2004){Blanc}, {Afonso}, {Alard}, {Albert},
  {Aldering}, {Amadon}, {Andersen}, {Ansari}, {Aubourg}, {Balland}, {Bareyre},
  {Beaulieu}, {Charlot}, {Conley}, {Coutures}, {Dahl{\'e}n}, {Derue}, {Fan},
  {Ferlet}, {Folatelli}, {Fouqu{\'e}}, {Garavini}, {Glicenstein}, {Goldman},
  {Goobar}, {Gould}, {Graff}, {Gros}, {Haissinski}, {Hamadache}, {Hardin},
  {Hook}, {de Kat}, {Kent}, {Kim}, {Lasserre}, {Le Guillou}, {Lesquoy}, {Loup},
  {Magneville}, {Marquette}, {Maurice}, {Maury}, {Milsztajn}, {Moniez},
  {Mouchet}, {Newberg}, {Nobili}, {Palanque-Delabrouille}, {Perdereau},
  {Pr{\'e}vot}, {Rahal}, {Regnault}, {Rich}, {Ruiz-Lapuente}, {Spiro},
  {Tisserand}, {Vidal-Madjar}, {Vigroux}, {Walton}, \& {Zylberajch}}]{Blanc_04}
{Blanc}, G., {Afonso}, C., {Alard}, C., {Albert}, J.~N., {Aldering}, G.,
  {Amadon}, A., {Andersen}, J., {Ansari}, R., {Aubourg}, {\'E}., {Balland}, C.,
  {Bareyre}, P., {Beaulieu}, J.~P., {Charlot}, X., {Conley}, A., {Coutures},
  C., {Dahl{\'e}n}, T., {Derue}, F., {Fan}, X., {Ferlet}, R., {Folatelli}, G.,
  {Fouqu{\'e}}, P., {Garavini}, G., {Glicenstein}, J.~F., {Goldman}, B.,
  {Goobar}, A., {Gould}, A., {Graff}, D., {Gros}, M., {Haissinski}, J.,
  {Hamadache}, C., {Hardin}, D., {Hook}, I.~M., {de Kat}, J., {Kent}, S.,
  {Kim}, A., {Lasserre}, T., {Le Guillou}, L., {Lesquoy}, {\'E}., {Loup}, C.,
  {Magneville}, C., {Marquette}, J.~B., {Maurice}, {\'E}., {Maury}, A.,
  {Milsztajn}, A., {Moniez}, M., {Mouchet}, M., {Newberg}, H., {Nobili}, S.,
  {Palanque-Delabrouille}, N., {Perdereau}, O., {Pr{\'e}vot}, L., {Rahal},
  Y.~R., {Regnault}, N., {Rich}, J., {Ruiz-Lapuente}, P., {Spiro}, M.,
  {Tisserand}, P., {Vidal-Madjar}, A., {Vigroux}, L., {Walton}, N.~A., \&
  {Zylberajch}, S. 2004, \aap, 423, 881

\bibitem[{{Blanton} {et~al.}(2003){Blanton}, {Hogg}, {Bahcall}, {Brinkmann},
  {Britton}, {Connolly}, {Csabai}, {Fukugita}, {Loveday}, {Meiksin}, {Munn},
  {Nichol}, {Okamura}, {Quinn}, {Schneider}, {Shimasaku}, {Strauss}, {Tegmark},
  {Vogeley}, \& {Weinberg}}]{Blanton_03}
{Blanton}, M.~R., {Hogg}, D.~W., {Bahcall}, N.~A., {Brinkmann}, J., {Britton},
  M., {Connolly}, A.~J., {Csabai}, I., {Fukugita}, M., {Loveday}, J.,
  {Meiksin}, A., {Munn}, J.~A., {Nichol}, R.~C., {Okamura}, S., {Quinn}, T.,
  {Schneider}, D.~P., {Shimasaku}, K., {Strauss}, M.~A., {Tegmark}, M.,
  {Vogeley}, M.~S., \& {Weinberg}, D.~H. 2003, \apj, 592, 819

\bibitem[{{Blondin} \& {Tonry}(2007)}]{Blondin_07}
{Blondin}, S. \& {Tonry}, J.~L. 2007, \apj, 666, 1024

\bibitem[{{Branch} {et~al.}(1995){Branch}, {Livio}, {Yungelson}, {Boffi}, \&
  {Baron}}]{Branch_95}
{Branch}, D., {Livio}, M., {Yungelson}, L.~R., {Boffi}, F.~R., \& {Baron}, E.
  1995, \pasp, 107, 1019

\bibitem[{{Cappellaro} {et~al.}(2007){Cappellaro}, {Botticella}, \&
  {Greggio}}]{Cappellaro_07}
{Cappellaro}, E., {Botticella}, M.~T., \& {Greggio}, L. 2007,
  astro-ph/0706.1299

\bibitem[{{Cappellaro} {et~al.}(1999){Cappellaro}, {Evans}, \&
  {Turatto}}]{Cappellaro_99}
{Cappellaro}, E., {Evans}, R., \& {Turatto}, M. 1999, \aap, 351, 459

\bibitem[{{Cardelli} {et~al.}(1989){Cardelli}, {Clayton}, \& {Mathis}}]{ccm89}
{Cardelli}, J.~A., {Clayton}, G.~C., \& {Mathis}, J.~S. 1989, \apj, 345, 245

\bibitem[{{Dahlen} {et~al.}(2004){Dahlen}, {Strolger}, {Riess}, {Mobasher},
  {Chary}, {Conselice}, {Ferguson}, {Fruchter}, {Giavalisco}, {Livio}, {Madau},
  {Panagia}, \& {Tonry}}]{Dahlen_04}
{Dahlen}, T., {Strolger}, L.-G., {Riess}, A.~G., {Mobasher}, B., {Chary},
  R.-R., {Conselice}, C.~J., {Ferguson}, H.~C., {Fruchter}, A.~S.,
  {Giavalisco}, M., {Livio}, M., {Madau}, P., {Panagia}, N., \& {Tonry}, J.~L.
  2004, \apj, 613, 189

\bibitem[{{F{\"o}rster} {et~al.}(2006){F{\"o}rster}, {Wolf}, {Podsiadlowski},
  \& {Han}}]{Forster_06}
{F{\"o}rster}, F., {Wolf}, C., {Podsiadlowski}, P., \& {Han}, Z. 2006, \mnras,
  368, 1893

\bibitem[{{Frieman} {et~al.}(2008){Frieman}, {Bassett}, {Becker}, {Choi},
  {Cinabro}, {DeJongh}, {Depoy}, {Dilday}, {Doi}, {Garnavich}, {Hogan},
  {Holtzman}, {Im}, {Jha}, {Kessler}, {Konishi}, {Lampeitl}, {Marriner},
  {Marshall}, {McGinnis}, {Miknaitis}, {Nichol}, {Prieto}, {Riess}, {Richmond},
  {Romani}, {Sako}, {Schneider}, {Smith}, {Takanashi}, {Tokita}, {van der
  Heyden}, {Yasuda}, {Zheng}, {Adelman-McCarthy}, {Annis}, {Assef},
  {Barentine}, {Bender}, {Blandford}, {Boroski}, {Bremer}, {Brewington},
  {Collins}, {Crotts}, {Dembicky}, {Eastman}, {Edge}, {Edmondson}, {Elson},
  {Eyler}, {Filippenko}, {Foley}, {Frank}, {Goobar}, {Gueth}, {Gunn},
  {Harvanek}, {Hopp}, {Ihara}, {Ivezi{\'c}}, {Kahn}, {Kaplan}, {Kent},
  {Ketzeback}, {Kleinman}, {Kollatschny}, {Kron}, {Krzesi{\'n}ski}, {Lamenti},
  {Leloudas}, {Lin}, {Long}, {Lucey}, {Lupton}, {Malanushenko}, {Malanushenko},
  {McMillan}, {Mendez}, {Morgan}, {Morokuma}, {Nitta}, {Ostman}, {Pan},
  {Rockosi}, {Romer}, {Ruiz-Lapuente}, {Saurage}, {Schlesinger}, {Snedden},
  {Sollerman}, {Stoughton}, {Stritzinger}, {Subba Rao}, {Tucker}, {Vaisanen},
  {Watson}, {Watters}, {Wheeler}, {Yanny}, \& {York}}]{Frieman_08}
{Frieman}, J.~A., {Bassett}, B., {Becker}, A., {Choi}, C., {Cinabro}, D.,
  {DeJongh}, F., {Depoy}, D.~L., {Dilday}, B., {Doi}, M., {Garnavich}, P.~M.,
  {Hogan}, C.~J., {Holtzman}, J., {Im}, M., {Jha}, S., {Kessler}, R.,
  {Konishi}, K., {Lampeitl}, H., {Marriner}, J., {Marshall}, J.~L., {McGinnis},
  D., {Miknaitis}, G., {Nichol}, R.~C., {Prieto}, J.~L., {Riess}, A.~G.,
  {Richmond}, M.~W., {Romani}, R., {Sako}, M., {Schneider}, D.~P., {Smith}, M.,
  {Takanashi}, N., {Tokita}, K., {van der Heyden}, K., {Yasuda}, N., {Zheng},
  C., {Adelman-McCarthy}, J., {Annis}, J., {Assef}, R.~J., {Barentine}, J.,
  {Bender}, R., {Blandford}, R.~D., {Boroski}, W.~N., {Bremer}, M.,
  {Brewington}, H., {Collins}, C.~A., {Crotts}, A., {Dembicky}, J., {Eastman},
  J., {Edge}, A., {Edmondson}, E., {Elson}, E., {Eyler}, M.~E., {Filippenko},
  A.~V., {Foley}, R.~J., {Frank}, S., {Goobar}, A., {Gueth}, T., {Gunn}, J.~E.,
  {Harvanek}, M., {Hopp}, U., {Ihara}, Y., {Ivezi{\'c}}, {\v Z}., {Kahn}, S.,
  {Kaplan}, J., {Kent}, S., {Ketzeback}, W., {Kleinman}, S.~J., {Kollatschny},
  W., {Kron}, R.~G., {Krzesi{\'n}ski}, J., {Lamenti}, D., {Leloudas}, G.,
  {Lin}, H., {Long}, D.~C., {Lucey}, J., {Lupton}, R.~H., {Malanushenko}, E.,
  {Malanushenko}, V., {McMillan}, R.~J., {Mendez}, J., {Morgan}, C.~W.,
  {Morokuma}, T., {Nitta}, A., {Ostman}, L., {Pan}, K., {Rockosi}, C.~M.,
  {Romer}, A.~K., {Ruiz-Lapuente}, P., {Saurage}, G., {Schlesinger}, K.,
  {Snedden}, S.~A., {Sollerman}, J., {Stoughton}, C., {Stritzinger}, M., {Subba
  Rao}, M., {Tucker}, D., {Vaisanen}, P., {Watson}, L.~C., {Watters}, S.,
  {Wheeler}, J.~C., {Yanny}, B., \& {York}, D. 2008, \aj, 135, 338

\bibitem[{{Fukugita} {et~al.}(1996){Fukugita}, {Ichikawa}, {Gunn}, {Doi},
  {Shimasaku}, \& {Schneider}}]{Fukugita_96}
{Fukugita}, M., {Ichikawa}, T., {Gunn}, J.~E., {Doi}, M., {Shimasaku}, K., \&
  {Schneider}, D.~P. 1996, \aj, 111, 1748

\bibitem[{{Greggio}(2005)}]{Greggio_05}
{Greggio}, L. 2005, \aap, 441, 1055

\bibitem[{{Gunn} {et~al.}(1998){Gunn}, {Carr}, {Rockosi}, {Sekiguchi}, {Berry},
  {Elms}, {de Haas}, {Ivezi{\'c}}, {Knapp}, {Lupton}, {Pauls}, {Simcoe},
  {Hirsch}, {Sanford}, {Wang}, {York}, {Harris}, {Annis}, {Bartozek},
  {Boroski}, {Bakken}, {Haldeman}, {Kent}, {Holm}, {Holmgren}, {Petravick},
  {Prosapio}, {Rechenmacher}, {Doi}, {Fukugita}, {Shimasaku}, {Okada}, {Hull},
  {Siegmund}, {Mannery}, {Blouke}, {Heidtman}, {Schneider}, {Lucinio}, \&
  {Brinkman}}]{Gunn_98}
{Gunn}, J.~E., {Carr}, M., {Rockosi}, C., {Sekiguchi}, M., {Berry}, K., {Elms},
  B., {de Haas}, E., {Ivezi{\'c}}, {\v Z}., {Knapp}, G., {Lupton}, R., {Pauls},
  G., {Simcoe}, R., {Hirsch}, R., {Sanford}, D., {Wang}, S., {York}, D.,
  {Harris}, F., {Annis}, J., {Bartozek}, L., {Boroski}, W., {Bakken}, J.,
  {Haldeman}, M., {Kent}, S., {Holm}, S., {Holmgren}, D., {Petravick}, D.,
  {Prosapio}, A., {Rechenmacher}, R., {Doi}, M., {Fukugita}, M., {Shimasaku},
  K., {Okada}, N., {Hull}, C., {Siegmund}, W., {Mannery}, E., {Blouke}, M.,
  {Heidtman}, D., {Schneider}, D., {Lucinio}, R., \& {Brinkman}, J. 1998, \aj,
  116, 3040

\bibitem[{{Gunn} {et~al.}(2006){Gunn}, {Siegmund}, {Mannery}, {Owen}, {Hull},
  {Leger}, {Carey}, {Knapp}, {York}, {Boroski}, {Kent}, {Lupton}, {Rockosi},
  {Evans}, {Waddell}, {Anderson}, {Annis}, {Barentine}, {Bartoszek}, {Bastian},
  {Bracker}, {Brewington}, {Briegel}, {Brinkmann}, {Brown}, {Carr},
  {Czarapata}, {Drennan}, {Dombeck}, {Federwitz}, {Gillespie}, {Gonzales},
  {Hansen}, {Harvanek}, {Hayes}, {Jordan}, {Kinney}, {Klaene}, {Kleinman},
  {Kron}, {Kresinski}, {Lee}, {Limmongkol}, {Lindenmeyer}, {Long}, {Loomis},
  {McGehee}, {Mantsch}, {Neilsen}, {Neswold}, {Newman}, {Nitta}, {Peoples},
  {Pier}, {Prieto}, {Prosapio}, {Rivetta}, {Schneider}, {Snedden}, \&
  {Wang}}]{SDSS_telescope}
{Gunn}, J.~E., {Siegmund}, W.~A., {Mannery}, E.~J., {Owen}, R.~E., {Hull},
  C.~L., {Leger}, R.~F., {Carey}, L.~N., {Knapp}, G.~R., {York}, D.~G.,
  {Boroski}, W.~N., {Kent}, S.~M., {Lupton}, R.~H., {Rockosi}, C.~M., {Evans},
  M.~L., {Waddell}, P., {Anderson}, J.~E., {Annis}, J., {Barentine}, J.~C.,
  {Bartoszek}, L.~M., {Bastian}, S., {Bracker}, S.~B., {Brewington}, H.~J.,
  {Briegel}, C.~I., {Brinkmann}, J., {Brown}, Y.~J., {Carr}, M.~A.,
  {Czarapata}, P.~C., {Drennan}, C.~C., {Dombeck}, T., {Federwitz}, G.~R.,
  {Gillespie}, B.~A., {Gonzales}, C., {Hansen}, S.~U., {Harvanek}, M., {Hayes},
  J., {Jordan}, W., {Kinney}, E., {Klaene}, M., {Kleinman}, S.~J., {Kron},
  R.~G., {Kresinski}, J., {Lee}, G., {Limmongkol}, S., {Lindenmeyer}, C.~W.,
  {Long}, D.~C., {Loomis}, C.~L., {McGehee}, P.~M., {Mantsch}, P.~M.,
  {Neilsen}, Jr., E.~H., {Neswold}, R.~M., {Newman}, P.~R., {Nitta}, A.,
  {Peoples}, J.~J., {Pier}, J.~R., {Prieto}, P.~S., {Prosapio}, A., {Rivetta},
  C., {Schneider}, D.~P., {Snedden}, S., \& {Wang}, S.-i. 2006, \aj, 131, 2332

\bibitem[{{Hamuy} {et~al.}(2003){Hamuy}, {Phillips}, {Suntzeff}, \&
  {Maza}}]{Hamuy_03}
{Hamuy}, M., {Phillips}, M., {Suntzeff}, N., \& {Maza}, J. 2003, \iaucirc,
  8151, 2

\bibitem[{{Hamuy} {et~al.}(1996){Hamuy}, {Phillips}, {Suntzeff}, {Schommer},
  {Maza}, \& {Aviles}}]{Hamuy_96a}
{Hamuy}, M., {Phillips}, M.~M., {Suntzeff}, N.~B., {Schommer}, R.~A., {Maza},
  J., \& {Aviles}, R. 1996, \aj, 112, 2391

\bibitem[{{Hardin} {et~al.}(2000){Hardin}, {Afonso}, {Alard}, {Albert},
  {Amadon}, {Andersen}, {Ansari}, {Aubourg}, {Bareyre}, {Bauer}, {Beaulieu},
  {Blanc}, {Bouquet}, {Char}, {Charlot}, {Couchot}, {Coutures}, {Derue},
  {Ferlet}, {Glicenstein}, {Goldman}, {Gould}, {Graff}, {Gros}, {Haissinski},
  {Hamilton}, {de Kat}, {Kim}, {Lasserre}, {Lesquoy}, {Loup}, {Magneville},
  {Mansoux}, {Marquette}, {Maurice}, {Milsztajn}, {Moniez},
  {Palanque-Delabrouille}, {Perdereau}, {Pr{\'e}vot}, {Regnault}, {Rich},
  {Spiro}, {Vidal-Madjar}, {Vigroux}, {Zylberajch}, \& {The EROS
  Collaboration}}]{Hardin_00}
{Hardin}, D., {Afonso}, C., {Alard}, C., {Albert}, J.~N., {Amadon}, A.,
  {Andersen}, J., {Ansari}, R., {Aubourg}, {\'E}., {Bareyre}, P., {Bauer}, F.,
  {Beaulieu}, J.~P., {Blanc}, G., {Bouquet}, A., {Char}, S., {Charlot}, X.,
  {Couchot}, F., {Coutures}, C., {Derue}, F., {Ferlet}, R., {Glicenstein},
  J.~F., {Goldman}, B., {Gould}, A., {Graff}, D., {Gros}, M., {Haissinski}, J.,
  {Hamilton}, J.~C., {de Kat}, J., {Kim}, A., {Lasserre}, T., {Lesquoy},
  {\'E}., {Loup}, C., {Magneville}, C., {Mansoux}, B., {Marquette}, J.~B.,
  {Maurice}, {\'E}., {Milsztajn}, A., {Moniez}, M., {Palanque-Delabrouille},
  N., {Perdereau}, O., {Pr{\'e}vot}, L., {Regnault}, N., {Rich}, J., {Spiro},
  M., {Vidal-Madjar}, A., {Vigroux}, L., {Zylberajch}, S., \& {The EROS
  Collaboration}. 2000, \aap, 362, 419

\bibitem[{{H{\"o}flich} {et~al.}(1996){H{\"o}flich}, {Khokhlov}, {Wheeler},
  {Phillips}, {Suntzeff}, \& {Hamuy}}]{Hoeflich_96}
{H{\"o}flich}, P., {Khokhlov}, A., {Wheeler}, J.~C., {Phillips}, M.~M.,
  {Suntzeff}, N.~B., \& {Hamuy}, M. 1996, \apjl, 472, L81+

\bibitem[{{H{\"o}flich} {et~al.}(1995){H{\"o}flich}, {Khokhlov}, \&
  {Wheeler}}]{Hoeflich_95}
{H{\"o}flich}, P., {Khokhlov}, A.~M., \& {Wheeler}, J.~C. 1995, \apj, 444, 831

\bibitem[{{Hogg} {et~al.}(2001){Hogg}, {Finkbeiner}, {Schlegel}, \&
  {Gunn}}]{Hogg_01}
{Hogg}, D.~W., {Finkbeiner}, D.~P., {Schlegel}, D.~J., \& {Gunn}, J.~E. 2001,
  \aj, 122, 2129

\bibitem[{{Holtzman et al.}(2008)}]{Holtzman_08}
{Holtzman et al.}, J. 2008, submitted to AJ

\bibitem[{{Hopkins} \& {Beacom}(2006)}]{Hopkins_06}
{Hopkins}, A.~M. \& {Beacom}, J.~F. 2006, \apj, 651, 142

\bibitem[{{Howell} {et~al.}(2007){Howell}, {Sullivan}, {Conley}, \&
  {Carlberg}}]{Howell_07}
{Howell}, D.~A., {Sullivan}, M., {Conley}, A., \& {Carlberg}, R. 2007,
  astro-ph/0701912

\bibitem[{{Ivezi{\'c}} {et~al.}(2004){Ivezi{\'c}}, {Lupton}, {Schlegel},
  {Boroski}, {Adelman-McCarthy}, {Yanny}, {Kent}, {Stoughton}, {Finkbeiner},
  {Padmanabhan}, {Rockosi}, {Gunn}, {Knapp}, {Strauss}, {Richards},
  {Eisenstein}, {Nicinski}, {Kleinman}, {Krzesinski}, {Newman}, {Snedden},
  {Thakar}, {Szalay}, {Munn}, {Smith}, {Tucker}, \& {Lee}}]{Ivezic_04}
{Ivezi{\'c}}, {\v Z}., {Lupton}, R.~H., {Schlegel}, D., {Boroski}, B.,
  {Adelman-McCarthy}, J., {Yanny}, B., {Kent}, S., {Stoughton}, C.,
  {Finkbeiner}, D., {Padmanabhan}, N., {Rockosi}, C.~M., {Gunn}, J.~E.,
  {Knapp}, G.~R., {Strauss}, M.~A., {Richards}, G.~T., {Eisenstein}, D.,
  {Nicinski}, T., {Kleinman}, S.~J., {Krzesinski}, J., {Newman}, P.~R.,
  {Snedden}, S., {Thakar}, A.~R., {Szalay}, A., {Munn}, J.~A., {Smith}, J.~A.,
  {Tucker}, D., \& {Lee}, B.~C. 2004, Astronomische Nachrichten, 325, 583

\bibitem[{{Ivezi{\'c}} {et~al.}(2007){Ivezi{\'c}}, {Smith}, {Miknaitis}, {Lin},
  {Tucker}, {Lupton}, {Gunn}, {Knapp}, {Strauss}, {Sesar}, {Doi}, {Tanaka},
  {Fukugita}, {Holtzman}, {Kent}, {Yanny}, {Schlegel}, {Finkbeiner},
  {Padmanabhan}, {Rockosi}, {Juri{\'c}}, {Bond}, {Lee}, {Stoughton}, {Jester},
  {Harris}, {Harding}, {Morrison}, {Brinkmann}, {Schneider}, \&
  {York}}]{Ivezic_07}
{Ivezi{\'c}}, {\v Z}., {Smith}, J.~A., {Miknaitis}, G., {Lin}, H., {Tucker},
  D., {Lupton}, R.~H., {Gunn}, J.~E., {Knapp}, G.~R., {Strauss}, M.~A.,
  {Sesar}, B., {Doi}, M., {Tanaka}, M., {Fukugita}, M., {Holtzman}, J., {Kent},
  S., {Yanny}, B., {Schlegel}, D., {Finkbeiner}, D., {Padmanabhan}, N.,
  {Rockosi}, C.~M., {Juri{\'c}}, M., {Bond}, N., {Lee}, B., {Stoughton}, C.,
  {Jester}, S., {Harris}, H., {Harding}, P., {Morrison}, H., {Brinkmann}, J.,
  {Schneider}, D.~P., \& {York}, D. 2007, \aj, 134, 973

\bibitem[{James \& Roos(1994)}]{minuit}
James, F. \& Roos, M. 1994, {\sc minuit}, CERN, Geneva

\bibitem[{{Jha} {et~al.}(2007){Jha}, {Riess}, \& {Kirshner}}]{Jha_07}
{Jha}, S., {Riess}, A.~G., \& {Kirshner}, R.~P. 2007, \apj, 659, 122

\bibitem[{{Kasen} \& {Woosley}(2007)}]{Kasan_07}
{Kasen}, D. \& {Woosley}, S.~E. 2007, \apj, 656, 661

\bibitem[{{Kuznetsova} {et~al.}(2007){Kuznetsova}, {Barbary}, {Connolly},
  {Kim}, {Pain}, {Roe}, {Aldering}, {Amanullah}, {Dawson}, {Doi}, {Fadeyev},
  {Fruchter}, {Gibbons}, {Goldhaber}, {Goobar}, {Gude}, {Knop}, {Kowalski},
  {Lidman}, {Morokuma}, {Meyers}, {Perlmutter}, {Rubin}, {Schlegel},
  {Spadafora}, {Stanishev}, {Strovink}, {Suzuki}, {Wang}, \&
  {Yasuda}}]{Kuznetsova_07}
{Kuznetsova}, N., {Barbary}, K., {Connolly}, B., {Kim}, A.~G., {Pain}, R.,
  {Roe}, N.~A., {Aldering}, G., {Amanullah}, R., {Dawson}, K., {Doi}, M.,
  {Fadeyev}, V., {Fruchter}, A.~S., {Gibbons}, R., {Goldhaber}, G., {Goobar},
  A., {Gude}, A., {Knop}, R.~A., {Kowalski}, M., {Lidman}, C., {Morokuma}, T.,
  {Meyers}, J., {Perlmutter}, S., {Rubin}, D., {Schlegel}, D.~J., {Spadafora},
  A.~L., {Stanishev}, V., {Strovink}, M., {Suzuki}, N., {Wang}, L., \&
  {Yasuda}, N. 2007, ArXiv e-prints, 710

\bibitem[{{Li} {et~al.}(2003){Li}, {Filippenko}, {Chornock}, {Berger},
  {Berlind}, {Calkins}, {Challis}, {Fassnacht}, {Jha}, {Kirshner}, {Matheson},
  {Sargent}, {Simcoe}, {Smith}, \& {Squires}}]{Li_03}
{Li}, W., {Filippenko}, A.~V., {Chornock}, R., {Berger}, E., {Berlind}, P.,
  {Calkins}, M.~L., {Challis}, P., {Fassnacht}, C., {Jha}, S., {Kirshner},
  R.~P., {Matheson}, T., {Sargent}, W.~L.~W., {Simcoe}, R.~A., {Smith}, G.~H.,
  \& {Squires}, G. 2003, \pasp, 115, 453

\bibitem[{{Lupton} {et~al.}(2001){Lupton}, {Gunn}, {Ivezi{\'c}}, {Knapp}, \&
  {Kent}}]{Lupton_01}
{Lupton}, R., {Gunn}, J.~E., {Ivezi{\'c}}, Z., {Knapp}, G.~R., \& {Kent}, S.
  2001, in ASP Conf. Ser. 238: Astronomical Data Analysis Software and Systems
  X, ed. F.~R. {Harnden}, Jr., F.~A. {Primini}, \& H.~E. {Payne}, 269--+

\bibitem[{{Lupton} {et~al.}(1999){Lupton}, {Gunn}, \& {Szalay}}]{Lupton_99}
{Lupton}, R.~H., {Gunn}, J.~E., \& {Szalay}, A.~S. 1999, \aj, 118, 1406

\bibitem[{{Madgwick} {et~al.}(2003){Madgwick}, {Hewett}, {Mortlock}, \&
  {Wang}}]{Madgwick_03}
{Madgwick}, D.~S., {Hewett}, P.~C., {Mortlock}, D.~J., \& {Wang}, L. 2003,
  \apjl, 599, L33

\bibitem[{{Mannucci} {et~al.}(2006){Mannucci}, {Della Valle}, \&
  {Panagia}}]{Mannucci_05}
{Mannucci}, F., {Della Valle}, M., \& {Panagia}, N. 2006, \mnras, 370, 773

\bibitem[{{Mannucci} {et~al.}(2005){Mannucci}, {Della Valle}, {Panagia},
  {Cappellaro}, {Cresci}, {Maiolino}, {Petrosian}, \& {Turatto}}]{Mannucci_05a}
{Mannucci}, F., {Della Valle}, M., {Panagia}, N., {Cappellaro}, E., {Cresci},
  G., {Maiolino}, R., {Petrosian}, A., \& {Turatto}, M. 2005, \aap, 433, 807

\bibitem[{{Matheson} {et~al.}(2005){Matheson}, {Blondin}, {Foley}, {Chornock},
  {Filippenko}, {Leibundgut}, {Smith}, {Sollerman}, {Spyromilio}, {Kirshner},
  {Clocchiatti}, {Aguilera}, {Barris}, {Becker}, {Challis}, {Covarrubias},
  {Garnavich}, {Hicken}, {Jha}, {Krisciunas}, {Li}, {Miceli}, {Miknaitis},
  {Prieto}, {Rest}, {Riess}, {Salvo}, {Schmidt}, {Stubbs}, {Suntzeff}, \&
  {Tonry}}]{Matheson_05}
{Matheson}, T., {Blondin}, S., {Foley}, R.~J., {Chornock}, R., {Filippenko},
  A.~V., {Leibundgut}, B., {Smith}, R.~C., {Sollerman}, J., {Spyromilio}, J.,
  {Kirshner}, R.~P., {Clocchiatti}, A., {Aguilera}, C., {Barris}, B., {Becker},
  A.~C., {Challis}, P., {Covarrubias}, R., {Garnavich}, P., {Hicken}, M.,
  {Jha}, S., {Krisciunas}, K., {Li}, W., {Miceli}, A., {Miknaitis}, G.,
  {Prieto}, J.~L., {Rest}, A., {Riess}, A.~G., {Salvo}, M.~E., {Schmidt},
  B.~P., {Stubbs}, C.~W., {Suntzeff}, N.~B., \& {Tonry}, J.~L. 2005, \aj, 129,
  2352

\bibitem[{{Neill} {et~al.}(2006){Neill}, {Sullivan}, {Balam}, {Pritchet},
  {Howell}, {Perrett}, {Astier}, {Aubourg}, {Basa}, {Carlberg}, {Conley},
  {Fabbro}, {Fouchez}, {Guy}, {Hook}, {Pain}, {Palanque-Delabrouille},
  {Regnault}, {Rich}, {Taillet}, {Aldering}, {Antilogus}, {Arsenijevic},
  {Balland}, {Baumont}, {Bronder}, {Ellis}, {Filiol}, {Gon{\c c}alves},
  {Hardin}, {Kowalski}, {Lidman}, {Lusset}, {Mouchet}, {Mourao}, {Perlmutter},
  {Ripoche}, {Schlegel}, \& {Tao}}]{Neill_06}
{Neill}, J.~D., {Sullivan}, M., {Balam}, D., {Pritchet}, C.~J., {Howell},
  D.~A., {Perrett}, K., {Astier}, P., {Aubourg}, E., {Basa}, S., {Carlberg},
  R.~G., {Conley}, A., {Fabbro}, S., {Fouchez}, D., {Guy}, J., {Hook}, I.,
  {Pain}, R., {Palanque-Delabrouille}, N., {Regnault}, N., {Rich}, J.,
  {Taillet}, R., {Aldering}, G., {Antilogus}, P., {Arsenijevic}, V., {Balland},
  C., {Baumont}, S., {Bronder}, J., {Ellis}, R.~S., {Filiol}, M., {Gon{\c
  c}alves}, A.~C., {Hardin}, D., {Kowalski}, M., {Lidman}, C., {Lusset}, V.,
  {Mouchet}, M., {Mourao}, A., {Perlmutter}, S., {Ripoche}, P., {Schlegel}, D.,
  \& {Tao}, C. 2006, \aj, 132, 1126

\bibitem[{{Nugent} {et~al.}(2002){Nugent}, {Kim}, \& {Perlmutter}}]{Nugent_02}
{Nugent}, P., {Kim}, A., \& {Perlmutter}, S. 2002, \pasp, 114, 803

\bibitem[{{Nugent} {et~al.}(1995){Nugent}, {Phillips}, {Baron}, {Branch}, \&
  {Hauschildt}}]{Nugent_95}
{Nugent}, P., {Phillips}, M., {Baron}, E., {Branch}, D., \& {Hauschildt}, P.
  1995, \apjl, 455, L147+

\bibitem[{{Oyaizu} {et~al.}(2007){Oyaizu}, {Lima}, {Cunha}, {Lin}, {Frieman},
  \& {Sheldon}}]{Oyaizu_07}
{Oyaizu}, H., {Lima}, M., {Cunha}, C.~E., {Lin}, H., {Frieman}, J., \&
  {Sheldon}, E.~S. 2007, astro-ph/0708.0030

\bibitem[{{Pain} {et~al.}(2002){Pain}, {Fabbro}, {Sullivan}, {Ellis},
  {Aldering}, {Astier}, {Deustua}, {Fruchter}, {Goldhaber}, {Goobar}, {Groom},
  {Hardin}, {Hook}, {Howell}, {Irwin}, {Kim}, {Kim}, {Knop}, {Lee}, {Lidman},
  {McMahon}, {Nugent}, {Panagia}, {Pennypacker}, {Perlmutter}, {Ruiz-Lapuente},
  {Schahmaneche}, {Schaefer}, \& {Walton}}]{Pain_02}
{Pain}, R., {Fabbro}, S., {Sullivan}, M., {Ellis}, R.~S., {Aldering}, G.,
  {Astier}, P., {Deustua}, S.~E., {Fruchter}, A.~S., {Goldhaber}, G., {Goobar},
  A., {Groom}, D.~E., {Hardin}, D., {Hook}, I.~M., {Howell}, D.~A., {Irwin},
  M.~J., {Kim}, A.~G., {Kim}, M.~Y., {Knop}, R.~A., {Lee}, J.~C., {Lidman}, C.,
  {McMahon}, R.~G., {Nugent}, P.~E., {Panagia}, N., {Pennypacker}, C.~R.,
  {Perlmutter}, S., {Ruiz-Lapuente}, P., {Schahmaneche}, K., {Schaefer}, B., \&
  {Walton}, N.~A. 2002, \apj, 577, 120

\bibitem[{{Phillips}(1993)}]{Phillips_93}
{Phillips}, M.~M. 1993, \apjl, 413, L105

\bibitem[{{Phillips} {et~al.}(2007){Phillips}, {Li}, {Frieman}, {Blinnikov},
  {DePoy}, {Prieto}, {Milne}, {Contreras}, {Folatelli}, {Morrell}, {Hamuy},
  {Suntzeff}, {Roth}, {Gonz{\'a}lez}, {Krzeminski}, {Filippenko}, {Freedman},
  {Chornock}, {Jha}, {Madore}, {Persson}, {Burns}, {Wyatt}, {Murphy}, {Foley},
  {Ganeshalingam}, {Serduke}, {Krisciunas}, {Bassett}, {Becker}, {Dilday},
  {Eastman}, {Garnavich}, {Holtzman}, {Kessler}, {Lampeitl}, {Marriner},
  {Frank}, {Marshall}, {Miknaitis}, {Sako}, {Schneider}, {van der Heyden}, \&
  {Yasuda}}]{Phillips_07}
{Phillips}, M.~M., {Li}, W., {Frieman}, J.~A., {Blinnikov}, S.~I., {DePoy}, D.,
  {Prieto}, J.~L., {Milne}, P., {Contreras}, C., {Folatelli}, G., {Morrell},
  N., {Hamuy}, M., {Suntzeff}, N.~B., {Roth}, M., {Gonz{\'a}lez}, S.,
  {Krzeminski}, W., {Filippenko}, A.~V., {Freedman}, W.~L., {Chornock}, R.,
  {Jha}, S., {Madore}, B.~F., {Persson}, S.~E., {Burns}, C.~R., {Wyatt}, P.,
  {Murphy}, D., {Foley}, R.~J., {Ganeshalingam}, M., {Serduke}, F.~J.~D.,
  {Krisciunas}, K., {Bassett}, B., {Becker}, A., {Dilday}, B., {Eastman}, J.,
  {Garnavich}, P.~M., {Holtzman}, J., {Kessler}, R., {Lampeitl}, H.,
  {Marriner}, J., {Frank}, S., {Marshall}, J.~L., {Miknaitis}, G., {Sako}, M.,
  {Schneider}, D.~P., {van der Heyden}, K., \& {Yasuda}, N. 2007, \pasp, 119,
  360

\bibitem[{{Pier} {et~al.}(2003){Pier}, {Munn}, {Hindsley}, {Hennessy}, {Kent},
  {Lupton}, \& {Ivezi{\'c}}}]{Pier_03}
{Pier}, J.~R., {Munn}, J.~A., {Hindsley}, R.~B., {Hennessy}, G.~S., {Kent},
  S.~M., {Lupton}, R.~H., \& {Ivezi{\'c}}, {\v Z}. 2003, \aj, 125, 1559

\bibitem[{{Poznanski} {et~al.}(2007){Poznanski}, {Maoz}, {Yasuda}, {Foley},
  {Doi}, {Filippenko}, {Fukugita}, {Gal-Yam}, {Jannuzi}, {Morokuma}, {Oda},
  {Schweiker}, {Sharon}, {Silverman}, \& {Totani}}]{Poznanski_07}
{Poznanski}, D., {Maoz}, D., {Yasuda}, N., {Foley}, R.~J., {Doi}, M.,
  {Filippenko}, A.~V., {Fukugita}, M., {Gal-Yam}, A., {Jannuzi}, B.~T.,
  {Morokuma}, T., {Oda}, T., {Schweiker}, H., {Sharon}, K., {Silverman}, J.~M.,
  \& {Totani}, T. 2007, \mnras, 382, 1169

\bibitem[{{Prieto} {et~al.}(2007){Prieto}, {Garnavich}, {Phillips}, {DePoy},
  {Parrent}, {Pooley}, {Dwarkadas}, {Baron}, {Bassett}, {Becker}, {Cinabro},
  {DeJongh}, {Dilday}, {Doi}, {Frieman}, {Hogan}, {Holtzman}, {Jha}, {Kessler},
  {Konishi}, {Lampeitl}, {Marriner}, {Marshall}, {Miknaitis}, {Nichol},
  {Riess}, {Richmond}, {Romani}, {Sako}, {Schneider}, {Smith}, {Takanashi},
  {Tokita}, {van der Heyden}, {Yasuda}, {Zheng}, {Wheeler}, {Barentine},
  {Dembicky}, {Eastman}, {Frank}, {Ketzeback}, {McMillan}, {Morrell},
  {Folatelli}, {Contreras}, {Burns}, {Freedman}, {Gonzalez}, {Hamuy},
  {Krzeminski}, {Madore}, {Murphy}, {Persson}, {Roth}, \&
  {Suntzeff}}]{Prieto_07}
{Prieto}, J.~L., {Garnavich}, P.~M., {Phillips}, M.~M., {DePoy}, D.~L.,
  {Parrent}, J., {Pooley}, D., {Dwarkadas}, V.~V., {Baron}, E., {Bassett}, B.,
  {Becker}, A., {Cinabro}, D., {DeJongh}, F., {Dilday}, B., {Doi}, M.,
  {Frieman}, J.~A., {Hogan}, C.~J., {Holtzman}, J., {Jha}, S., {Kessler}, R.,
  {Konishi}, K., {Lampeitl}, H., {Marriner}, J., {Marshall}, J.~L.,
  {Miknaitis}, G., {Nichol}, R.~C., {Riess}, A.~G., {Richmond}, M.~W.,
  {Romani}, R., {Sako}, M., {Schneider}, D.~P., {Smith}, M., {Takanashi}, N.,
  {Tokita}, K., {van der Heyden}, K., {Yasuda}, N., {Zheng}, C., {Wheeler},
  J.~C., {Barentine}, J., {Dembicky}, J., {Eastman}, J., {Frank}, S.,
  {Ketzeback}, W., {McMillan}, R.~J., {Morrell}, N., {Folatelli}, G.,
  {Contreras}, C., {Burns}, C.~R., {Freedman}, W.~L., {Gonzalez}, S., {Hamuy},
  M., {Krzeminski}, W., {Madore}, B.~F., {Murphy}, D., {Persson}, S.~E.,
  {Roth}, M., \& {Suntzeff}, N.~B. 2007, astro-ph/0706.4088

\bibitem[{{Pskovskii}(1977)}]{Psk_77}
{Pskovskii}, I.~P. 1977, Soviet Astronomy, 21, 675

\bibitem[{{Quimby} {et~al.}(2005){Quimby}, {H{\"o}flich}, {Kannappan},
  {Odewahn}, \& {Terrazas}}]{CBET_266}
{Quimby}, R., {H{\"o}flich}, P., {Kannappan}, S.~J., {Odewahn}, S.~C., \&
  {Terrazas}, E. 2005, Central Bureau Electronic Telegrams, 266, 1

\bibitem[{{Riess} {et~al.}(1996){Riess}, {Press}, \& {Kirshner}}]{Riess_96}
{Riess}, A.~G., {Press}, W.~H., \& {Kirshner}, R.~P. 1996, \apj, 473, 88

\bibitem[{{Sako} {et~al.}(2008){Sako}, {Bassett}, {Becker}, {Cinabro},
  {DeJongh}, {Depoy}, {Dilday}, {Doi}, {Frieman}, {Garnavich}, {Hogan},
  {Holtzman}, {Jha}, {Kessler}, {Konishi}, {Lampeitl}, {Marriner}, {Miknaitis},
  {Nichol}, {Prieto}, {Riess}, {Richmond}, {Romani}, {Schneider}, {Smith},
  {Subba Rao}, {Takanashi}, {Tokita}, {van der Heyden}, {Yasuda}, {Zheng},
  {Barentine}, {Brewington}, {Choi}, {Dembicky}, {Harnavek}, {Ihara}, {Im},
  {Ketzeback}, {Kleinman}, {Krzesi{\'n}ski}, {Long}, {Malanushenko},
  {Malanushenko}, {McMillan}, {Morokuma}, {Nitta}, {Pan}, {Saurage}, \&
  {Snedden}}]{Sako_08}
{Sako}, M., {Bassett}, B., {Becker}, A., {Cinabro}, D., {DeJongh}, F., {Depoy},
  D.~L., {Dilday}, B., {Doi}, M., {Frieman}, J.~A., {Garnavich}, P.~M.,
  {Hogan}, C.~J., {Holtzman}, J., {Jha}, S., {Kessler}, R., {Konishi}, K.,
  {Lampeitl}, H., {Marriner}, J., {Miknaitis}, G., {Nichol}, R.~C., {Prieto},
  J.~L., {Riess}, A.~G., {Richmond}, M.~W., {Romani}, R., {Schneider}, D.~P.,
  {Smith}, M., {Subba Rao}, M., {Takanashi}, N., {Tokita}, K., {van der
  Heyden}, K., {Yasuda}, N., {Zheng}, C., {Barentine}, J., {Brewington}, H.,
  {Choi}, C., {Dembicky}, J., {Harnavek}, M., {Ihara}, Y., {Im}, M.,
  {Ketzeback}, W., {Kleinman}, S.~J., {Krzesi{\'n}ski}, J., {Long}, D.~C.,
  {Malanushenko}, E., {Malanushenko}, V., {McMillan}, R.~J., {Morokuma}, T.,
  {Nitta}, A., {Pan}, K., {Saurage}, G., \& {Snedden}, S.~A. 2008, \aj, 135,
  348

\bibitem[{{Scannapieco} \& {Bildsten}(2005)}]{ScanBildsten}
{Scannapieco}, E. \& {Bildsten}, L. 2005, \apjl, 629, L85

\bibitem[{{Schechter} {et~al.}(1993){Schechter}, {Mateo}, \&
  {Saha}}]{Schecter_93}
{Schechter}, P.~L., {Mateo}, M., \& {Saha}, A. 1993, \pasp, 105, 1342

\bibitem[{{Schlegel} {et~al.}(1998){Schlegel}, {Finkbeiner}, \&
  {Davis}}]{Schlegel_98}
{Schlegel}, D.~J., {Finkbeiner}, D.~P., \& {Davis}, M. 1998, \apj, 500, 525

\bibitem[{{Shimasaku} {et~al.}(2001){Shimasaku}, {Fukugita}, {Doi}, {Hamabe},
  {Ichikawa}, {Okamura}, {Sekiguchi}, {Yasuda}, {Brinkmann}, {Csabai},
  {Ichikawa}, {Ivezi{\'c}}, {Kunszt}, {Schneider}, {Szokoly}, {Watanabe}, \&
  {York}}]{shimasaku_01}
{Shimasaku}, K., {Fukugita}, M., {Doi}, M., {Hamabe}, M., {Ichikawa}, T.,
  {Okamura}, S., {Sekiguchi}, M., {Yasuda}, N., {Brinkmann}, J., {Csabai}, I.,
  {Ichikawa}, S.-I., {Ivezi{\'c}}, Z., {Kunszt}, P.~Z., {Schneider}, D.~P.,
  {Szokoly}, G.~P., {Watanabe}, M., \& {York}, D.~G. 2001, \aj, 122, 1238

\bibitem[{{Smith} {et~al.}(2002){Smith}, {Tucker}, {Kent}, {Richmond},
  {Fukugita}, {Ichikawa}, {Ichikawa}, {Jorgensen}, {Uomoto}, {Gunn}, {Hamabe},
  {Watanabe}, {Tolea}, {Henden}, {Annis}, {Pier}, {McKay}, {Brinkmann}, {Chen},
  {Holtzman}, {Shimasaku}, \& {York}}]{Smith_02}
{Smith}, J.~A., {Tucker}, D.~L., {Kent}, S., {Richmond}, M.~W., {Fukugita}, M.,
  {Ichikawa}, T., {Ichikawa}, S.-i., {Jorgensen}, A.~M., {Uomoto}, A., {Gunn},
  J.~E., {Hamabe}, M., {Watanabe}, M., {Tolea}, A., {Henden}, A., {Annis}, J.,
  {Pier}, J.~R., {McKay}, T.~A., {Brinkmann}, J., {Chen}, B., {Holtzman}, J.,
  {Shimasaku}, K., \& {York}, D.~G. 2002, \aj, 123, 2121

\bibitem[{{Stoughton} {et~al.}(2002){Stoughton}, {Lupton}, {Bernardi},
  {Blanton}, {Burles}, {Castander}, {Connolly}, {Eisenstein}, {Frieman},
  {Hennessy}, {Hindsley}, {Ivezi{\'c}}, {Kent}, {Kunszt}, {Lee}, {Meiksin},
  {Munn}, {Newberg}, {Nichol}, {Nicinski}, {Pier}, {Richards}, {Richmond},
  {Schlegel}, {Smith}, {Strauss}, {SubbaRao}, {Szalay}, {Thakar}, {Tucker},
  {Vanden Berk}, {Yanny}, {Adelman}, {Anderson}, {Anderson}, {Annis},
  {Bahcall}, {Bakken}, {Bartelmann}, {Bastian}, {Bauer}, {Berman},
  {B{\"o}hringer}, {Boroski}, {Bracker}, {Briegel}, {Briggs}, {Brinkmann},
  {Brunner}, {Carey}, {Carr}, {Chen}, {Christian}, {Colestock}, {Crocker},
  {Csabai}, {Czarapata}, {Dalcanton}, {Davidsen}, {Davis}, {Dehnen},
  {Dodelson}, {Doi}, {Dombeck}, {Donahue}, {Ellman}, {Elms}, {Evans}, {Eyer},
  {Fan}, {Federwitz}, {Friedman}, {Fukugita}, {Gal}, {Gillespie}, {Glazebrook},
  {Gray}, {Grebel}, {Greenawalt}, {Greene}, {Gunn}, {de Haas}, {Haiman},
  {Haldeman}, {Hall}, {Hamabe}, {Hansen}, {Harris}, {Harris}, {Harvanek},
  {Hawley}, {Hayes}, {Heckman}, {Helmi}, {Henden}, {Hogan}, {Hogg}, {Holmgren},
  {Holtzman}, {Huang}, {Hull}, {Ichikawa}, {Ichikawa}, {Johnston}, {Kauffmann},
  {Kim}, {Kimball}, {Kinney}, {Klaene}, {Kleinman}, {Klypin}, {Knapp},
  {Korienek}, {Krolik}, {Kron}, {Krzesi{\'n}ski}, {Lamb}, {Leger},
  {Limmongkol}, {Lindenmeyer}, {Long}, {Loomis}, {Loveday}, {MacKinnon},
  {Mannery}, {Mantsch}, {Margon}, {McGehee}, {McKay}, {McLean}, {Menou},
  {Merelli}, {Mo}, {Monet}, {Nakamura}, {Narayanan}, {Nash}, {Neilsen},
  {Newman}, {Nitta}, {Odenkirchen}, {Okada}, {Okamura}, {Ostriker}, {Owen},
  {Pauls}, {Peoples}, {Peterson}, {Petravick}, {Pope}, {Pordes}, {Postman},
  {Prosapio}, {Quinn}, {Rechenmacher}, {Rivetta}, {Rix}, {Rockosi}, {Rosner},
  {Ruthmansdorfer}, {Sandford}, {Schneider}, {Scranton}, {Sekiguchi}, {Sergey},
  {Sheth}, {Shimasaku}, {Smee}, {Snedden}, {Stebbins}, {Stubbs}, {Szapudi},
  {Szkody}, {Szokoly}, {Tabachnik}, {Tsvetanov}, {Uomoto}, {Vogeley}, {Voges},
  {Waddell}, {Walterbos}, {Wang}, {Watanabe}, {Weinberg}, {White}, {White},
  {Wilhite}, {Wolfe}, {Yasuda}, {York}, {Zehavi}, \& {Zheng}}]{SDSS_EDR}
{Stoughton}, C., {Lupton}, R.~H., {Bernardi}, M., {Blanton}, M.~R., {Burles},
  S., {Castander}, F.~J., {Connolly}, A.~J., {Eisenstein}, D.~J., {Frieman},
  J.~A., {Hennessy}, G.~S., {Hindsley}, R.~B., {Ivezi{\'c}}, {\v Z}., {Kent},
  S., {Kunszt}, P.~Z., {Lee}, B.~C., {Meiksin}, A., {Munn}, J.~A., {Newberg},
  H.~J., {Nichol}, R.~C., {Nicinski}, T., {Pier}, J.~R., {Richards}, G.~T.,
  {Richmond}, M.~W., {Schlegel}, D.~J., {Smith}, J.~A., {Strauss}, M.~A.,
  {SubbaRao}, M., {Szalay}, A.~S., {Thakar}, A.~R., {Tucker}, D.~L., {Vanden
  Berk}, D.~E., {Yanny}, B., {Adelman}, J.~K., {Anderson}, Jr., J.~E.,
  {Anderson}, S.~F., {Annis}, J., {Bahcall}, N.~A., {Bakken}, J.~A.,
  {Bartelmann}, M., {Bastian}, S., {Bauer}, A., {Berman}, E., {B{\"o}hringer},
  H., {Boroski}, W.~N., {Bracker}, S., {Briegel}, C., {Briggs}, J.~W.,
  {Brinkmann}, J., {Brunner}, R., {Carey}, L., {Carr}, M.~A., {Chen}, B.,
  {Christian}, D., {Colestock}, P.~L., {Crocker}, J.~H., {Csabai}, I.,
  {Czarapata}, P.~C., {Dalcanton}, J., {Davidsen}, A.~F., {Davis}, J.~E.,
  {Dehnen}, W., {Dodelson}, S., {Doi}, M., {Dombeck}, T., {Donahue}, M.,
  {Ellman}, N., {Elms}, B.~R., {Evans}, M.~L., {Eyer}, L., {Fan}, X.,
  {Federwitz}, G.~R., {Friedman}, S., {Fukugita}, M., {Gal}, R., {Gillespie},
  B., {Glazebrook}, K., {Gray}, J., {Grebel}, E.~K., {Greenawalt}, B.,
  {Greene}, G., {Gunn}, J.~E., {de Haas}, E., {Haiman}, Z., {Haldeman}, M.,
  {Hall}, P.~B., {Hamabe}, M., {Hansen}, B., {Harris}, F.~H., {Harris}, H.,
  {Harvanek}, M., {Hawley}, S.~L., {Hayes}, J.~J.~E., {Heckman}, T.~M.,
  {Helmi}, A., {Henden}, A., {Hogan}, C.~J., {Hogg}, D.~W., {Holmgren}, D.~J.,
  {Holtzman}, J., {Huang}, C.-H., {Hull}, C., {Ichikawa}, S.-I., {Ichikawa},
  T., {Johnston}, D.~E., {Kauffmann}, G., {Kim}, R.~S.~J., {Kimball}, T.,
  {Kinney}, E., {Klaene}, M., {Kleinman}, S.~J., {Klypin}, A., {Knapp}, G.~R.,
  {Korienek}, J., {Krolik}, J., {Kron}, R.~G., {Krzesi{\'n}ski}, J., {Lamb},
  D.~Q., {Leger}, R.~F., {Limmongkol}, S., {Lindenmeyer}, C., {Long}, D.~C.,
  {Loomis}, C., {Loveday}, J., {MacKinnon}, B., {Mannery}, E.~J., {Mantsch},
  P.~M., {Margon}, B., {McGehee}, P., {McKay}, T.~A., {McLean}, B., {Menou},
  K., {Merelli}, A., {Mo}, H.~J., {Monet}, D.~G., {Nakamura}, O., {Narayanan},
  V.~K., {Nash}, T., {Neilsen}, Jr., E.~H., {Newman}, P.~R., {Nitta}, A.,
  {Odenkirchen}, M., {Okada}, N., {Okamura}, S., {Ostriker}, J.~P., {Owen}, R.,
  {Pauls}, A.~G., {Peoples}, J., {Peterson}, R.~S., {Petravick}, D., {Pope},
  A., {Pordes}, R., {Postman}, M., {Prosapio}, A., {Quinn}, T.~R.,
  {Rechenmacher}, R., {Rivetta}, C.~H., {Rix}, H.-W., {Rockosi}, C.~M.,
  {Rosner}, R., {Ruthmansdorfer}, K., {Sandford}, D., {Schneider}, D.~P.,
  {Scranton}, R., {Sekiguchi}, M., {Sergey}, G., {Sheth}, R., {Shimasaku}, K.,
  {Smee}, S., {Snedden}, S.~A., {Stebbins}, A., {Stubbs}, C., {Szapudi}, I.,
  {Szkody}, P., {Szokoly}, G.~P., {Tabachnik}, S., {Tsvetanov}, Z., {Uomoto},
  A., {Vogeley}, M.~S., {Voges}, W., {Waddell}, P., {Walterbos}, R., {Wang},
  S.-i., {Watanabe}, M., {Weinberg}, D.~H., {White}, R.~L., {White}, S.~D.~M.,
  {Wilhite}, B., {Wolfe}, D., {Yasuda}, N., {York}, D.~G., {Zehavi}, I., \&
  {Zheng}, W. 2002, \aj, 123, 485

\bibitem[{{Strateva} {et~al.}(2001){Strateva}, {Ivezi{\'c}}, {Knapp},
  {Narayanan}, {Strauss}, {Gunn}, {Lupton}, {Schlegel}, {Bahcall}, {Brinkmann},
  {Brunner}, {Budav{\'a}ri}, {Csabai}, {Castander}, {Doi}, {Fukugita}, {Gy{\H
  o}ry}, {Hamabe}, {Hennessy}, {Ichikawa}, {Kunszt}, {Lamb}, {McKay},
  {Okamura}, {Racusin}, {Sekiguchi}, {Schneider}, {Shimasaku}, \&
  {York}}]{Strateva_01}
{Strateva}, I., {Ivezi{\'c}}, {\v Z}., {Knapp}, G.~R., {Narayanan}, V.~K.,
  {Strauss}, M.~A., {Gunn}, J.~E., {Lupton}, R.~H., {Schlegel}, D., {Bahcall},
  N.~A., {Brinkmann}, J., {Brunner}, R.~J., {Budav{\'a}ri}, T., {Csabai}, I.,
  {Castander}, F.~J., {Doi}, M., {Fukugita}, M., {Gy{\H o}ry}, Z., {Hamabe},
  M., {Hennessy}, G., {Ichikawa}, T., {Kunszt}, P.~Z., {Lamb}, D.~Q., {McKay},
  T.~A., {Okamura}, S., {Racusin}, J., {Sekiguchi}, M., {Schneider}, D.~P.,
  {Shimasaku}, K., \& {York}, D. 2001, \aj, 122, 1861

\bibitem[{{Strolger} \& {Riess}(2006)}]{Strolger_06}
{Strolger}, L.-G. \& {Riess}, A.~G. 2006, \aj, 131, 1629

\bibitem[{{Strolger} {et~al.}(2004){Strolger}, {Riess}, {Dahlen}, {Livio},
  {Panagia}, {Challis}, {Tonry}, {Filippenko}, {Chornock}, {Ferguson},
  {Koekemoer}, {Mobasher}, {Dickinson}, {Giavalisco}, {Casertano}, {Hook},
  {Blondin}, {Leibundgut}, {Nonino}, {Rosati}, {Spinrad}, {Steidel}, {Stern},
  {Garnavich}, {Matheson}, {Grogin}, {Hornschemeier}, {Kretchmer}, {Laidler},
  {Lee}, {Lucas}, {de Mello}, {Moustakas}, {Ravindranath}, {Richardson}, \&
  {Taylor}}]{Strolger_04}
{Strolger}, L.-G., {Riess}, A.~G., {Dahlen}, T., {Livio}, M., {Panagia}, N.,
  {Challis}, P., {Tonry}, J.~L., {Filippenko}, A.~V., {Chornock}, R.,
  {Ferguson}, H., {Koekemoer}, A., {Mobasher}, B., {Dickinson}, M.,
  {Giavalisco}, M., {Casertano}, S., {Hook}, R., {Blondin}, S., {Leibundgut},
  B., {Nonino}, M., {Rosati}, P., {Spinrad}, H., {Steidel}, C.~C., {Stern}, D.,
  {Garnavich}, P.~M., {Matheson}, T., {Grogin}, N., {Hornschemeier}, A.,
  {Kretchmer}, C., {Laidler}, V.~G., {Lee}, K., {Lucas}, R., {de Mello}, D.,
  {Moustakas}, L.~A., {Ravindranath}, S., {Richardson}, M., \& {Taylor}, E.
  2004, \apj, 613, 200

\bibitem[{{Sullivan} {et~al.}(2006){Sullivan}, {Le Borgne}, {Pritchet},
  {Hodsman}, {Neill}, {Howell}, {Carlberg}, {Astier}, {Aubourg}, {Balam},
  {Basa}, {Conley}, {Fabbro}, {Fouchez}, {Guy}, {Hook}, {Pain},
  {Palanque-Delabrouille}, {Perrett}, {Regnault}, {Rich}, {Taillet}, {Baumont},
  {Bronder}, {Ellis}, {Filiol}, {Lusset}, {Perlmutter}, {Ripoche}, \&
  {Tao}}]{Sullivan_06}
{Sullivan}, M., {Le Borgne}, D., {Pritchet}, C.~J., {Hodsman}, A., {Neill},
  J.~D., {Howell}, D.~A., {Carlberg}, R.~G., {Astier}, P., {Aubourg}, E.,
  {Balam}, D., {Basa}, S., {Conley}, A., {Fabbro}, S., {Fouchez}, D., {Guy},
  J., {Hook}, I., {Pain}, R., {Palanque-Delabrouille}, N., {Perrett}, K.,
  {Regnault}, N., {Rich}, J., {Taillet}, R., {Baumont}, S., {Bronder}, J.,
  {Ellis}, R.~S., {Filiol}, M., {Lusset}, V., {Perlmutter}, S., {Ripoche}, P.,
  \& {Tao}, C. 2006, \apj, 648, 868

\bibitem[{{Tonry} \& {Davis}(1979)}]{Tonry_79}
{Tonry}, J. \& {Davis}, M. 1979, \aj, 84, 1511

\bibitem[{{Tonry} {et~al.}(2003){Tonry}, {Schmidt}, {Barris}, {Candia},
  {Challis}, {Clocchiatti}, {Coil}, {Filippenko}, {Garnavich}, {Hogan},
  {Holland}, {Jha}, {Kirshner}, {Krisciunas}, {Leibundgut}, {Li}, {Matheson},
  {Phillips}, {Riess}, {Schommer}, {Smith}, {Sollerman}, {Spyromilio},
  {Stubbs}, \& {Suntzeff}}]{Tonry_03}
{Tonry}, J.~L., {Schmidt}, B.~P., {Barris}, B., {Candia}, P., {Challis}, P.,
  {Clocchiatti}, A., {Coil}, A.~L., {Filippenko}, A.~V., {Garnavich}, P.,
  {Hogan}, C., {Holland}, S.~T., {Jha}, S., {Kirshner}, R.~P., {Krisciunas},
  K., {Leibundgut}, B., {Li}, W., {Matheson}, T., {Phillips}, M.~M., {Riess},
  A.~G., {Schommer}, R., {Smith}, R.~C., {Sollerman}, J., {Spyromilio}, J.,
  {Stubbs}, C.~W., \& {Suntzeff}, N.~B. 2003, \apj, 594, 1

\bibitem[{{Tucker} {et~al.}(2006){Tucker}, {Kent}, {Richmond}, {Annis},
  {Smith}, {Allam}, {Rodgers}, {Stute}, {Adelman-McCarthy}, {Brinkmann}, {Doi},
  {Finkbeiner}, {Fukugita}, {Goldston}, {Greenway}, {Gunn}, {Hendry}, {Hogg},
  {Ichikawa}, {Ivezi{\'c}}, {Knapp}, {Lampeitl}, {Lee}, {Lin}, {McKay},
  {Merrelli}, {Munn}, {Neilsen}, {Newberg}, {Richards}, {Schlegel},
  {Stoughton}, {Uomoto}, \& {Yanny}}]{Tucker_06}
{Tucker}, D.~L., {Kent}, S., {Richmond}, M.~W., {Annis}, J., {Smith}, J.~A.,
  {Allam}, S.~S., {Rodgers}, C.~T., {Stute}, J.~L., {Adelman-McCarthy}, J.~K.,
  {Brinkmann}, J., {Doi}, M., {Finkbeiner}, D., {Fukugita}, M., {Goldston}, J.,
  {Greenway}, B., {Gunn}, J.~E., {Hendry}, J.~S., {Hogg}, D.~W., {Ichikawa},
  S.-I., {Ivezi{\'c}}, {\v Z}., {Knapp}, G.~R., {Lampeitl}, H., {Lee}, B.~C.,
  {Lin}, H., {McKay}, T.~A., {Merrelli}, A., {Munn}, J.~A., {Neilsen}, Jr.,
  E.~H., {Newberg}, H.~J., {Richards}, G.~T., {Schlegel}, D.~J., {Stoughton},
  C., {Uomoto}, A., \& {Yanny}, B. 2006, Astronomische Nachrichten, 327, 821

\bibitem[{{Yamauchi} {et~al.}(2005){Yamauchi}, {Ichikawa}, {Doi}, {Yasuda},
  {Yagi}, {Fukugita}, {Okamura}, {Nakamura}, {Sekiguchi}, \&
  {Goto}}]{yamauchi_05}
{Yamauchi}, C., {Ichikawa}, S.-i., {Doi}, M., {Yasuda}, N., {Yagi}, M.,
  {Fukugita}, M., {Okamura}, S., {Nakamura}, O., {Sekiguchi}, M., \& {Goto}, T.
  2005, \aj, 130, 1545

\bibitem[{{York} {et~al.}(2000){York}, {Adelman}, {Anderson}, {Anderson},
  {Annis}, {Bahcall}, {Bakken}, {Barkhouser}, {Bastian}, {Berman}, {Boroski},
  {Bracker}, {Briegel}, {Briggs}, {Brinkmann}, {Brunner}, {Burles}, {Carey},
  {Carr}, {Castander}, {Chen}, {Colestock}, {Connolly}, {Crocker}, {Csabai},
  {Czarapata}, {Davis}, {Doi}, {Dombeck}, {Eisenstein}, {Ellman}, {Elms},
  {Evans}, {Fan}, {Federwitz}, {Fiscelli}, {Friedman}, {Frieman}, {Fukugita},
  {Gillespie}, {Gunn}, {Gurbani}, {de Haas}, {Haldeman}, {Harris}, {Hayes},
  {Heckman}, {Hennessy}, {Hindsley}, {Holm}, {Holmgren}, {Huang}, {Hull},
  {Husby}, {Ichikawa}, {Ichikawa}, {Ivezi{\'c}}, {Kent}, {Kim}, {Kinney},
  {Klaene}, {Kleinman}, {Kleinman}, {Knapp}, {Korienek}, {Kron}, {Kunszt},
  {Lamb}, {Lee}, {Leger}, {Limmongkol}, {Lindenmeyer}, {Long}, {Loomis},
  {Loveday}, {Lucinio}, {Lupton}, {MacKinnon}, {Mannery}, {Mantsch}, {Margon},
  {McGehee}, {McKay}, {Meiksin}, {Merelli}, {Monet}, {Munn}, {Narayanan},
  {Nash}, {Neilsen}, {Neswold}, {Newberg}, {Nichol}, {Nicinski}, {Nonino},
  {Okada}, {Okamura}, {Ostriker}, {Owen}, {Pauls}, {Peoples}, {Peterson},
  {Petravick}, {Pier}, {Pope}, {Pordes}, {Prosapio}, {Rechenmacher}, {Quinn},
  {Richards}, {Richmond}, {Rivetta}, {Rockosi}, {Ruthmansdorfer}, {Sandford},
  {Schlegel}, {Schneider}, {Sekiguchi}, {Sergey}, {Shimasaku}, {Siegmund},
  {Smee}, {Smith}, {Snedden}, {Stone}, {Stoughton}, {Strauss}, {Stubbs},
  {SubbaRao}, {Szalay}, {Szapudi}, {Szokoly}, {Thakar}, {Tremonti}, {Tucker},
  {Uomoto}, {Vanden Berk}, {Vogeley}, {Waddell}, {Wang}, {Watanabe},
  {Weinberg}, {Yanny}, \& {Yasuda}}]{York_00}
{York}, D.~G., {Adelman}, J., {Anderson}, Jr., J.~E., {Anderson}, S.~F.,
  {Annis}, J., {Bahcall}, N.~A., {Bakken}, J.~A., {Barkhouser}, R., {Bastian},
  S., {Berman}, E., {Boroski}, W.~N., {Bracker}, S., {Briegel}, C., {Briggs},
  J.~W., {Brinkmann}, J., {Brunner}, R., {Burles}, S., {Carey}, L., {Carr},
  M.~A., {Castander}, F.~J., {Chen}, B., {Colestock}, P.~L., {Connolly}, A.~J.,
  {Crocker}, J.~H., {Csabai}, I., {Czarapata}, P.~C., {Davis}, J.~E., {Doi},
  M., {Dombeck}, T., {Eisenstein}, D., {Ellman}, N., {Elms}, B.~R., {Evans},
  M.~L., {Fan}, X., {Federwitz}, G.~R., {Fiscelli}, L., {Friedman}, S.,
  {Frieman}, J.~A., {Fukugita}, M., {Gillespie}, B., {Gunn}, J.~E., {Gurbani},
  V.~K., {de Haas}, E., {Haldeman}, M., {Harris}, F.~H., {Hayes}, J.,
  {Heckman}, T.~M., {Hennessy}, G.~S., {Hindsley}, R.~B., {Holm}, S.,
  {Holmgren}, D.~J., {Huang}, C.-h., {Hull}, C., {Husby}, D., {Ichikawa},
  S.-I., {Ichikawa}, T., {Ivezi{\'c}}, {\v Z}., {Kent}, S., {Kim}, R.~S.~J.,
  {Kinney}, E., {Klaene}, M., {Kleinman}, A.~N., {Kleinman}, S., {Knapp},
  G.~R., {Korienek}, J., {Kron}, R.~G., {Kunszt}, P.~Z., {Lamb}, D.~Q., {Lee},
  B., {Leger}, R.~F., {Limmongkol}, S., {Lindenmeyer}, C., {Long}, D.~C.,
  {Loomis}, C., {Loveday}, J., {Lucinio}, R., {Lupton}, R.~H., {MacKinnon}, B.,
  {Mannery}, E.~J., {Mantsch}, P.~M., {Margon}, B., {McGehee}, P., {McKay},
  T.~A., {Meiksin}, A., {Merelli}, A., {Monet}, D.~G., {Munn}, J.~A.,
  {Narayanan}, V.~K., {Nash}, T., {Neilsen}, E., {Neswold}, R., {Newberg},
  H.~J., {Nichol}, R.~C., {Nicinski}, T., {Nonino}, M., {Okada}, N., {Okamura},
  S., {Ostriker}, J.~P., {Owen}, R., {Pauls}, A.~G., {Peoples}, J., {Peterson},
  R.~L., {Petravick}, D., {Pier}, J.~R., {Pope}, A., {Pordes}, R., {Prosapio},
  A., {Rechenmacher}, R., {Quinn}, T.~R., {Richards}, G.~T., {Richmond}, M.~W.,
  {Rivetta}, C.~H., {Rockosi}, C.~M., {Ruthmansdorfer}, K., {Sandford}, D.,
  {Schlegel}, D.~J., {Schneider}, D.~P., {Sekiguchi}, M., {Sergey}, G.,
  {Shimasaku}, K., {Siegmund}, W.~A., {Smee}, S., {Smith}, J.~A., {Snedden},
  S., {Stone}, R., {Stoughton}, C., {Strauss}, M.~A., {Stubbs}, C., {SubbaRao},
  M., {Szalay}, A.~S., {Szapudi}, I., {Szokoly}, G.~P., {Thakar}, A.~R.,
  {Tremonti}, C., {Tucker}, D.~L., {Uomoto}, A., {Vanden Berk}, D., {Vogeley},
  M.~S., {Waddell}, P., {Wang}, S.-i., {Watanabe}, M., {Weinberg}, D.~H.,
  {Yanny}, B., \& {Yasuda}, N. 2000, \aj, 120, 1579

\bibitem[{{Zheng et al.}(2008)}]{Zheng_08}
{Zheng et al.}, C. 2008, submitted to AJ

\end{thebibliography}
